\documentclass[12pt]{article}
\usepackage{graphicx} 
\def\laq{\raise 0.4ex\hbox{$<$}\kern -0.8em\lower 0.62
ex\hbox{$\sim$}}
\def\gaq{\raise 0.4ex\hbox{$>$}\kern -0.7em\lower 0.62
ex\hbox{$\sim$}}

\begin{document}

\begin{titlepage}
\begin{flushright}
CERN-PH-TH/2008-078
\end{flushright}
\vspace*{1cm}

\begin{center}

{\large{\bf Faraday rotation, stochastic magnetic fields and CMB maps}}
\vskip1.cm

Massimo Giovannini$^{a,b}$\footnote{Electronic address: massimo.giovannini@cern.ch} and Kerstin E. Kunze$^{c}$\footnote{Electronic address: kkunze@usal.es}

\vskip1.5cm
{\sl $^a$INFN, Section of Milan-Bicocca, 20126 Milan, Italy}
\vskip 0.2cm 
{\sl $^b$  Department of Physics, Theory Division, CERN, 1211 Geneva 23, Switzerland}
\vskip 0.2cm 
{\sl $^c$ Departamento de F\'\i sica Fundamental, \\
 Universidad de Salamanca, Plaza de la Merced s/n, E-37008 Salamanca, Spain}
\vspace*{1cm}

\begin{abstract}
The high- and low-frequency descriptions of the pre-decoupling plasma 
are deduced from the Vlasov-Landau treatment generalized to curved space-times and in the 
presence of the relativistic fluctuations of the geometry. It is demonstrated that 
the interplay between one-fluid and  two-fluid treatments is mandatory for a complete and reliable calculation of the polarization observables.  The Einstein-Boltzmann hierarchy is generalized to handle the dispersive propagation of the electromagnetic disturbances in the pre-decoupling plasma. Given the improved physical and numerical framework, 
the polarization observables are computed within the magnetized $\Lambda$CDM paradigm (m$\Lambda$CDM). 
In particular, the Faraday-induced B-mode is consistently estimated by taking into account the effects of the magnetic fields on the initial conditions of the Boltzmann hierarchy, on the dynamical equations and on the dispersion relations.
The complete calculations of the angular power spectra constitutes the first step for 
the derivation of magnetized maps of the  CMB temperature and polarization which are 
here obtained for the first time and within the minimal m$\Lambda$CDM model.   
The obtained results set the ground for direct experimental scrutiny of large-scale magnetism via the low and high frequency instruments of the Planck explorer satellite. 
\end{abstract}

\end{center}
\end{titlepage}
\newpage
\renewcommand{\theequation}{1.\arabic{equation}}
\setcounter{equation}{0}
\section{Plasma hierarchies and plasma descriptions}
\label{sec1}
Prior to matter-radiation equality and throughout decoupling the number of charge carriers  present within the Debye sphere is, overall, inversely  proportional to the baryonic concentration. 
The corresponding (dimensionless) plasma parameter \cite{spitzer, krall,stix} is ${\mathcal O}(10^{-7})$  and, more specifically\footnote{ Typical plasma parameters in glow discharges are of the order of $10^{-2}$. The units 
used in the present investigation will be such that $\hbar = c = k_{\mathrm{B}} =1$. To facilitate 
the comparison with experimental data, the angular power spectra will be however assigned 
in units of $(\mu \mathrm{K})^2$ as it will be explicitly pointed out.} 
\begin{eqnarray}
&&g_{\mathrm{plasma}} = \frac{1}{V_{\mathrm{D}} n_{0} x_{\mathrm{e}}} = 24 e^{3}  \sqrt{\frac{\zeta(3)}{\pi}} 
\sqrt{x_{\mathrm{e}} \eta_{\mathrm{b}0}}
= 2.308 \times 10^{-7} \sqrt{x_{\mathrm{e}}} \biggl(\frac{h_{0}^2\Omega_{\mathrm{b0}}}{0.02273}\biggr)^{1/2},
\nonumber\\
&&V_{\mathrm{D}} = \frac{4}{3} \pi \lambda_{\mathrm{D}}^3,\qquad 
\lambda_{\mathrm{D}} =\sqrt{\frac{T}{8\pi e^2 n_{0} x_{\mathrm{e}}}},
\label{gplasma}
\end{eqnarray}
where $x_{\mathrm{e}}$ is the ionization fraction and 
\begin{equation}
\eta_{\mathrm{b}0} = 6.219 \times 10^{-10} \biggl(\frac{h_{0}^2\Omega_{\mathrm{b}0}}{0.02773}\biggr) \biggl(\frac{T_{\gamma 0}}{2.725}\biggr)^{-3}, 
\label{etab0}
\end{equation}
is the ratio between the baryonic concentration and the photon concentration. 
Equations (\ref{gplasma})--(\ref{etab0})  
assume the values of the cosmological parameters implied by the 5-year WMAP data alone
\cite{WMAP51,WMAP52,WMAP53,WMAP54,WMAP55} and analyzed in the light of the 
conventional $\Lambda$CDM paradigm \footnote{For the explicit estimates,
 the 5-year WMAP data alone will be consistently adopted. In the numerical study, however, 
different data sets will also be discussed.}.  Since the plasma 
is globally neutral  (i.e. $n_{\mathrm{e}} = n_{\mathrm{i}} = 
n_{0} = \eta_{\mathrm{b0}} \,n_{\gamma 0}$),  the concentration of charge carriers entering Eq. (\ref{gplasma}) 
 will be  ten billion times smaller than the photon concentration:  this is the ultimate rationale for the hierarchy 
provided by Eq. (\ref{gplasma}) and for the intrinsic validity of the plasma 
approximation.

The minuteness of $g_{\mathrm{plasma}}$  determines various hierarchies  between the physical 
quantities characterizing the pre-decoupling plasma. Every time a hierarchy arises, a potentially interesting 
(approximate) physical description is at our disposal. For instance, the hierarchy between the 
Hubble rate and the collision frequencies of the Thompson and Coulomb scattering permits, before 
equality, to treat the baryon-lepton-photon system as a unique dynamical entity. The latter 
approximation is always implemented, in standard Boltzmann solvers, to avoid 
the stiffness of the numerical system prior to equality.  In the present paper it will be argued that 
effective (i.e. one-fluid) descriptions of the baryon-lepton system are not adequate for the 
calculation of the polarization observables when large-scale magnetic fields
\footnote{By large-scales we shall mean here typical length-scales $L$ at least of  order of the Hubble radius at equality, i.e. $r_{\mathrm{H}}(\tau_{\mathrm{eq}}) = H_{\mathrm{eq}}^{-1}$. } intervene in the physics of the 
 pre-decoupling plasma.  In different words, when the large-scale magnetic fields are present 
 it is certainly allowed to solve numerically the system by making use of the hierarchy 
 of different physical scales. These hierarchies are, however, less conventional 
 from the ones arising in the standard case (i.e. when magnetic fields are absent). 
The calculation must then treat appropriately  
the specific hierarchies of the problem to avoid that hugely different physical scales 
appear simultaneously in the same numerical integration.  In this 
introduction we first intend to make the various hierarchies explicit. In the second 
place the main strategy of the present calculation will be summarized and contrasted 
with previous attempts. 

To introduce the aspects of the problem in physical terms, it is appropriate to show how the main 
hierarchies of the pre-decoupling plasma are all controlled, directly or indirectly, by $g_{\mathrm{plasma}}$.
This exercise will pin down, implicitly, the different plasma descriptions emerging in the analysis.
In short the different hierarchies determined by $g_{\mathrm{plasma}}$ are:
\begin{itemize}
\item{} the hierarchy between the Debye length (i.e. $\lambda_{\mathrm{D}}$) and the Coulomb mean free path 
(i.e. $\lambda_{\mathrm{Coul}}$);
\item{} the hierarchy between the plasma frequency (of the electrons) and the collision frequency;
\item{} the largeness of the conductivity in units of the plasma frequency.
\end{itemize}
To appreciate the validity of this statement, it suffices to write the aforementioned quantities 
in the form of dimensionless ratios depending solely upon $g_{\mathrm{plasma}}$.
By doing this it can be concluded, for instance, that $\lambda_{\mathrm{Coul}}$,  is not the shortest scale of the problem: 
\begin{equation}
\frac{\lambda_{\mathrm{D}}}{\lambda_{\mathrm{Coul}}} = \frac{g_{\mathrm{plasma}}}{48\pi}
\ln{\Lambda_{\mathrm{C}}}, \qquad \Lambda_{\mathrm{C}} = \frac{18 \sqrt{2}}{g_{\mathrm{plasma}}},
\label{hier1}
\end{equation}
where $\ln{\Lambda}_{\mathrm{C}}$ is the Coulomb logarithm \cite{krall,stix}.  With similar manipulations, 
the plasma frequency of the electrons, i.e. $\omega_{\mathrm{pe}}$,  turns out to be  much larger 
than the collision frequency (related to the Coulomb rate  of interactions), i.e. 
\footnote{In this introduction we shall not dwell on the distinction between comoving and physical 
frequencies. In curved space-times of Friedmann-Robertson-Walker type, 
the electron and proton masses break the conformal invariance and, therefore, the plasma and Larmor 
frequencies will also depend upon the scale factor, as it will be explicitly shown.}
\begin{equation}
\frac{\Gamma_{\mathrm{Coul}}}{\omega_{\mathrm{pe}}} = \frac{\ln{\Lambda_{\mathrm{C}}}}{24 \sqrt{2} \pi } 
g_{\mathrm{plasma}}, \qquad \omega_{\mathrm{pe}} = \sqrt{\frac{4\pi n_{0} x_{\mathrm{e}}}{m_{\mathrm{e}}}}.
\label{hier2}
\end{equation}
Equation (\ref{hier2}) implies that the electrons will oscillate many times before undergoing a collision.  
The third of the aforementioned hierarchies stipulates that the conductivity (denoted by $\sigma$) 
is be parametrically larger than $1$ in units of the plasma frequency:
\begin{equation}
\sigma = \frac{\omega_{\mathrm{pe}}^2}{4 \pi \Gamma_{\mathrm{Coul}}} = \frac{6 \sqrt{2}}{\ln{\Lambda_{\mathrm{C}}}} \frac{\omega_{\mathrm{pe}}}{g_{\mathrm{plasma}}}.
\label{hier3}
\end{equation}
Recalling Eq. (\ref{gplasma}),  Eq. (\ref{hier3}) implies, indeed, $\sigma/\omega_{\mathrm{pe}} \gg 1$ 
and also that $\sigma/T \gg 1$.

According to Eq. (\ref{gplasma}),  the Debye length around equality is of the order of $10$ to $100$ cm  in comparison with $r_{\mathrm{H}}= H^{-1}$ which is, around equality, $20$ orders of magnitude larger.
Large-scale electric (rather than magnetic) fields are highly suppressed by powers of $\sigma^{-1}$ in the baryon rest frame. 
At the same reference time, magnetic fields can be present over typical length-scales $L > L_{\sigma}$ where $L_{\sigma}$ 
\begin{equation}
L_{\sigma} \simeq (4\pi \sigma H_{\mathrm{eq}})^{-1},\qquad 
\sigma = \frac{T}{e^2  \ln{\Lambda_{\mathrm{C}}}} \biggl(\frac{T}{m_{\mathrm{e}}}\biggr)^{1/2},
\label{Lsigma}
\end{equation}
(see also Eq. (\ref{hier3})) is called  magnetic diffusivity length.
For typical values of the cosmological parameters, around equality, 
$L_{\sigma} \simeq 10^{-17} r_{\mathrm{H}}$. Magnetic fields 
over typical length-scales $L \simeq {\mathcal O}(r_{\mathrm{H}})$ (and possibly 
larger) can be present without suffering appreciable diffusion.
Since the magnetic fields touched by the present discussion will 
have, at most, nG strength at the onset of galaxy formation, the Larmor radius around equality will be much smaller 
than the range of variation of the magnetic field, i.e. 
\begin{equation}
r_{\mathrm{Be}} \ll  L \simeq r_{\mathrm{H}},
 \qquad r_{\mathrm{Be}} = \frac{v_{\perp}}{\omega_{\mathrm{Be}}},\qquad v_{\perp} \simeq v_{\mathrm{th}},
\label{hier4}
\end{equation}
where $v_{\mathrm{th}} \simeq \sqrt{T/m_{\mathrm{e}}}$ and $\omega_{\mathrm{Be}}$ is 
the Larmor frequency. 
As pointed out after Eq. (\ref{Lsigma})  $L > L_{\sigma}$ and it is always much 
larger than $r_{\mathrm{Be}}$.  

The plasma hierarchies introduced in Eqs. (\ref{gplasma}), (\ref{hier1})--(\ref{hier3}) and (\ref{Lsigma})--(\ref{hier4})
 imply  different descriptions valid for  complementary branches of the spectrum of plasma excitations. More specifically:
\begin{itemize}
\item{} for typical length scales much larger than the Debye length and for typical times 
exceeding the inverse of the plasma frequency, a single-fluid theory naturally emerges and it often dubbed 
magnetohydrodynamics (MHD in what follows);
\item{} in the opposite limits the two-fluid nature of the system cannot be ignored 
and, unlike in the MHD description, the electromagnetic disturbances 
propagate in a dispersive medium which has, in our case, a finite 
concentration of charge carriers and a large-scale magnetic field.
\end{itemize}
The MHD limit possesses various sub-limits which make the whole dynamics rather rich (see, for instance, \cite{jones}).
The stochastic magnetic field evolving according to MHD and coupled 
to the fluctuations of the geometry affects, by its presence, 
the propagation of electromagnetic disturbances in the plasma: electromagnetic waves with positive 
(or negative) helicities will propagate with different phase (and group) velocities. 
The latter effect cannot be treated within a one-fluid approximation where the displacement current is consistently neglected and the Ohmic current is solenoidal \cite{krall}. This is the strategy followed in the present 
paper and it corresponds to the logic used in the analysis of laboratory plasmas. 
The physical and technical challenges of the problem reside in the occurrence that, prior to photon decoupling, 
 the metric is a dynamical quantity and that it can fluctuate. It is easily imaginable that, in spite 
 of the initial physical analogy with laboratory plasmas, the actual problem will be technically more difficult 
 than in Minkowskian space-time.

Accurate calculations of magnetized CMB anisotropies 
are an essential tools for the scrutiny of the origin and evolution of large-scale 
magnetism which is observed in the largest gravitationally 
bound systems such as clusters \cite{clusters1,clusters2}, galaxies \cite{galaxies1,galaxies2} and even in some 
superculster. For the interplay between CMB physics and large-scale magnetic fields see, for instance, \cite{rev1,rev2}.
More general reviews on the problems and challenges of large-scale 
magnetism can be found in \cite{rev3,rev4,rev5}. In simple words, we want to embark 
in these calculations because we want to give reasonable predictions of 
the potential effects of large-scale magnetic fields on the CMB anisotropies. 

Large-scale magnetic fields present prior to equality affect the evolution of the 
fluctuations of the scalar modes of the geometry. 
The evolution equations of the 
Boltzmann hierarchy (as well as the initial conditions) must be appropriately modified. 
This theoretical problem has been scrutinized in  a number of 
recent papers \cite{mg1,mg2,mg3} (see also \cite{mg4}).  In 
\cite{mg1} the estimate of the Sachs-Wolfe plateau has been carried on 
by employing the technique of the transfer matrices. In \cite{mg2} 
the temperature autocorrelations and the polarization cross-correlations 
have been computed numerically in the tight-coupling approximation.
In \cite{mg3} a semi-analytical evaluation of the TT angular 
power spectra has been carried on \footnote{Following the current terminology 
we will denote by TT the angular power spectrum of the temperature 
autocorrelation. By TE we shall denote  the power spectrum 
of the cross-correlation between temperature and polarization. Finally the EE spectrum 
denotes the autocorrelation of the polarization. The precise definition of the various power spectra 
is discussed in Section \ref{sec5}.}.

A dedicated numerical approach for the calculation of magnetized 
temperature and polarization observables has been 
devised in a series of recent papers \cite{gk1,gk2,gk3}. Such a numerical approach in constructed 
from one of the standard Boltzmann solvers i.e. CMBFAST 
\cite{cmbfast1,cmbfast2} which is, in turn, based on the COSMICS
package by E. Bertschinger \cite{cosmics1,cosmics2}. 
In \cite{gk1} the minimal framework for the analysis of the effects of  large-scale magnetic fields on the CMB anisotropies has been spelled out and dubbed magnetized $\Lambda$CDM model (m$\Lambda$CDM model in what follows). 
In the minimal realization of the m$\Lambda$CDM scenario the inclusion of large-scale magnetic fields 
amounts to the addition of two extra parameters. In \cite{gk2} it has been shown that 
the numerical analysis leads to shapes of the TT angular power spectra which are 
exactly the ones computed in \cite{mg3}. The numerical approach is intrinsically 
more accurate especially at high multipoles. In \cite{gk3} all the possible (non-adiabatic) initial conditions 
of the magnetized Einstein-Boltzmann hierarchy have been worked out analytically 
and scrutinized numerically. The results reported in \cite{mg1,mg2,mg3} and in \cite{gk1,gk2,gk3}
bring the treatment of magnetized CMB anisotropies to the same standards employed in the case 
when large-scale magnetic fields are absent from the very beginning. 

The problem left out from previous analyses, as stressed in \cite{gk2}, has to do with a more consistent calculation 
of the polarization observables. The problem is, in short, the following.  The large-scale 
description of temperature anisotropies demands a coarse grained (one-fluid) approach for the 
electron-ion system: this is the so called baryon fluid which is treated (with no exceptions) as a single 
fluid in popular Boltzmann solvers such as COSMICS, CMBFAST and their descendants. On the other hand the dispersive propagation of electromagnetic disturbances demands to treat separately electrons and ions, at least at high frequencies. 
Faraday rotation is one of the situations where the inadequacy of the one-fluid approximation (for the baryon-lepton fluid) is manifest. The positive and negative helicities composing the (linear) CMB polarization experience,
in a background magnetic field, two different phase velocities, two different dielectric contants and, ultimately, two different 
refractive indices. The mismatch between the refractive index of the positive and negative helicities  induces, effectively, a rotation of the CMB polarization and, hence, a B-mode. The inclusion of the Faraday effect in the treatment implies, physically, that the proton-electron fluid (sometimes dubbed as baryon fluid) should be treated 
as effectively composed by two different species, i.e. the electrons and the ions. 

Various studies were concerned, in the past, with the estimate 
of Faraday rotation effects in the framework of CMB physics. None of the 
previous studies, however, could profit of a dedicated numerical approach 
where the effects of the magnetic fields could be included at the level of the initial 
conditions and at the level of the dynamical equations. 
Before outlining the analytical and numerical strategies used in the present paper 
the main results obtained so far will be summarized.

In \cite{far1,far2,far3} it has been noted, within slightly 
different perspectives,  that the (linear) CMB polarization 
can be Faraday rotated. A common aspect of the attempts of \cite{far1,far2,far3}
was that the magnetic field was assumed to be {\em uniform} (i.e. homogeneous 
in space) and described within a simplified magnetohydrodynamical 
description which consisted, effectively, to enforce the conservation 
of the magnetic flux. In \cite{far3}, on the basis of an explicit model, it was argued that the magnetic fields 
should be treated, in fact, as stochastically distributed if we do not want 
to break (explicitly) the spatial isotropy of the background geometry: 
the large-scale magnetic fields arise typically from the parametric amplification 
of vacuum fluctuations or from some phase transition (see, e.g. \cite{rev3}) and in both cases
the produced fields are not uniform but rather stochastically distributed.

The first convincing measurements of CMB polarization 
\cite{WMAP11,WMAP12,WMAP13} (see also \cite{WMAP31,WMAP32}) 
clearly suggested the adiabaticity of the initial conditions (because of the location of the first anticorrelation 
peak in the TE power spectra). 
In \cite{far4} the TB correlations have been computed from the 
initial TE correlations provided by the adiabatic mode. The main 
assumption of \cite{far4} has been that the magnetic field is, once more,  uniform 
and that it does not affect, in any way, the initial TE angular power spectra.
In \cite{far8} and \cite{far5} it was recognized that 
the uniform field case is not realistic.  The interplay between stochastic 
magnetic fields and Faraday rotation 
has been more directly scrutinized. The work of Ref. \cite{far5} 
suggested to treat the effect by applying a Faraday screen to the 
CMB polarization. Also in \cite{far5} the main 
assumption has been to neglect any possible effect of the 
magnetic fields on the TE and EE angular power spectra.

In the present paper, it will not be assumed that the large-scale magnetic fields
are uniform. Hence, the direct effects of the magnetic fields 
on the TE and EE angular power spectra will not be neglected.
The line of reasoning pursued in the present paper 
 is based on the plasma hierarchies outlined in Eqs. (\ref{gplasma}) 
and in Eqs. (\ref{hier1})--(\ref{hier3}). The strategy will be to generalize, around matter-radiation equality 
and throughout decoupling, the standard treatment of weakly coupled plasmas
in the different branches of the spectrum of plasma excitations.
The smallness of $g_{\mathrm{plasma}}$ (together with the smallness 
of the electrons and ions kinetic temperatures in comparison with the corresponding masses)
allows to enforce the (cold) plasma approximation to an excellent 
degree.   The flat space-equations cannot be simply 
employed around decoupling or even equality since space-time 
is not flat. Furthermore,  the fluctuations of the geometry must be properly 
taken into account since they are still relativistic at the time when 
the initial conditions of the temperature and polarization anisotropies 
are customarily set \cite{mg1,mg2,mg3}.The physical rationale for the strategy employed here is rooted on the simple 
observation of Eq. (\ref{hier4}), which stipulates that the Larmor radius is always 
smaller than the typical inhomogeneity scale of the magnetic field. The approximation which
will be adopted here is called, in plasma physics, the guiding center approximation and it is due 
to the pioneering work of Alfv\'en \cite{ALF1,ALF2}.

A relevant class of results of the present investigation concerns 
the problem of simulating maps of magnetized CMB anisotropies. 
Magnetized maps of the temperature and polarization observables 
will be reported to illustrate the viability of our numerical approach\footnote{Even if temperature 
autocorrelations are not central to the present discussion (which is focussed on magnetized 
polarization observables) we will also report maps of the temperature anisotropies, mainly 
for completeness.}.
In the light of the Planck explorer mission \cite{planck} it will 
be particularly important to have magnetized maps both for the 
temperature autocorrelations and maps 
for the polarization observables. This step is mandatory 
once the corresponding angular power spectra can be numerically 
computed. With our numerical code, appropriately extended 
to include the Faraday mixing term and the two-fluid effects, we are able to compute accurately 
all the required power spectra. Armed with all these necessary theoretical tools,  
magnetized CMB maps can be obtained.

The present paper is organized as follows. In Section 2 the generalized
two-fluid description for the charged species (electrons and ions) is introduced.  
In Section 3 the two-fluid treatment is shown to be equivalent to a single fluid description at large-scales. 
In Sections 4 and 5 the high-frequency branch of the spectrum 
of plasma excitations is discussed. Sections 6 and  7
are devoted to the calculation of the TE, EE and BB angular power spectra.
Illustrative examples of magnetized CMB maps 
are collected in Section 8. Section 9 contains our concluding 
remarks. To avoid excessive technicalities, relevant results and derivations  have been included in the Appendix. 

\renewcommand{\theequation}{2.\arabic{equation}}
\setcounter{equation}{0}
\section{Electrons and ions}
\label{sec2}
There is the custom, in CMB studies, to treat ions and electrons as a single 
dynamical entity (see, for instance, \cite{cosmics2} and also \cite{an1,an2,an3}). 
This is certainly justified for typical scales $L \gg \lambda_{\mathrm{D}}$ (see also Eq. (\ref{gplasma})) 
 and for frequencies parametrically smaller than the plasma 
frequency. Owing to the largeness of the Coulomb rate ions and electrons 
are tightly coupled and the effective degree of freedom which should be studied is 
the so-called baryon velocity, i.e. the centre-of-mass velocity of the electron-ion system.
Furthermore, owing to the largeness of Thompson scattering, baryons 
and photons are also tightly coupled (but just well before equality). This 
observation is the basis of the tight-coupling expansion which was pioneered 
in \cite{an1} (see also \cite{peebook}) and since then widely used 
for semi-analytical estimates of the CMB temperature autocorrelations.

If large-scale magnetic fields are included in the game some of the considerations 
of the previous paragraph apply \cite{mg1,mg2,mg3} (see also \cite{gk1,gk2}). However,
if we ought to treat phenomena related to the propagation 
of electromagnetic disturbances in a plasma, the one-fluid description 
is known to be insufficient \cite{stix}. The insufficiency of the 
one-fluid description is already apparent in the calculation of the magnetized polarization observables. This caveat has been made explicit in \cite{gk2}.

In what follows the pre-decoupling plasma will be treated as a laboratory plasma\footnote{It is amusing 
to notice that the plasma parameter of the decoupling plasma (i.e. 
 ${\mathcal O}(10^{-7})$, see Eq. (\ref{gplasma})) is of the same order of magnitude of the 
 plasma parameter of a  tokamak. Of course, in a tokamak, the 
 density of charge carries will be much larger. However, the Debye length will be comparatively
 much smaller. For instance, in a tokamak the concentration 
 of charge carriers will be typically $10^{14}\,\,\mathrm{cm}^{-3}$; the typical temperature will be of the order
 of $10$ keV. Consequently, the Debye scale is, in this situation, $7\times 10^{-5}$ m; the corresponding 
 plasma parameter (defined in Eq. (\ref{gplasma})) will then be ${\mathcal O}(10^{-8})$. This, as anticipated, 
 is the same figure one obtains around equality.} with the following notable differences:
\begin{itemize}
\item{} around radiation-matter equality and thought photon decoupling space-time 
is curved;
\item{} since electrons and ions are non-relativistic 
conformal invariance is explicitly 
broken and this fact has various physical implications, as we shall see;
\item{} relativistic fluctuations of the geometry, immaterial for  plasmas in the 
laboratory, have to be included in our discussion.
\end{itemize}
The conformally flat line element characterizing the space-time around equality and decoupling will be written as
\begin{equation}
ds^2 = g_{\mu\nu} dx^{\mu} dx^{\nu} = a^2(\tau) [ d\tau^2 - d\vec{x}^2],
\qquad g_{\mu\nu} = a^2(\tau) \eta_{\mu\nu},
\label{M1}
\end{equation}
where $\tau$ is the conformal time coordinate 
to distinguish it from the cosmic time $t$ (of course $dt = a(\tau) d\tau$).
The perturbed metric will be parametrized, without loss 
of generality, in the synchronous coordinate system where \footnote{
For practical purposes the pivotal description will be taken to be the 
one of the synchronous gauge \cite{gk1,gk2,gk3}. Different
descriptions (in complementary gauges) can be straightforwardly deduced 
using the fully gauge-invariant approaches of \cite{mg1,mg2} and \cite{mg5}.}
\begin{equation}
\delta_{\mathrm{s}} g_{ij}(\vec{x},\tau) = a^2(\tau) h_{ij}(\vec{x},\tau).
\label{M2}
\end{equation}
Equation (\ref{M2}) holds in real space. It is often useful to separate the fluctuation 
of the geometry in a trace part supplemented by a traceless 
contribution. For practical reasons this is done in Fourier space where the perturbed 
metric becomes, 
\begin{equation}
\delta_{\mathrm{s}} g_{ij}(k,\tau) = a^2(\tau)\biggl[ \hat{k}_{i} \hat{k}_{j}  h(k,\tau) + 6 \xi(k,\tau) \biggl( \hat{k}_{i} \hat{k}_{j} - \frac{\delta_{ij}}{3}\biggr)\biggr],
\label{M3}
\end{equation}
which makes clear that the $h_{ij}(\vec{x},\tau)$ of Eq. (\ref{M2}) carries, effectively, only 
two (scalar) degrees of freedom. If a fully inhomogeneous 
magnetic field is present, the evolution equations of the scalar 
inhomogeneities will be necessarily affected. 
The homogeneous evolution of the background geometry (see Eq. (\ref{M1}))  is dictated by the Friedmann-Lema\^itre equations:
\begin{equation}
3{\mathcal H}^2 = 8\pi G a^2 \rho_{\mathrm{t}},\qquad 
{\mathcal H}^2 -{\mathcal H}' = 4\pi G a^2 (p_{\mathrm{t}} + \rho_{\mathrm{t}}), \qquad \rho_{\mathrm{t}}' + 3 {\mathcal H}(p_{\mathrm{t}} + \rho_{\mathrm{t}})=0,
\label{M4}
\end{equation}
where the prime denotes a derivation with respect to the conformal 
time coordinate and where ${\mathcal H} = a'/a$.
The quantities $\rho_{\mathrm{t}}$ and $p_{\mathrm{t}}$ denote the 
total energy density and the total pressure of the plasma 
which consists, in the present situation, by two charged species 
(electrons and ions) and  by neutral species three neutral species 
(neutrinos, photons and CDM particles). The dark-energy 
component will be simply parametrized, as in the case of the 
$\Lambda$CDM scenario, by a cosmological constant. 

The evolution equations 
of the problem will now be introduced with particular attention to the electromagnetic part.
There is the custom of treating dispersive phenomena within
a single fluid plasma description. This habit is, strictly speaking, 
misleading as already pointed out in \cite{birefringence}.  The logic 
followed in the present treatment can be summarized as follows:
\begin{itemize}
\item{} start with a bona fide two-fluid description in curved space 
taking into account all the relevant degrees of freedom and, in particular,
the electrons and the ions as separated components;
\item{} derive the one-fluid description which will allow 
to follow the evolution of the inhomogeneous magnetic field;
\item{} derive the relevant dispersion relations.
\end{itemize}
A reasonably general description of the electron-ion system
can be achieved from the evolution equation of the one-body 
distribution functions which have the characteristic Vlasov form (see also Appendix \ref{APPA} for a derivation):
\begin{eqnarray}
&&\frac{\partial f_{\mathrm{i}}}{\partial \tau} + \vec{v} \cdot\vec{\nabla}_{\vec{x}} f_{\mathrm{i}} + e (\vec{E} + \vec{v} \times \vec{B})\cdot \vec{\nabla}_{\vec{q}} f_{\mathrm{i}} = \biggl( \frac{\partial f_{\mathrm{i}}}{\partial \tau} \biggr)_{\mathrm{coll}},
\label{BZ1}\\
&& \frac{\partial f_{\mathrm{e}}}{\partial \tau} + \vec{v} \cdot\vec{\nabla}_{\vec{x}} f_{\mathrm{e}} - e (\vec{E} + \vec{v} \times \vec{B})\cdot \vec{\nabla}_{\vec{q}} f_{\mathrm{e}}  = \biggl( \frac{\partial f_{\mathrm{e}}}{\partial \tau} \biggr)_{\mathrm{coll}},
\label{BZ2}
\end{eqnarray}
where the collision terms are provided by Coulomb scattering and where 
the electric and magnetic fields are rescaled as:
\begin{equation}
\vec{B} = a^2 \vec{{\mathcal B}},\qquad \vec{E} = a^2 \vec{{\mathcal E}}, 
\qquad \vec{v} = \frac{\vec{q}}{\sqrt{|\vec{q}|^2 + m^2 a^2}}.
\label{BZ3}
\end{equation}
In Eqs. (\ref{BZ1}) and (\ref{BZ2}),  $\vec{q}$ is the comoving three-momentum 
which is customarily defined in the curved-space Boltzmann treatment, i.e. 
$\vec{q} = a \vec{p}$. For a massive particle, such as 
the electron or the proton, the conjugate momenta satisfy 
$g_{\alpha\beta} P^{\alpha}P^{\beta} = m^2$. By identifying $g_{ij} P^{i} P^{j} = - \delta_{ij} p^{i} p^{j}$, the comoving three momentum is introduced\footnote{See Appendix \ref{APPA}  for further details on the geodesics 
of massive particles endowed with an electric charge.}. If we would be in the 
ultra-relativistic limit (where electrons and protons are effectively massless)
Eqs. (\ref{BZ1}) and (\ref{BZ2}) (written in terms of the conformal time 
coordinate $\tau$ and in terms of the appropriately rescaled fields of Eq. 
(\ref{BZ3})) would be exactly the same equations we would have in flat space-time. 
This is a consequence of the conformallly flat nature 
of the background (see Eq. (\ref{M1})).  Prior to equality, i.e. 
when the calculation of the CMB anisotropies is initialized, the electrons 
and ions are non-relativistic  and the presence of a mass term will break 
(explicitly) conformal invariance. This observation is clear by noticing that, in the non-relativistic limit, Eq. (\ref{BZ3}) implies that $\vec{v} = \vec{q}/(m a)$. 

It is useful, for immediate convenience, to introduce here the distinction between 
cold and warm plasmas\footnote{This distinction is customarily 
 employed in plasma literature (see, e. g. \cite{stix,krall})}. The pre-equality plasma is cold in the sense 
 that the kinetic temperatures of the charged species are much smaller than the corresponding masses. 
 In the opposite case the plasma is warm. If the initial conditions of the Boltzmann hierarchy are set at a temperature which is 
say one tenth of the temperature of neutrino decoupling, the plasma will be already cold to a good approximation. 
In can be shown that, in a warm plasma approach, the corrections to the Faraday rate will be suppressed, to leading order, by powers of $T_{\mathrm{e}}/m_{\mathrm{e}}$ \cite{birefringence,skilling}.

The (approximate) equilibrium solution of the Boltzmann equation will be a 
Maxwellian velocity distribution and, from this observation, the Boltzmann 
equations can be perturbed to obtain the evolution equations 
of the various moments of the distribution functions such as 
the evolution of the charge concentrations (from the zeroth-order moment),
the evolution equations for the velocities (i.e. the first-order moment) and so on.  While more detailed 
considerations can be found in Appendix \ref{APPA},  the same two-fluid equations obtainable from the Vlasov description 
can be  recovered from the charge and four-momentum conservation in curved space-time, as it will be discussed in a moment.

The evolution equations of the gauge fields will be given, in the 
present context, by the appropriate Maxwell equations which can 
be written as\footnote{The 
covariant and controvariant indices of the various vectors and tensors 
must not be confused with the subscripts (always in roman style) which
denote the corresponding species.}
\begin{eqnarray}
&&\nabla_{\mu} F^{\mu\nu} = 4\pi (j_{\mathrm{i}}^{\nu} + j_{\mathrm{e}}^{\nu}),
\qquad \nabla_{\mu} \tilde{F}^{\mu\nu} =0, 
\label{mxg1}\\
&&\nabla_{\mu}j_{\mathrm{i}}^{\nu}=0,\qquad \nabla_{\mu} j_{\mathrm{e}}^{\nu}=0,
\label{mxg2}
\end{eqnarray}
where $\nabla_{\mu}$ denotes the covariant derivative. The electron and ion currents are given, respectively, by 
\begin{equation}
j_{\mathrm{e}}^{\mu} = - e\,\tilde{n}_{\mathrm{e}} \,u_{\mathrm{e}}^{\mu},
\qquad  j_{\mathrm{i}}^{\mu} = e\,\tilde{n}_{\mathrm{i}} \,u_{\mathrm{i}}^{\mu},
\label{mxg3}
\end{equation}
where $\tilde{n}_{\mathrm{i}}$ and $\tilde{n}_{\mathrm{e}}$ are, respectively, the ion and electron concentrations. The velocity 
fields satisfy the conditions
\begin{equation}
g_{\alpha\beta} u^{\alpha}_{\mathrm{i}} u^{\beta}_{\mathrm{i}}=1,\qquad 
g_{\alpha\beta} u^{\alpha}_{\mathrm{e}} u^{\beta}_{\mathrm{e}}=1,
\label{mxg4}
\end{equation}
which imply, in the non-relativistic limit, that $u^{0}_{\mathrm{e}} = u^{0}_{\mathrm{i}} = 1/a$.
The Maxwell field strengths and their duals are simply 
\begin{equation}
F_{0i} = a^2 {\mathcal E}_{i}, \qquad F_{ij} = - a^2 \epsilon_{ijk} {\mathcal B}^{k},\qquad \tilde{F}^{0i} = -\frac{{\mathcal B}^{i}}{a^2},\qquad 
\tilde{F}^{ij} = \frac{\epsilon^{ijk}}{a^2} {\mathcal B}_{k}.
\label{mxg6}
\end{equation}
Using Eqs. (\ref{mxg3}), (\ref{mxg4}) and (\ref{mxg6}) inside 
Eq. (\ref{mxg1}), the following set of equations can be obtained:
\begin{eqnarray}
&&\vec{\nabla} \cdot \vec{E} = 4\pi e ( n_{\mathrm{i}} - n_{\mathrm{e}}),
\label{MX1}\\
&&\vec{\nabla}\cdot \vec{B} =0,
\label{MX2}\\
&& \vec{\nabla}\times \vec{E} + \vec{B}' =0,
\label{MX3}\\
&& \vec{\nabla}\times \vec{B} = 4\pi e (n_{\mathrm{i}} \vec{v}_{\mathrm{i}} - n_{\mathrm{e}} \vec{v}_{\mathrm{e}}) + \vec{E}'.
\label{MX4}
\end{eqnarray}
where,  $\vec{E} = a^2 \vec{{\mathcal E}}$ and  
$\vec{B} = a^2 \vec{{\mathcal B}}$ are the rescaled electric and magnetic fields.
The  concentrations   and velocities appearing in the four-currents (see Eq. (\ref{mxg3})) 
have been rescaled as:
\begin{eqnarray}
&&n_{\mathrm{i}} = a^3 \tilde{n}_{\mathrm{i}},\qquad n_{\mathrm{e}} = a^3 \tilde{n}_{\mathrm{e}},
\nonumber\\
&&\vec{v}_{\mathrm{i}} = a \vec{u}_{\mathrm{i}}, 
\qquad \vec{v}_{\mathrm{e}} = a \vec{u}_{\mathrm{e}}.
\label{MX5}
\end{eqnarray}
where $\vec{v}_{\mathrm{i}}$ and $\vec{v}_{\mathrm{e}}$ are the comoving three-velocities \footnote{These
 variables emerge naturally from  the geodesics of charged particles 
in the gravitational field. They are relevant for a correct derivation of the Vlasov-Landau equation 
in the configuration-velocity space.}.
In terms of the comoving concentrations defined in Eq. (\ref{MX5}), the charge conservation equation, i.e. Eq. (\ref{mxg2}) will imply,  for electrons and ions, 
\begin{eqnarray}
&&  \frac{\partial n_{\mathrm{i}}}{\partial \tau} + \theta_{\mathrm{i}} n_{\mathrm{i}} + \vec{v}_{\mathrm{i}}\cdot \vec{\nabla} n_{\mathrm{i}} =0,
\label{MX6}\\
&&\frac{\partial n_{\mathrm{e}}}{\partial \tau} + \theta_{\mathrm{e}} n_{\mathrm{e}} + \vec{v}_{\mathrm{e}}\cdot \vec{\nabla} n_{\mathrm{e}} =0.
\label{MX7}
\end{eqnarray}
where $\theta_{\mathrm{i}} = \vec{\nabla}\cdot \vec{v}_{\mathrm{i}}$ and  
$\theta_{\mathrm{e}} = \vec{\nabla}\cdot \vec{v}_{\mathrm{e}}$ are the 
three-divergences of the comoving three-velocities. 
Since the plasma is globally neutral, the electron and ion 
concentrations are equal and are approximately ten billion times 
smaller than the photon concentration, i.e.
$n_{\mathrm{e}} = n_{\mathrm{i}} = n_{\mathrm{0}} = \eta_{\mathrm{b}0} n_{\gamma0}$ where $\eta_{\mathrm{b}0}$ 
has been given by Eq. (\ref{etab0}). Owing to this figure the metric fluctuations can be neglected in Eqs. (\ref{MX6}) and (\ref{MX7}) 
(see, however, Eqs. (\ref{VL13}) and (\ref{VL14}) for their explicit inclusion).

Equations (\ref{MX6}) and (\ref{MX7}) (derived from Eq. (\ref{mxg2})) can be obtained directly by taking 
the zeroth-order moment of the evolution equations for the one-body distribution functions. In particular, defining 
\begin{eqnarray}
&&n_{\mathrm{e}}(\vec{x},\tau) = n_{0} \int d^{3} v f_{\mathrm{e}}(\vec{x},\vec{v},\tau),\qquad 
n_{\mathrm{i}}(\vec{x},\tau) = n_{0} \int d^{3} v f_{\mathrm{i}}(\vec{x},\vec{v},\tau),
\label{BZ4}\\
&& \vec{v}_{\mathrm{e}}(\vec{x},\tau) = n_{0} \int d^3 v\, \vec{v}\,f_{\mathrm{e}}(\vec{x},\vec{v},\tau),\qquad 
\vec{v}_{\mathrm{i}}(\vec{x},\tau) = n_{0} \int d^{3} v\, \vec{v}\, f_{\mathrm{i}}(\vec{x},\vec{v},\tau),
\label{BZ5}
\end{eqnarray}
we indeed obtain, from Eqs. (\ref{BZ1}) and (\ref{BZ2}), 
\begin{equation}
\frac{\partial n_{\mathrm{i}}}{\partial \tau} + \vec{\nabla}\cdot(n_{\mathrm{i}} \vec{v}_{\mathrm{i}})=0, \qquad \frac{\partial n_{\mathrm{e}}}{\partial \tau} + \vec{\nabla}\cdot(n_{\mathrm{e}} \vec{v}_{\mathrm{e}})=0,
\label{BZ6}
\end{equation}
which are equivalent to Eqs.(\ref{MX6}) and (\ref{MX7}).

The evolution equations for the velocity fields and the density 
contrasts can also be derived by perturbing, to first 
order and in the synchronous gauge (see Eq. (\ref{M3})), the covariant momentum conservation:
\begin{eqnarray}
&& \partial_{\mu} \delta_{\mathrm{s}} T^{\mu\nu}_{(\mathrm{i})} + 
\delta_{\mathrm{s}} \Gamma^{\mu}_{\mu\alpha} \overline{T}_{(\mathrm{i})}^{\alpha\nu} + \overline{\Gamma}^{\mu}_{\mu\alpha} \delta_{\mathrm{s}} T^{\alpha\nu}_{(\mathrm{i})} 
+ \delta_{\mathrm{s}} \Gamma^{\nu}_{\alpha\beta} \overline{T}^{\alpha\beta}_{(\mathrm{i})} + 
\overline{\Gamma}^{\nu}_{\alpha\beta} \delta_{\mathrm{s}} T^{\alpha\beta}_{(\mathrm{i})}
= F^{\nu\alpha} j_{\alpha}^{(\mathrm{i})},
\label{cion}\\
&& \partial_{\mu} \delta_{\mathrm{s}} T^{\mu\nu}_{(\mathrm{e})} + 
\delta_{\mathrm{s}} \Gamma^{\mu}_{\mu\alpha} \overline{T}_{(\mathrm{e})}^{\alpha\nu}+ \overline{\Gamma}^{\mu}_{\mu\alpha} \delta_{\mathrm{s}} T^{\alpha\nu}_{(\mathrm{e})}
+ \delta_{\mathrm{s}} \Gamma^{\nu}_{\alpha\beta} \overline{T}^{\alpha\beta}_{(\mathrm{e})} + 
\overline{\Gamma}^{\nu}_{\alpha\beta} \delta_{\mathrm{s}} T^{\alpha\beta}_{(\mathrm{e})}
 = F^{\nu\alpha} j_{\alpha}^{(\mathrm{e})},
\label{cel}
\end{eqnarray}
where the barred symbols denote the background 
values of the corresponding quantity and where $\delta_{\mathrm{s}}$
represents the scalar fluctuation of a given tensor or connection.
Recalling that the electrons and ions are both non-relativistic, the corresponding 
energy-momentum tensors will be given as \footnote{Since the electron and ion pressures are given by 
$p_{\mathrm{e}} = \tilde{n}_{\mathrm{e}} T_{\mathrm{e}}$ and by 
$p_{\mathrm{i}} = \tilde{n}_{\mathrm{i}} T_{\mathrm{i}}$ they are suppressed
as $T/m_{\mathrm{e},\mathrm{p}}$ and shall be neglected. }
\begin{eqnarray}
&&\overline{T}_{(\mathrm{e})}^{00} =\frac{\rho_{\mathrm{e}}}{a^2},\qquad 
\overline{T}_{(\mathrm{i})}^{00} =\frac{\rho_{\mathrm{i}}}{a^2},
\label{dt1}\\
&&\delta_{\mathrm{s}}T_{(\mathrm{e})}^{00}= \frac{\rho_{\mathrm{e}} \delta_{\mathrm{e}}}{a^2},\qquad 
\delta_{\mathrm{s}}T_{(\mathrm{i})}^{00}= \frac{\rho_{\mathrm{i}}\delta_{\mathrm{i}}}{a^2},
\label{dt2}\\
&& \delta_{\mathrm{s}} T_{(\mathrm{e})}^{0i} = \frac{\rho_{\mathrm{e}}}{a^2}
v^{i}_{\mathrm{e}},\qquad \delta_{\mathrm{s}} T_{(\mathrm{i})}^{0i} = \frac{\rho_{\mathrm{i}}}{a^2}
v^{i}_{\mathrm{i}},
\label{dt3}
\end{eqnarray}
where 
\begin{equation}
\rho_{\mathrm{e}} = m_{\mathrm{e}} \tilde{n}_{\mathrm{e}}, \qquad \rho_{\mathrm{i}} = m_{\mathrm{i}} \tilde{n}_{\mathrm{i}},\qquad 
\delta_{\mathrm{e}} = \frac{\delta_{\mathrm{s}} \rho_{\mathrm{e}}}{\rho_{\mathrm{e}}}, \qquad \delta_{\mathrm{i}} = \frac{\delta_{\mathrm{s}} \rho_{\mathrm{i}}}{\rho_{\mathrm{i}}}.
\label{dt4}
\end{equation}
The masses of the electrons and ions are given, respectively, by   $m_{\mathrm{e}} =0.511$ MeV 
and $m_{\mathrm{i}} \simeq m_{\mathrm{p}} =0.938$ GeV.
The binary collision between electrons and protons are 
rather efficient in bringing the whole system to an approximate 
common temperature which will coincide with the photon temperature 
because of the strength of Thompson scattering (see also \cite{gk2}).

Within the same conventions of Eqs. (\ref{dt1}), (\ref{dt2}) and (\ref{dt3}), 
the time-like and space-like components of Eqs. (\ref{cion}) and (\ref{cel}) 
lead, respectively, to the following equations
\begin{eqnarray}
&& \delta_{\mathrm{e}}' = - \theta_{\mathrm{e}} + \frac{h'}{2} - \frac{e n_{\mathrm{e}}}{\rho_{\mathrm{e}} a^4} \vec{E} \cdot \vec{v}_{\mathrm{e}},
\label{dte1a}\\
&&\delta_{\mathrm{i}}' = - \theta_{\mathrm{i}} + \frac{h'}{2} + \frac{e n_{\mathrm{i}}}{\rho_{\mathrm{i}} a^4} \vec{E} \cdot \vec{v}_{\mathrm{i}},
\label{dte1}\\
&&\vec{v}_{\mathrm{e}}\,' + {\mathcal H} \vec{v}_{\mathrm{e}} = - 
\frac{e n_{\mathrm{e}}}{\rho_{\mathrm{e}} a^4} [\vec{E} + \vec{v}_{\mathrm{e}}\times \vec{B}] + {\mathcal C}_{\mathrm{ep}},
\label{dte2a}\\
&&\vec{v}_{\mathrm{i}}\,' + {\mathcal H} \vec{v}_{\mathrm{i}} =  
\frac{e n_{\mathrm{i}}}{\rho_{\mathrm{i}} a^4} [\vec{E} + \vec{v}_{\mathrm{i}}\times \vec{B}] + {\mathcal C}_{\mathrm{pe}},
\label{dte2}
\end{eqnarray}
where ${\mathcal C}_{\mathrm{ep}}$ and ${\mathcal C}_{\mathrm{pe}}$ 
follow from the collision terms provided by Coulomb scattering. Needless to say that Eqs. (\ref{dte1a})--(\ref{dte1}) and (\ref{dte2a})--(\ref{dte2}) can also be directly obtained from the moments of the Vlasov 
equation (written in the synchronous gauge) as discussed in Appendix \ref{APPA}.

For typical length-scales 
much larger than the Debye length and for angular frequencies $\omega \ll \omega_{\mathrm{pe}}$, 
Eqs. (\ref{dte1a})--(\ref{dte1}) and (\ref{dte2a})--(\ref{dte2}) reduce to an effective one-fluid theory.
In the one-fluid limit, by definition, the propagation 
of the electromagnetic (i.e. high frequency) disturbances 
is negligible and the displacement current vanishes. The 
Ohmic electric fields are then vanishing in the baryon 
rest frame and the (fully inhomogeneous) magnetic fields 
affect the Boltzmann hierarchy through the magnetic pressure, the magnetic energy density, the anisotropic stress and the Lorentz force.

In the opposite limit (i.e. angular frequencies larger than the plasma frequency), 
the two-fluid nature of the problem becomes physically relevant: 
the one-fluid approximation breaks down and electromagnetic waves will not simply travel at the 
speed of light (i.e. $1$ in our units). Their group velocity 
will be affected both by the plasma and Larmor frequencies of electrons 
and ions. Finally, since the motion of electrons and ions is non-relativistic (i.e. 
$T\ll m_{\mathrm{e},\mathrm{p}}$) we also have, in general terms, that
$\lambda_{\mathrm{D}} \omega_{\mathrm{pe}}= \sqrt{T/(2 m_{\mathrm{e}})}\ll 1$.
In the following two sections the one-fluid and two-fluid 
treatments will be physically compared in the light of the numerical results which will be later reported.

\renewcommand{\theequation}{3.\arabic{equation}}
\setcounter{equation}{0}
\section{Large-scale magnetic field and one-fluid treatment}
\label{sec3}
Defining, respectively, the centre of mass velocity of the electron and ion system $\vec{v}_{\mathrm{b}}$, 
the total current $\vec{J}$ and the total density contrast $\delta_{\mathrm{b}}$
\begin{equation}
\vec{v}_{\mathrm{b}} = \frac{m_{\mathrm{e}} \vec{v}_{\mathrm{e}} + m_{\mathrm{p}} \vec{v}_{\mathrm{i}}}{m_{\mathrm{e}} + m_{\mathrm{p}}},
\qquad \vec{J} = e (n_{\mathrm{i}} \vec{v}_{\mathrm{i}} - n_{\mathrm{e}} 
\vec{v}_{\mathrm{e}}),\qquad \delta_{\mathrm{b}} = \frac{\rho_{\mathrm{e}} \delta_{\mathrm{e}} + 
\rho_{\mathrm{i}} \delta_{\mathrm{i}}}{\rho_{\mathrm{e}} + \rho_{\mathrm{i}}},
\label{global1}
\end{equation}
the generalized MHD reduction can be also implemented in curved space-times.
In Eq. (\ref{global1}) $\vec{v}_{\mathrm{b}}$ is, effectively, the  bulk velocity of the plasma:
 $\vec{v}_{\mathrm{b}}$ (and its 
three-divergence $\theta_{\mathrm{b}}$) is exactly what 
it is normally employed to describe, in standard Boltzmann 
solvers \cite{cmbfast1,cmbfast2,cosmics1,cosmics2}, the baryon velocity.  Conversely 
$\delta_{\mathrm{b}}$ denotes the total density 
contrast of the electron-ion system and it is often naively identified with the baryon density contrast. 
The evolution equation for $\theta_{\mathrm{b}}$ 
can be derived by taking the three-divergence 
of the evolution equation of the electrons and of the 
ions. From Eq. (\ref{dte2}), since $\vec{\nabla}\cdot \vec{E} = 0$ we obtain
\begin{eqnarray}
&& \theta_{\mathrm{e}}' + {\mathcal H} \theta_{\mathrm{e}} = - \frac{e n_{\mathrm{e}}}{\rho_{\mathrm{e}} a^4} \vec{\nabla} 
\cdot (\vec{v}_{\mathrm{e}}\times \vec{B}),\qquad \theta_{\mathrm{e}} = \vec{\nabla}\cdot \vec{v}_{\mathrm{e}},
\label{velel}\\
&& \theta_{\mathrm{i}}' + {\mathcal H} \theta_{\mathrm{i}} =  \frac{e n_{\mathrm{i}}}{\rho_{\mathrm{i}} a^4 }\vec{\nabla} 
\cdot (\vec{v}_{\mathrm{i}}\times \vec{B}),\qquad \theta_{\mathrm{i}} = \vec{\nabla}\cdot \vec{v}_{\mathrm{i}}.
\label{velion}
\end{eqnarray}
Summing up Eq. (\ref{velel}) (multiplied by $m_{\mathrm{e}}$)
and Eq. (\ref{velion}) (multiplied by $m_{\mathrm{p}}$) the evolution equation 
for $\theta_{\mathrm{b}}$ immediately follows. Similarly, the 
evolution equation for the total density contrast of the electron ion system can be derived.  The resulting equations 
for $\theta_{\mathrm{b}}$ and $\delta_{\mathrm{b}}$ is finally given by: 
\begin{eqnarray}
&&\theta_{\mathrm{b}}' + {\mathcal H} \theta_{\mathrm{b}} = \frac{\vec{\nabla} \cdot [ \vec{J} \times \vec{B}]}{\rho_{\mathrm{b}}a^4 \biggl( 1 + \frac{m_{\mathrm{e}}}{m_{\mathrm{p}}}\biggr)} + \frac{4}{3} \frac{\rho_{\gamma}}{\rho_{\mathrm{b}}} \epsilon' (\theta_{\gamma} - \theta_{\mathrm{b}}),\qquad 
\label{global2}\\
&& \delta_{\mathrm{b}}' = - \theta_{\mathrm{b}} + \frac{h'}{2}  + \frac{\vec{E} \cdot\vec{J}}{\rho_{\mathrm{b}} a^4\biggl(1 + \frac{m_{\mathrm{e}}}{m_{\mathrm{p}}}\biggr)},
\label{global3}
\end{eqnarray}
where the Thomposn drag term has been included as:
\begin{equation}
\epsilon' = \tilde{n}_{\mathrm{e}} x_{\mathrm{e}} \sigma_{\mathrm{Th}} a,\qquad \sigma_{\mathrm{Th}} = \frac{8}{3} \pi \biggl(\frac{e^2}{m_{\mathrm{e}}^2}\biggr)^2,
\label{drag}
\end{equation}
where $\epsilon'$ is the differential optical depth ($x_{\mathrm{e}}$ is the ionization fraction). 
Note that $\epsilon'$ has been correctly written in terms 
of $\tilde{n}_{\mathrm{e}}$ (and not in terms of our comoving concentration $n_{\mathrm{e}} = a^3 \tilde{n}_{\mathrm{e}}$).
The Thompson drag is dominated by the electron-photon cross-section since the photon-proton cross section is, roughly, six orders of magnitude smaller. In Eq. (\ref{global2}), $\theta_{\gamma}$ is the three-divergence 
of the photon velocity. The photons are neutral species and, therefore, they will be described, in the first 
approximation, by the perturbed covariant conservation equation which leads to:
\begin{equation}
\theta_{\gamma}' = - \frac{1}{4} \nabla^2 \delta_{\gamma} + \epsilon' (\theta_{\mathrm{b}} - \theta_{\gamma}),\qquad \delta_{\gamma}' = - \frac{4}{3} \theta_{\gamma} + \frac{h'}{2}.
\label{photon1}
\end{equation}
Well before photon decoupling the photon and baryon velocity fields can be further combined. 
Indeed, during the radiation epoch, the efficiency of the Thompson drag synchronizes the baryon 
and photon velocities almost exactly, i.e. $\theta_{\gamma} \simeq \theta_{\mathrm{b}} = \theta_{\gamma\mathrm{b}}$. Indeed, by subtracting Eqs. (\ref{global2}) and (\ref{photon1}), the evolution equation for 
$(\theta_{\gamma} - \theta_{\mathrm{b}})$ can be obtained and it has the form:
\begin{equation}
(\theta_{\mathrm{b}} - \theta_{\gamma})' + \epsilon'\biggl(\frac{R_{\mathrm{b}} +1}{R_{\mathrm{b}}}\biggr) (\theta_{\mathrm{b}} - \theta_{\gamma}) = \frac{\vec{\nabla} \cdot [\vec{J} \times \vec{B}]}{a^4 \rho_{\mathrm{b}}} + \frac{\nabla^2 \delta_{\gamma}}{4} - {\mathcal H} \theta_{\mathrm{b}},
\label{photon2}
\end{equation}
where the $m_{\mathrm{e}}/m_{\mathrm{p}} \ll 1$ has been consistently neglected and where
\begin{equation}
R_{\mathrm{b}} = \frac{3}{4} \frac{\rho_{\mathrm{b}}}{\rho_{\gamma}} = 
\biggl(\frac{690.18}{1+ z}\biggr) \biggl(\frac{\omega_{\mathrm{b}0}}{0.02273}\biggr)
\label{RB}
\end{equation}
is the baryon to photon ratio. The ratio $m_{\mathrm{e}}/m_{\mathrm{p}}$ 
is customarily used as an expansion parameter in the standard 
MHD approximation scheme \cite{krall}. Equation (\ref{photon2}) shows that, in spite of the pre-equality 
differences in $\theta_{\gamma}$ and $\theta_{\mathrm{b}}$ 
the two velocities will be inevitably driven towards 
a common value, i.e. $\theta_{\gamma\mathrm{b}}$ which follows 
from the sum of Eqs. (\ref{global2}) and (\ref{photon1}):
\begin{equation}
\theta_{\gamma\mathrm{b}}' + {\mathcal H} \frac{R_{\mathrm{b}}}{R_{\mathrm{b}} + 1} \theta_{\gamma\mathrm{b}} = \frac{3}{4 \overline{\rho}_{\gamma} ( 1 + R_{\mathrm{b}})} \vec{\nabla}\cdot [\vec{J} \times \vec{B}] - 
\frac{\nabla^2 \delta_{\gamma}}{4 ( 1 + R_{\mathrm{b}})},
\label{TCvel}
\end{equation}
where $\overline{\rho}_{\gamma} = a^4 \rho_{\gamma}$.
While the sum of the momentum conservation for the ions and for the 
electrons led to the bulk velocity of the plasma, another linear 
combination of the same set of equations leads to the Ohm's law, i.e. 
\begin{equation}
\biggl(\frac{e^2 n_{0}}{m_{\mathrm{e}} a}\biggr)\biggl[ \vec{E} + \vec{v_{\mathrm{b}}} \times \vec{B}\biggr] - \Gamma_{\mathrm{Coul}} \vec{J} =0,
\label{Ohm1}
\end{equation}
where the Hall terms (proportional to $\vec{J} \times \vec{B}$) has been 
neglected and where the collision terms have been estimated in the 
linear approximation around a Maxwellian distribution of velocities. 
The neglect of the Hall term is perfectly consistent since we will keep
only terms which are quadratic in the magnetic field intensities. 
 This term would lead in the Einstein equations to terms 
that are cubic in the magnetic field intensities and shall then be negligible. 
Equation (\ref{Ohm1}) leads then to the canonical form 
of the Ohmic current \footnote{For some applications is useful to recall that 
$\sigma = a \overline{\sigma}$ where $\overline{\sigma}$ is the physical 
conductivity and $\sigma$ is the comoving conductivity.  
The expression of $\overline{\sigma}$ is obtained from $\sigma$ by writing $T= a \overline{T}$, i.e. by going back from the comoving to the 
physical temperatures.}:
\begin{equation}
\vec{J} = \sigma (\vec{E} + \vec{v}_{\mathrm{b}}\times \vec{B}), \qquad \sigma = 
\frac{e^2 n_{0}}{m_{\mathrm{e}} a \Gamma_{\mathrm{Coul}}} = \frac{T}{e^2}\sqrt{\frac{T}{m_{\mathrm{e}} a}}.
\label{Ohm2}
\end{equation}
It can be easily checked that the expression of the conductivity of Eq. 
(\ref{Ohm2}) is exactly the one anticipated in Eq. (\ref{hier3}). 
In this one-fluid limit the displacement current vanishes and, therefore, 
\begin{equation}
\vec{J} = \frac{1}{4\pi} \vec{\nabla}\times \vec{B}, \qquad \vec{E} + \vec{v}_{\mathrm{b}} \times \vec{B} = \frac{\vec{\nabla}\times \vec{B}}{4\pi \sigma},
\label{Ohm3}
\end{equation}
which demonstrates  that large-scale electric fields 
are suppressed in the baryon rest frame by one power of the conductivity.
Furthermore,  collisions are sufficiently frequent 
to keep the isotropy and this fit with the stochastic nature of the 
pre-decoupling magnetic field. Equation (\ref{Ohm3}) 
can be used to simplify the evolution equations of the 
photon-baryon velocity. In particular, using simple 
vector identities it can be easily shown that the following 
relations are satisfied by the large-scale magnetic fields:
\begin{equation}
\frac{3}{4\overline{\rho}_{\gamma}} \vec{\nabla}\cdot [\vec{J}\times 
\vec{B}] =  \nabla^2 \sigma_{\mathrm{B}} -\frac{ \nabla^2 \Omega_{\mathrm{B}}}{4},
\label{vect1}
\end{equation}
where $\Omega_{\mathrm{B}}$ and $\sigma_{\mathrm{B}}$ are 
two dimensionless quantities which are related, respectively, to the 
magnetic energy density and to the anisotropic stress. Because of the large value of the conductivity, 
the electric component of the MHD energy-momentum tensor 
are suppressed by two powers of the conductivity:
\begin{eqnarray}
&& \delta_{\mathrm{s}} {\mathcal T}_{0}^{0} = 
\delta_{\mathrm{s}} \rho_{\mathrm{B}}, \qquad \delta_{\mathrm{s}} {\mathcal T}_{i}^{j} = - \delta_{\mathrm{s}} p_{\mathrm{B}} + \tilde{\Pi}_{i}^{j}, \qquad 
\delta_{\mathrm{s}} {\mathcal T}_{0}^{i} = \frac{(\vec{E}\times \vec{B})^{i}}{4\pi a^4}
\label{EMT1}\\
&& \delta_{\mathrm{s}} \rho_{\mathrm{B}} = \frac{B^2}{8 \pi a^4}, \qquad 
\delta_{\mathrm{s}} p_{\mathrm{B}} = \frac{\delta_{\mathrm{s}}\rho_{\mathrm{B}}}{3},\qquad \tilde{\Pi}_{i}^{j} = \frac{1}{4\pi a^4} \biggl( B_{i} B^{j} - \frac{1}{3} \delta_{i}^{j} B^2\biggr),
\label{EMT2}
\end{eqnarray}
where $B^2 = B_{i} B^{i}$. Defining then 
\begin{equation}
\Omega_{\mathrm{B}} = \frac{\delta_{\mathrm{s}} \rho_{\mathrm{B}}}{\rho_{\gamma}}, \qquad \partial_{j}\partial^{i}\tilde{\Pi}_{i}^{j} = (p_{\gamma}+ \rho_{\gamma}) \nabla^2 \sigma_{\mathrm{B}},
\label{EMT3}
\end{equation}
Equation (\ref{vect1}) can be recovered by repeated use of known 
vector identities. Note that the Poynting vector of Eq. (\ref{EMT1}) 
is only suppressed by one power of the conductivity since it 
contains the Ohmic electric field which vanishes as $\sigma^{-1}$ 
in the baryon rest frame.

The MHD description allows then to reduce the problem 
of the evolution of large-scale magnetic fields to a 
rather well defined system where the plasma is globally 
neutral and the Ohmic current is solenoidal. This description 
is ideal in order to study the large-scale effects of the 
magnetic fields, i.e. exactly when gravity  is important. 
Indeed the MHD energy-momentum tensor must be consistently 
included in the perturbed Einstein equations 
which take then the form:
\begin{eqnarray}
&& 2 \nabla^2 \xi + {\mathcal H} h' = - 8\pi G a^2 [\delta_{\mathrm{s}}
\rho_{\mathrm{t}} + \delta_{\mathrm{s}} \rho_{\mathrm{B}}],
\label{00}\\
&& \nabla^2 \xi' = 4\pi a^2 (p_{\mathrm{t}} + \rho_{\mathrm{t}}) \theta_{\mathrm{t}}, 
\label{0i}\\
&& h'' + 2 {\mathcal H} h' + 2 \nabla^2 \xi = 24 \pi G a^2 [\delta_{\mathrm{s}} p_{\mathrm{t}} + \delta p_{\mathrm{B}}],
\label{i=j}\\
&& ( h + 6 \xi)'' + 2 {\mathcal H} (h+ 6 \xi)' + 2 \nabla^2 \xi = 
24 \pi G a^2 [(p_{\nu} + \rho_{\nu}) \sigma_{\nu} + (p_{\gamma} + 
\rho_{\gamma})\sigma_{\mathrm{B}}],
\label{ineqj}
\end{eqnarray}
where Eqs. (\ref{00}) and (\ref{0i}) are, respectively, the Hamiltonian 
and the momentum constraints. Equations (\ref{i=j}) and (\ref{ineqj}) 
follow, respectively, from the $(i=j)$ and $(i\neq j)$ components 
of the perturbed Einstein equations. The term containing 
$\sigma_{\nu}$ is simply related to the neutrino 
anisotropic stress as $\partial_{i}\partial^{j} \Pi_{j}^{i} = (p_{\nu} + \rho_{\nu})
\nabla^2\sigma_{\nu}$.
In Eqs. (\ref{00})--(\ref{ineqj}) $\delta_{\mathrm{s}}\rho_{\mathrm{t}}$ 
and $\theta_{\mathrm{t}}$ denote the total density fluctuation
and the total velocity field:
\begin{eqnarray}
&&\delta_{\mathrm{s}} \rho_{\mathrm{t}} = 
\delta_{\mathrm{s}}\rho_{\gamma} +\delta_{\mathrm{s}} \rho_{\nu} + 
\delta_{\mathrm{s}} \rho_{\mathrm{i}} + \delta\rho_{\mathrm{e}} 
+ \delta\rho_{\mathrm{c}},
\label{drhot}\\
&& (p_{\mathrm{t}} + \rho_{\mathrm{t}}) \theta_{\mathrm{t}} = 
\frac{4}{3} \rho_{\gamma} \theta_{\gamma} +
\frac{4}{3} \rho_{\nu} \theta_{\nu} + \rho_{\mathrm{e}} \theta_{\mathrm{e}}
+ \rho_{\mathrm{i}}\theta_{\mathrm{i}} + \rho_{\mathrm{c}} \theta_{\mathrm{c}}.
\label{tht}
\end{eqnarray}
In Eqs. (\ref{drhot}) and (\ref{tht}) the various subscripts 
refer, with obvious notation, to the various species of the plasma 
and the different numerical weights in Eq. (\ref{tht}) 
are a simple reflection of the difference of the various pressures.
The only species not introduced so far are the CDM particles 
and the neutrinos. The CDM component obeys, in the synchronous 
gauge, the following pair of equations
\begin{equation}
\delta_{c}' = - \theta_{\mathrm{c}} + \frac{h'}{2},\qquad \theta_{\mathrm{c}}' + {\mathcal H} \theta_{\mathrm{c}} =0.
\label{CDM}
\end{equation}
The neutrino component decouples at temperatures 
of the order of the MeV and it is therefore described by the appropriate 
Boltzmann equation which reads, in the synchronous gauge (see last part of Appendix \ref{APPA}), 
\begin{equation}
{\mathcal F}_{\nu}' + i k \mu {\mathcal F}_{\nu} = - 4 \xi' + 2 \mu^2 (h' + 6 \xi').
\label{BZnu}
\end{equation}
The evolution equations of lowest moments of the Boltzmann equation can be obtained with standard techniques and they are:
\begin{eqnarray}
&&\delta_{\nu}' = - \frac{4}{3} \theta_{\nu} + \frac{2}{3} h',
\label{nu1}\\
&& \theta_{\nu}' = \frac{k^2}{4} \delta_{\nu} - k^2 \sigma_{\nu},
\label{nu2}\\
&& \sigma_{\nu}' = \frac{4}{15} \theta_{\nu} - \frac{3k}{10} {\mathcal F}_{\nu 3} - \frac{2}{15}(h' + 6 \xi'),
\label{nu3}
\end{eqnarray}
where $\sigma_{\nu} = {\mathcal F}_{\nu 2}/2$, i.e. the neutrino anisotropic stress appearing in Eq. (\ref{ineqj}) is simply 
related by a numerical factor to the quadrupole moment of the neutrino 
phase space distribution.

Most of the considerations related to the presence of the magnetic fields before equality 
can be conducted in rather general terms. For instance, the analytical form of the initial conditions
(reported for completeness in Appendix \ref{APPB})  
can be deduced without specifying the magnetic field configuration. 
Prior to equality magnetic fields can certainly be present even if they 
are physically constrained by the isotropy of the background 
geometry:  the magnetic field 
cannot be uniform. If the magnetic field is stochastically 
distributed, its two-point function in Fourier space must be divergenceless 
and it is given by 
\begin{equation}
\langle B_{i}(\vec{k}) B_{j}(\vec{p}) \rangle = \frac{2\pi^2}{k^3} P_{ij}(k) {\mathcal P}_{B}(k) \delta^{(3)}(\vec{k} + \vec{p}), \qquad P_{ij}(k) = \biggl( \delta_{ij} - \frac{k_{i} k_{j}}{k^2}\biggr),
\label{F0}
\end{equation}
where ${\mathcal P}_{B}(k)$ denotes the magnetic power spectrum. 
In principle there could be a second contribution to the two-point function 
of Eq. (\ref{F0}).  Such a contribution would lead to a non-vanishing 
magnetic helicity. Configurations of this type might arise 
at the electroweak epoch and have been named hypermagnetic knots  \cite{hk1,hk2} (see also \cite{hk3,hk4}).
As far as the scalar modes are concerned the helical part of the field plays no role.
The same conclusion holds in the case of the Faraday effect by a stochastic magnetic field, as it has 
been explicitly discussed in \cite{far8} (see also \cite{far5}).

It is useful to mention, at this point, that large-scale (tangled) magnetic fields 
might have also specific effects related to the vector and tensor modes of the geometry (which are minute at large scales). These effects have been analyzed in \cite{bs1,bs2,bs3}. 
The analysis of large-scale magnetic fields  can also be conducted within fully 
covariant approaches \cite{cs1,cs2} which are, when appropriately handled, equivalent to the one devised in this paper and based on earlier analyses 
\cite{mg1,mg4}.
\renewcommand{\theequation}{4.\arabic{equation}}
\setcounter{equation}{0}
\section{High-frequency waves and two-fluid effects}
\label{sec4}

As already noticed in \cite{gk2}, the tight Thompson coupling, well 
verified prior to equality, effectively breaks down, by definition, 
around photon decoupling so the baryon and photon velocities 
(i.e. $\theta_{\mathrm{b}} \neq \theta_{\gamma}$) 
will have to be numerically followed, as in \cite{gk2}, by means
of a generalized slip equation. 
In similar terms, when the frequency of the polarized CMB 
photons greatly exceeds the plasma frequency, we can expect 
that the two-fluid nature of the baryon-lepton system will be somehow 
``resolved". By resolved we mean that it will make actually 
a difference that instead of one (globally neutral) fluid 
the plasma is formed by two (intrinsically charged) fluids (i.e. $\theta_{\mathrm{i}} \neq \theta_{\mathrm{e}}$ and $\vec{v}_{\mathrm{e}} \neq \vec{v}_{\mathrm{i}}$).

Dispersive phenomena in the pre-decoupling plasma arise exactly in this limit
 where the angular frequencies of the propagating electromagnetic waves 
are larger than the plasma frequency.  
Faraday rotation (as well as other dispersive phenomena at high frequencies)
can be derived by studying the evolution of electromagnetic disturbances 
when the electromagnetic background contains a finite 
concentration of charges and a stochastic magnetic field.

This analysis mirrors what is customarily done 
in the case of a magnetized plasmas in the laboratory:  MHD is used to 
describe the large-scale magnetic field while the two fluid 
treatment is enforced to scrutinize the propagation of electromagnetic disturbances.
In this sense, the large-scale magnetic field plays effectively the role of a {\em background }
field. This terminology is a bit ambiguous in the sense that the magnetic field is not part of the 
{\em geometric background} (i.e. it does not determine the expansion rate at equality or decoupling), but it is rather 
part of the   {\em electromagnetic background } determining the dispersion relations for the waves propagating in the plasma.

Equations (\ref{MX6}) and (\ref{MX7}) together with Equations (\ref{dte2})
and Eqs. (\ref{MX1})--(\ref{MX4}) can be  linearized in the 
presence of the magnetic field $\vec{B}(\vec{x})$, i.e. 
\begin{eqnarray}
&& n_{\rm e,\,i}(\tau, \vec{x}) = n_{0} + \delta n_{\rm e,\,i}( \tau,\vec{x}),
\qquad \vec{B}(\tau,\vec{x}) = \vec{B}(\vec{x}) + \vec{b}(\tau,\vec{x}),
\nonumber\\
&& \vec{v}_{\rm e,\,i}(\tau,\vec{x}) = 
\delta \vec{v}_{\rm e,\,i} (\tau,\vec{x}),\qquad
\vec{E}(\tau,\vec{x})=  \vec{e}(\tau,\vec{x}).
\label{fluct}
\end{eqnarray}
where $\vec{B}(\vec{x})$ is the large-scale magnetic 
obeying the MHD equations discussed in the previous section and characterized by the two-point 
function given in Eq. (\ref{F0}).  The field variables  
$\vec{e}$ and $\vec{b}$ denote, respectively,
the electric and magnetic fields of the wave. The absence 
of a background electric field is a reflection of the largeness of the Coulomb conductivity in units 
of the plasma frequency (see Eq. (\ref{hier3}) and also Section \ref{sec3}).
Using the notations of Eq. (\ref{fluct}), Eqs. (\ref{MX6}), (\ref{MX7}) 
and (\ref{dte2}) lead, respectively, to the  following set of equations:
\begin{eqnarray}
&& \delta n_{\rm e}' + n_{0} \vec{\nabla} \cdot \delta \vec{v}_{\rm e} =0,
\,\,\,\,\,\,\,\,\,\, \delta n_{\rm i}' + n_{0} \vec{\nabla} \cdot \delta \vec{v}_{\rm i} =0,
\label{deltanep}\\
&& \delta\vec{v}_{\rm e}' + {\cal H} \delta \vec{v}_{\rm e} = 
- \frac{e }{m_{\rm e} a} \biggl[\vec{e} 
+ \delta\vec{v_{\rm e}} \times \vec{B}\biggr],
\qquad \delta\vec{v}_{\rm i}^{\,\prime} + {\cal H} \delta \vec{v}_{\rm i} = \frac{e}{m_{\rm i}a} \biggl[ \vec{e} 
+ \delta\vec{v_{\rm i}} \times \vec{B}\biggr],
\label{deltavep}\\
&&\vec{\nabla} \times \vec{e} = -  \vec{b}^{\,\prime},
\,\,\,\,\,\,\,\, \vec{\nabla} \cdot \vec{e} = 0,
\label{deltadiv}\\
&& \vec{\nabla}\times  \vec{b} =  \vec{e}^{\,\prime}+ 4\pi\,e\,n_{0} ( \delta \vec{v}_{\rm i} - \delta \vec{v}_{\rm e}).
\label{deltacurl}
\end{eqnarray}
As the structure of the system 
shows, the waves parallel (i.e. $\parallel$)  and orthogonal to the magnetic field $\vec{B}$ (i.e. 
$\perp$)will then obey different dispersion relations.  The important proviso is that 
the magnetic field is inhomogeneous over typical scales that are much larger than the 
Larmor radius. This observation, already mentioned in Eq. (\ref{hier4}), will now be made 
more explicit. Let us consider, for instance, the decoupling time which takes place, according to the 
WMAP 5-year data, for $z_{\mathrm{dec}} = 1089.9$. At this epoch $T_{\mathrm{dec}}\simeq 0.25$ eV 
and $r_{\mathrm{H}}(\tau_{\mathrm{dec}}) = 286$ Mpc. The typical scale of variation 
of the magnetic field at decoupling, i.e. $L(\tau_{\mathrm{dec}})$ will be of the order of the Hubble radius and certainly 
larger than the magnetic diffusivity length defined in Eq. (\ref{Lsigma}):
\begin{eqnarray}
&& r_{\mathrm{Be}}(\tau_{\mathrm{dec}}) = \frac{v_{\mathrm{th}}}{\omega_{\mathrm{Be}}(\tau_{\mathrm{dec}})} \simeq 1.19 \times 
10^{10} \biggl( \frac{\vec{B}\cdot \hat{n}}{\mathrm{nG}}\biggr)^{-1}\,\, \mathrm{cm},
\nonumber\\
&& L_{\sigma}(\tau_{\mathrm{dec}}) = ( 4 \pi \sigma_{\mathrm{dec}} H_{\mathrm{dec}})^{-1/2} \simeq 1.4 \times 10^{11}\,\,\mathrm{cm}.
\label{var1}
\end{eqnarray}
These figures imply that, indeed, $L(\tau_{\mathrm{dec}}) > r_{\mathrm{Be}}(\tau_{\mathrm{dec}})$ since $L(\tau_{\mathrm{dec}})\gg 
L_{\sigma}(\tau_{\mathrm{dec}})$. The magnetic 
field, inhomogeneous over scales much larger than the Larmor radius, can therefore 
be considered uniform for the purpose of deriving the dispersion relations. What has been introduced here in 
terms of the typical scales of our problem is indeed one of the most useful approximations 
of the physics of cold plasmas and it has been pioneered by Alfv\'en \cite{ALF1,ALF2}.

 In general the background magnetic fields will be, as in our 
case, solution of the Maxwell's equations and will therefore be both space and time 
dependent. Let us restrict ourselves, for the present, to spatially inhomogeneous magnetic fields\footnote{This 
is the physical situation for $\vec{B}(\vec{x})$ if the magnetic flux is strictly conserved. In other words, it can be 
shown that $\vec{B}(\vec{x})$ is a solution of the MHD equations in the limit of infinite conductivity. In the opposite case 
a power series in $1/\sigma$ will arise in  the background solution.} $\vec{B}(\vec{x})$. If the inhomogeneity 
is small, it is possible to determine the trajectory of the charged species as a perturbation around the center 
of the particle orbit:
\begin{equation}
\vec{B}(\vec{x}) = \vec{B}(\vec{x}_{0}) +  (\delta \vec{x}\cdot \vec{\nabla}) \vec{B}|_{\vec{x}=\vec{x_{0}}}, \qquad \delta\vec{x} = 
\vec{x} - \vec{x}_{0},
\label{guide}
\end{equation}
where $\vec{x}_{0}$ is the instantaneous position of the guiding center. Equation (\ref{guide}) 
holds provided $\delta \vec{B}$ (i.e. the change in the magnetic field over a typical distance 
of the order of the Larmor radius $r_{\mathrm{Be}}$)  is such that $|\delta\vec{B}| = |(\delta \vec{x}\cdot \vec{\nabla}) \vec{B}|
\ll \vec{B}$. But this implies exactly Eq. (\ref{var1}), i.e. $r_{\mathrm{Be}}\ll L$ where $L$ is the distance over which 
the magnetic field changes significantly.
As already mentioned this perturbative approach was extensively applied by Alfv\'en  and it often dubbed 
as the guiding centre approximation in the plasma literature.
This point is rather important and it is precisely where our treatment diverges from the studies 
reported in Ref. \cite{far1,far2,far3,far4}. In those treatments the magnetic fields have been 
considered just globally uniform. This would break spatial isotropy and it is not true 
for a stochastic magnetic field. The correct physical statement is that the magnetic field is uniform 
over the typical scale of the particle orbit, i.e. the field experienced by the electron in traversing a Larmor 
orbit is almost constant. It is then 
practical to Fourier transform with respect to the conformal time 
variable \footnote{In the present and in the following sections 
$\overline{\omega}$ and $k$ denote, respectively, the comoving angular frequency and the 
comoving wavenumber.}
\begin{eqnarray}
\vec{b}(\vec{x}, \tau) = \vec{b}_{\overline{\omega}}(\vec{x}) e^{- i \int \overline{\omega} d\tau},\qquad 
\vec{e}(\vec{x},\tau) = \vec{e}_{\overline{\omega}}(\vec{x}) e^{- i \int \overline{\omega} d\tau},
\end{eqnarray}
The solution of Eq. (\ref{deltavep}) is trivial along the direction 
of $\vec{B}$ and the total current can be written as 
\begin{equation}
\vec{j}_{\parallel, \overline{\omega}}(\vec{x}) = \frac{i}{4\pi} \frac{\overline{\omega}_{\rm p\,i}^2 
+ \overline{\omega}_{\rm p\,e}^2}{\overline{\omega} ( 1 + \alpha)} \vec{e}_{\parallel,\overline{\omega}}(\vec{x}),
\label{jpar2}
\end{equation}
where $\alpha = i {\mathcal H}/\overline{\omega} = H/\omega\ll 1$ accounts for the curved-space corrections and where
\begin{eqnarray}
&&\overline{\omega}_{\mathrm{pi}} = \sqrt{\frac{4\pi n_{0}x_{\mathrm{e}} e^2}{m_{\mathrm{p}} a}}  
\equiv \omega_{\mathrm{pi}} a,
\qquad \omega_{\mathrm{pi}} = \sqrt{\frac{4\pi \tilde{n}_{\mathrm{i}}x_{\mathrm{e}} e^2}{m_{\mathrm{p}}}},
\label{freq1}\\
&&\overline{\omega}_{\mathrm{pe}} = \sqrt{\frac{4\pi n_{0} x_{\mathrm{e}} e^2}{m_{\mathrm{e}} a}}  
\equiv  \omega_{\mathrm{pe}} a,
\qquad \omega_{\mathrm{pe}} =\sqrt{\frac{4\pi \tilde{n}_{\mathrm{e}} x_{\mathrm{e}} e^2}{m_{\mathrm{e}}}}.  
\label{freq2}
\end{eqnarray}
In Eqs. (\ref{freq1}) and (\ref{freq2}) 
the comoving (angular) frequencies  $\overline{\omega}_{\mathrm{pe}}$ and $\overline{\omega}_{\mathrm{Be}}$ 
have been related to the corresponding physical frequencies  $\omega_{\mathrm{pe}}$ and $\omega_{\mathrm{Be}}$ by recalling 
that $n_{0} = a^{3} \tilde{n}_{\mathrm{e}} =  a^{3} \tilde{n}_{\mathrm{i}}$. It is relevant 
to mention, once more, that the presence of the scale factor in $\overline{\omega}_{\mathrm{pi}}$ and 
$\overline{\omega}_{\mathrm{pe}}$ is a direct consequence of the breaking of conformal invariance induced by the 
masses of the electrons and ions.
In the direction orthogonal to $\vec{B}$, 
the evolution equations of the electromagnetic waves can be recast in the following 
handy form:
\begin{equation}
(\vec{\nabla}\times \vec{b})_{\overline{\omega}} = - i \overline{\omega} \epsilon(\overline{\omega},\alpha)
\vec{e}_{\overline{\omega}}, \qquad (\vec{\nabla}\times \vec{e})_{\overline{\omega}} - i \overline{\omega} \vec{b}_{\overline{\omega}}  =0,
\label{freq3}
\end{equation}
where  the dielectric tensor $\epsilon(\overline{\omega},\alpha)$ can be written in a generalized matrix notation as:
\begin{equation}
\epsilon(\overline{\omega},\alpha) 
= \left(\matrix{\epsilon_{\perp 1}(\overline{\omega},\alpha)
& i\epsilon_{\perp 2}(\overline{\omega},\alpha) & 0&\cr
-i \epsilon_{\perp 2}(\overline{\omega},\alpha) & \epsilon_{\perp 1}(\overline{\omega},\alpha) &0&\cr
0&0&\epsilon_{\parallel}(\overline{\omega},\alpha) }\right),
\label{epstens}
\end{equation}
having defined the corresponding components as 
\begin{eqnarray}
&& \epsilon_{\parallel}(\overline{\omega},\alpha) = 1 - \frac{ \overline{\omega}_{\rm p\,i}^2}{\overline{\omega}^2 (1 +\alpha)} - \frac{ \overline{\omega}_{\rm p\,e}^2}{\overline{\omega}^2 (1 +\alpha)},
\label{epspar}\\
&& \epsilon_{\perp 1}(\overline{\omega},\alpha) = 1 - \frac{\overline{\omega}^2_{\rm p\, i} (\alpha + 1) }{\overline{\omega}^2 (\alpha+ 1)^2 - \overline{\omega}_{\rm B\, i}^2} -
\frac{\overline{\omega}^2_{\rm p\, e} (\alpha + 1) }{\overline{\omega}^2 (\alpha+ 1)^2 - \overline{\omega}_{\rm B\, e}^2},
\label{epsperp1}\\
&& \epsilon_{\perp 2}(\omega,\alpha) = 
\frac{\overline{\omega}_{\rm B\, e}}{\overline{\omega}} \frac{\overline{\omega}^2_{\rm p\, e } }{\overline{\omega}^2 (\alpha+ 1)^2 - \overline{\omega}^2_{\rm B\, e}} 
-  \frac{\overline{\omega}_{\rm B\, i}}{\overline{\omega}} \frac{\overline{\omega}^2_{\rm p\, i } }{\overline{\omega}^2 (\alpha+ 1)^2 - \overline{\omega}^2_{\rm B\, i}}.
\label{epsperp2}
\end{eqnarray}
In Eqs. (\ref{epsperp1}) and (\ref{epsperp2}), on top of the comoving plasma frequencies introduced in 
Eqs. (\ref{freq1})--(\ref{freq2}), there appear also the comoving Larmor frequencies for the electrons and for the ions:
\begin{eqnarray}
&& \overline{\omega}_{\mathrm{Be}} = \frac{e \vec{B}\cdot\hat{n}}{m_{\mathrm{e}} a} = \omega_{\mathrm{Be}} a,\qquad  
\omega_{\mathrm{Be}} = \frac{ e \vec{{\mathcal B}}\cdot \hat{n}}{m_{\mathrm{e}}}
\label{freq4}\\
&&\overline{\omega}_{\mathrm{Bi}} = \frac{e \vec{B}\cdot\hat{n}}{m_{\mathrm{i}}a} = \omega_{\mathrm{Bi}} a,\qquad  
\omega_{\mathrm{Bi}} = \frac{ e \vec{{\mathcal B}}\cdot \hat{n}}{m_{\mathrm{p}}}.
\label{freq5}
\end{eqnarray}
where the relation between the comoving and the physical magnetic fields, i.e.  $\vec{B} = a^2 \vec{{\mathcal B}}$, has been 
exploited.
Equations (\ref{freq3}) and (\ref{epstens}) can be finally combined leading to the compact relation
\begin{equation}
k^2 \vec{e}_{\vec{k},\overline{\omega}} - (\vec{k}\cdot\vec{e}_{\vec{k},\overline{\omega}}) \vec{k} =\epsilon(\overline{\omega},\alpha) \overline{\omega}^2 \vec{e}_{\vec{k},\overline{\omega}}.
\label{relation}
\end{equation}
Defining
\begin{equation}
\vec{k}_{\parallel} = k (\hat{k}\cdot\hat{n}) = k \cos{\vartheta},\qquad 
\vec{k}_{\perp} = k \sin{\vartheta},
\end{equation}
 the dispersion relations can be read-off from Eq. (\ref{relation}), in matrix form\footnote{ As usual the refractive 
index is defined as $\overline{\omega}/k= 1/n$. The refractive index $n$ should not be confused, of course,  with the unit 
vector $\hat{n}$.}:
\begin{equation}
 \left(\matrix{\bigl[1 - \frac{\epsilon_{\perp1}}{n^2}\bigr]
& - i\frac{\epsilon_{\perp 2}}{n^2} 
 & 0 &\cr
i\frac{\epsilon_{\perp 2}}{n^2} 
& \bigl[ \cos^2{\vartheta} - \frac{\epsilon_{\perp 1}}{n^2} \bigr] 
& - \sin{\vartheta} \cos{\vartheta}&\cr
0  & - \sin{\vartheta} \cos{\vartheta}
&\bigl[ \sin^2{\vartheta} -\frac{\epsilon_{\parallel}}{n^2}\bigr] }\right) \left(\matrix{e_{k,\omega,\perp 1}\cr e_{k,\omega,\perp 2}\cr
 e_{k,\omega,\parallel}}\right)=0.
\label{Amatrix}
\end{equation}
Requiring that the determinant of the matrix (\ref{Amatrix}) vanishes we get what is 
sometimes called Appleton-Hartree dispersion relation \cite{stix}:
\begin{equation}
2 \epsilon_{\parallel} \cos^2{\theta}[ (n^2 - \epsilon_{-}) (n^2 - \epsilon_{+})]
= \sin^{2}{\theta} (\epsilon_{\parallel} - n^2) [ n^2 (\epsilon_{+} + \epsilon_{-})
- 2 \epsilon_{+} \epsilon_{-}],
\label{dispersion}
\end{equation}
where 
\begin{equation}
\epsilon_{+}(\omega,\alpha) = \epsilon_{\perp 1}(\omega,\alpha) + \epsilon_{\perp 2}(\omega,\alpha) ,
\qquad 
\epsilon_{-}(\omega,\alpha) = \epsilon_{\perp 1}(\omega,\alpha) - \epsilon_{\perp 2}(\omega,\alpha).
\label{dispersion2}
\end{equation}
If $\theta =0$ in Eq. (\ref{dispersion}) (i.e. the wave propagates along the magnetic field) the waves with positive helicity  (i.e. $\hat{e}_{+}$) and negative helicity (i.e. 
$\hat{e}_{-}$) experience two different phase velocities given, respectively, by
\begin{equation}
v_{\pm}(\omega,\alpha) = \frac{1}{n_{\pm}(\overline{\omega},\alpha)},\qquad n_{\pm}(\overline{\omega},\alpha) = 
\sqrt{\epsilon_{\pm}(\overline{\omega},\alpha)}.
\label{dispersion3}
\end{equation}
If the propagation of the wave is orthogonal to the magnetic field direction, there are two 
possible modes the {\em ordinary}  and the {\em extraordinary} wave with 
dispersion relations given, respectively, by
\begin{equation}
n^2(\overline{\omega},\alpha) = \epsilon_{\parallel}(\overline{\omega},\alpha),\qquad n^2(\overline{\omega},\alpha) = 
\frac{2 \epsilon_{+}(\overline{\omega},\alpha)\epsilon_{-}(\overline{\omega},\alpha)}{\epsilon_{+}(\overline{\omega},\alpha) + 
\epsilon_{-}(\overline{\omega},\alpha)}.
\label{dispersion4}
\end{equation}
The dispersion relations of Eq. (\ref{dispersion4}) are simply $\overline{\omega}^2 = k^2$ in the physical 
range of parameters. The reason for this conclusion is that, in the physical system under study, 
\begin{itemize}
\item{} the plasma and ion frequencies of the electrons are always much larger 
than the corresponding frequencies of the ions;
\item{} the CMB angular frequency is, in turn, much larger than, both, the 
Larmor and plasma frequencies of the electrons.
\end{itemize}
The comoving Larmor and plasma frequencies for electrons and ions are, respectively:
\begin{eqnarray}
&& \overline{\omega}_{\mathrm{Be}} =0.01759 \biggl(\frac{\hat{n}\cdot\vec{B}}{\mathrm{nG}}\biggr) \,\, \mathrm{Hz},
\qquad \overline{\omega}_{\mathrm{Bi}} = 9.578\times 10^{-6} \biggl(\frac{\hat{n}\cdot\vec{B}}{\mathrm{nG}}\biggr) \mathrm{Hz},
\label{freq6}\\
&& \overline{\omega}_{\mathrm{pe}} = 0.285 \, \sqrt{x_{\mathrm{e}}} \, \biggl( \frac{h_{0}^2 \Omega_{\mathrm{b}0}}{0.02773}\biggr)^{1/2} \,\, \mathrm{MHz}, \qquad  \overline{\omega}_{\mathrm{pi}} = 6.652 \, \sqrt{x_{\mathrm{e}}} \, \biggl( \frac{h_{0}^2 \Omega_{\mathrm{b}0}}{0.02273}\biggr)^{1/2} \,\, \mathrm{kHz}.
\label{freq7}
\end{eqnarray}
Since in our code the present value of the scale factor is normalized to $1$, the present value of the physical 
frequency coincides, by definition, with the comoving frequencies. The figures of Eqs. (\ref{freq6}) and (\ref{freq7}) should 
be compared with the typical frequency of CMB photons which is clearly much larger: the maximum 
of the CMB emission is located for a comoving angular frequency $\overline{\omega}_{\mathrm{max}} = 2\pi \overline{\nu}_{\mathrm{max}}$ where $\overline{\nu}_{\mathrm{max}} = 222.617\,\, \mathrm{GHz}$.  We are 
therefore in the physical regime specified by the following hierarchies between 
frequencies
\begin{equation}
\frac{\overline{\omega}_{\mathrm{pe}}}{\overline{\omega}_{\mathrm{Be}}}\gg 1, \qquad \frac{\overline{\omega}_{\mathrm{pi}}}{\overline{\omega}_{\mathrm{Bi}}} \gg 1, 
\qquad \frac{\overline{\omega}_{\mathrm{pe}}}{\overline{\omega}_{\mathrm{pi}}} \gg 1, \qquad \frac{\overline{\omega}}{\overline{\omega}_{\mathrm{pe}}} \gg 1,
\label{freq8}
\end{equation}
where $\overline{\omega}$ denotes a typical CMB angular frequency. Under the conditions expressed by Eq. (\ref{freq8}) 
the expressions of $\epsilon_{\pm}(\overline{\omega},\alpha)$ greatly simplifies and the result is, quite 
straightforwardly, 
\begin{eqnarray}
&&\epsilon_{+}(\overline{\omega},\alpha) = 1 - \frac{\overline{\omega}_{\mathrm{pe}}^2}{\overline{\omega}[ \overline{\omega}(\alpha + 1) + \overline{\omega}_{\mathrm{Be}}]}, 
\label{freq9}\\
&&\epsilon_{-}(\overline{\omega},\alpha) = 1 - \frac{\overline{\omega}_{\mathrm{pe}}^2}{\overline{\omega}[ \overline{\omega}(\alpha + 1) - \overline{\omega}_{\mathrm{Be}}]}. 
\label{freq10}
\end{eqnarray}
Equations (\ref{freq9}) and (\ref{freq10}) imply, together with the hierarchies deduced in Eq. (\ref{freq8}), that 
the dispersion relations of the ordinary and extraordinary wave are $\overline{\omega}^2 =k^2$.
Indeed, the relation between the comoving wavenumber 
and the comoving angular frequency for the ordinary (i.e. $k_{\mathrm{O}}$) and for the extraordinary (i.e. $k_{\mathrm{E}}$) waves):
\begin{eqnarray}
&& k_{\mathrm{O}} = \overline{\omega}\sqrt{1 - \frac{\overline{\omega}_{\mathrm{pe}}^2}{\overline{\omega}^2 (1 + \alpha)}},
\label{OW}\\
&& k_{\mathrm{E}} = \overline{\omega} \sqrt{\frac{\overline{\omega}
[\overline{\omega}^2 (\alpha + 1)^2  - \overline{\omega}_{\mathrm{Be}}^2] - 2 \overline{\omega}_{\mathrm{pe}}^2 \overline{\omega}^2 (\alpha + 1) + \overline{\omega}_{\mathrm{pe}}^4}{\overline{\omega}^2 [ \overline{\omega}^2 (\alpha+1)^2 - \overline{\omega}_{\mathrm{Be}}^2 - \overline{\omega}_{\mathrm{pe}}^2 (\alpha + 1)]}}.
\label{EW}
\end{eqnarray}
It is clear that Eq. (\ref{OW}) leads immediately to $k_{\mathrm{O}} = \overline{\omega}$ since, according to Eqs. (\ref{freq7}) 
and (\ref{freq8}), $(\overline{\omega}_{\mathrm{pe}}^2/\overline{\omega}^2) \simeq {\mathcal O}(10^{-12})$. 
Similar conclusion can be drawn, after some algebra, for the extraordinary wave. 
Equations (\ref{freq10}) and (\ref{EW}) do not seem to forbid the existence 
of resonances. The resonance of Eq. (\ref{freq10}) is the well known cyclotron resonance 
which is here avoided since the typical CMB frequencies are much larger than $\overline{\omega}_{\mathrm{Be}}$.
For the same reason also the resonance arising in Eq. (\ref{EW}) is avoided for the physical range 
of frequencies.

\renewcommand{\theequation}{5.\arabic{equation}}
\setcounter{equation}{0}
\section{CMB polarization and Faraday screening}
\label{sec5}
 The different dispersive 
behaviour of the positive and negative helicities implies a rotation of the components of the electric (or magnetic)
field of the wave. Thus, also the related Stokes parameters will be rotated.
Within the adopted set of conventions, the Faraday-rotated 
Stokes parameters are the same ones we would obtain by the appropriate rotation of the coordinate system.  
Consider, for sake of concreteness, a monochromatic wave propagating along the magnetic field direction an linearly 
polarized along $\hat{e}_{1}$ at $\tau =0$ and $z=0$:
\begin{equation}
\vec{e}(z, \tau) = E_{0} \hat{e}_{1} e^{- i (\overline{\omega}\tau - k z)}.
\label{pol1}
\end{equation}
Since the positive and negative helicities are defined as
\begin{equation}
\hat{e}_{+} = \frac{\hat{e}_{1} + i \hat{e}_{2}}{\sqrt{2}}, \qquad 
\hat{e}_{-} = \frac{\hat{e}_{1} - i \hat{e}_{2}}{\sqrt{2}}, 
\label{pol1a}
\end{equation}
the linear polarization is simply composed 
of two circularly polarized waves, one with positive helicity (propagating with wavenumber 
$k_{+} = \sqrt{\epsilon_{+}(\overline{\omega},\alpha)} \,\,\overline{\omega}$) the other with negative helicity (propagating with wavenumber $k_{-} = \sqrt{\epsilon_{-}(\overline{\omega},\alpha)} \,\, \overline{\omega}$).
Equation (\ref{pol1}) can be rephrased in terms of the two propagating helicities:
\begin{equation}
\vec{e}(z,\tau) = \frac{E_{0}}{\sqrt{2}} \biggl[ \hat{e}_{+} e^{- i (\overline{\omega}\tau - k_{+} z)} + \hat{e}_{-} e^{- i (\overline{\omega}\tau - k_{+} z)}\biggr] =
\frac{E_{0}}{2} \biggl[ \hat{e}_{1} \biggl( e^{i k_{+} z} + e^{i k_{-} z}\biggr) + i \hat{e}_{2}  \biggl( e^{i k_{+} z} - e^{i k_{-} z}\biggr)\biggl]e^{- i \overline{\omega}\tau},
\label{pol2}
\end{equation}
where the second equality follows from the definitions of $\hat{e}_{\pm}$; the 
refractive indices $n_{\pm}(\overline{\omega},\alpha) = \sqrt{\epsilon_{\pm}(\overline{\omega},\alpha)}$ have been introduced in Eq. (\ref{dispersion3})  (see also Eqs. (\ref{freq9}) and (\ref{freq10})).
Since the polarization plane of the incoming wave is rotated, two out of four Stokes parameter 
\begin{eqnarray}
&& I = |\vec{e}\cdot\hat{e}_{1}|^2 + |\vec{e}\cdot\hat{e}_{2}|^2,
\label{Istokes}\\
&& Q =  |\vec{e}\cdot\hat{e}_{1}|^2 - |\vec{e}\cdot\hat{e}_{2}|^2,
\label{Qstokes}\\
&& U=  2 \mathrm{Re}[(\vec{e}\cdot\hat{e}_{1})^{*} (\vec{e} \cdot\hat{e}_{2})],
\label{Ustokes}\\
&& V = 2 \mathrm{Im}[(\vec{e}\cdot\hat{e}_{1})^{*} (\vec{e} \cdot\hat{e}_{2})].
\label{Vstokes}
\end{eqnarray}
will be rotated: while $I$ and $V$ will be left invariant, $Q$ and $U$ are rotated.
Suppose then the initial wave is linearly polarized along $\hat{e}_{1}$ and $Q^{(\mathrm{in})} = E_{0}^2$. 
Inserting Eq. (\ref{pol2}) into 
Eqs. (\ref{Qstokes}) and (\ref{Ustokes}), the Faraday rotated Stokes parameters 
are
\begin{equation}
Q^{(\mathrm{F})} = Q^{(\mathrm{in})} \cos{(2 \Delta \varphi^{\mathrm{F}})},\qquad 
U^{(\mathrm{F})} = -  Q^{(\mathrm{in})} \sin{(2 \Delta \varphi^{\mathrm{F}})},
\label{rot1}
\end{equation}
where
\begin{equation}
\Delta \varphi^{(\mathrm{F})} = \frac{\overline{\omega}}{2}\biggl[ \sqrt{\epsilon_{+}(\overline{\omega}, \alpha)} - 
\sqrt{\epsilon_{-}(\overline{\omega}, \alpha)}\biggr] \Delta z.
\label{rot2}
\end{equation}
Equation (\ref{rot1}) implies also that $Q^{(\mathrm{F})}/U^{(\mathrm{F})} = - \cot{(2 \Delta \varphi^{(\mathrm{F})})}$.
In more general terms, if the initial wave is not polarized along a specific Cartesian direction,  Eq. (\ref{rot1}) 
becomes:
\begin{eqnarray}
&&Q^{(\mathrm{F})} = Q^{(\mathrm{in})} \cos{( 2 \Delta \varphi^{(\mathrm{F})})} +  U^{(\mathrm{in})} 
\sin{( 2 \Delta \varphi^{(\mathrm{F})})},
\nonumber\\
&& U^{(\mathrm{F})} = - Q^{(\mathrm{in})} \sin{( 2 \Delta \varphi^{(\mathrm{F})})} +  U^{(\mathrm{in})} 
\cos{( 2 \Delta \varphi^{(\mathrm{F})})}.
\label{rot3}
\end{eqnarray}
In similar terms,  while $I$ and $V$ are invariant under rotations of the polarization plane, $Q$ and 
$U$ do rotate if $\hat{e}_{1}$ and $\hat{e}_{2}$ rotate. In fact, defining 
$\varphi = \Delta\varphi^{(\mathrm{F})}$, the rotation of the two unit vectors $\hat{e}_{1}$ and $\hat{e}_{2}$ by $\varphi$
\begin{equation}
\hat{e}_{1}\,' = \hat{e}_{1} \cos{\varphi} + \hat{e}_{2} \sin{\varphi}, \qquad 
\hat{e}_{2}\,' = - \hat{e}_{1} \sin{\varphi} + \hat{e}_{2} \cos{\varphi},
\label{rot4}
\end{equation}
leads to the same rotated  Stokes parameters which have been computed in Eq. (\ref{rot3}).

The rate of rotation per unit time is called Faraday rotation rate  and it is given by:
\begin{equation}
{\mathcal F}(\hat{n}) = \frac{d\varphi^{(\mathrm{F})}}{d\tau} = \frac{\overline{\omega}}{2}\biggl[\sqrt{\epsilon_{+}(\overline{\omega},\alpha)} -\sqrt{\epsilon_{-}(\overline{\omega},\alpha)}\biggr],
\label{varphi}
\end{equation}
where we used that, in our units, $dz = d\tau$.
To compute the difference of the two refraction indices appearing in Eq. (\ref{varphi}) we can expand $\epsilon_{\pm}(\overline{\omega},\alpha)$ given in Eqs. (\ref{freq9}) and (\ref{freq10}) for 
$|\overline{\omega}/\overline{\omega}_{\mathrm{pe}}|\ll 1$ and for $ |\overline{\omega}/\overline{\omega}_{\mathrm{Be}}|\ll 1$.
As already discussed (see Eqs. (\ref{freq6}), (\ref{freq7}) and (\ref{freq8})) this is fully justified by the physical values of the aforementioned angular frequencies.  The Faraday rotation rate then becomes:
\begin{equation}
{\mathcal F}(\hat{n}) = \frac{\overline{\omega}_{\mathrm{Be}}}{2} \biggl(\frac{\overline{\omega}_{\mathrm{pe}}}{\overline{\omega}}\biggl)^2 \equiv  \frac{e^3}{2\pi m_{\mathrm{e}}^2} a \tilde{n}_{\mathrm{e}} x_{\mathrm{e}} \frac{\vec{B}\cdot \hat{n}}{\overline{\nu}^2},
\label{rate2}
\end{equation}
where we used that $n_{0}= a^3 \tilde{n}_{\mathrm{e}}$ and that $\overline{\omega} = 2 \pi \overline{\nu}$.
Equation (\ref{rate2}) can be simplified even further by taking 
into account Eq. (\ref{drag}) and the definition of the differential optical depth, i.e. $\epsilon' = a \tilde{n}_{\mathrm{e}} x_{\mathrm{e}}
\sigma_{\mathrm{Th}}$:
\begin{equation}
{\mathcal F}(\hat{n}) = \epsilon' \mathrm{F}(\hat{n}), \qquad \mathrm{F}(\hat{n}) = \frac{ 3}{16 \pi^2 e} 
 \frac{\hat{n} \cdot \vec{B}}{\overline{\nu}^2}.  
\label{F0a}
\end{equation}
The numerical code used for the calculation is the extended version of the code 
used in \cite{gk1,gk2,gk3} and it will be dubbed, in the following two sections, as MAGcmb. As already acknowledged in the introduction, MAGcmb is based on CMBFAST \cite{cmbfast1,cmbfast2} (which is, in turn, based on COSMICS).
We shall not dwell here on the problem of the initial conditions 
of the Einstein-Boltzmann hierarchy which has been throughly discussed
in Refs. \cite{mg1,mg2} (see also \cite{mg4})
 and in \cite{gk2,gk3}.  By initial conditions 
of the Einstein-Boltzmann hierarchy we simply mean a consistent 
solution of the Einstein equations (see Eqs. (\ref{00})--(\ref{ineqj})) and of the 
lowest multipoles of the evolution equations of neutrinos, photons, baryons 
and CDM particles in the approximation of tight Coulomb and Thompson scattering.
Initial conditions are set well before matter-radiation equality.  To avoid possible 
confusions and to make the present script self-consistent,  the analytic form 
of the magnetized adiabatic mode has been reported in Appendix \ref{APPB} (see Eqs. (\ref{S1})--(\ref{S10})).
Of course, other initial conditions are, in principle, at our disposal: they include 
the magnetized version of all the non-adiabatic modes which have 
been introduced in \cite{mg2,mg3,mg4} and discussed, within different perspectives, in \cite{mg5,gk3}.
For purposes 
of illustration we will stick here on the minimal m$\Lambda$CDM 
model where the initial conditions are given in terms of a single magnetized adiabatic 
mode.

The polarization is only generated very near the surface 
of last scattering as the photons begin to decouple from the electrons 
and generate a quadrupole moment through free-streaming. The effects of the magnetic field on the generation 
of the linear polarization will be first computed according to the following set of equations 
derived in \cite{mg2,gk2}:
\begin{eqnarray}
&&  \Delta_{\mathrm{I}}' + i k \mu \Delta_{\mathrm{I}} = - \biggl[ \xi' - \frac{\mu^2}{2}( h' + 6 \xi')\biggr] +
\epsilon' \biggl[  - \Delta_{\mathrm{I}} + \Delta_{\mathrm{I}0} + \mu v_{\mathrm{b}} - \frac{1}{2} P_{2}(\mu) S_{\mathrm{Q}}\biggr],
\label{BR1}\\
&& \Delta_{\mathrm{Q}}' + i k \mu \Delta_{\mathrm{Q}} = \epsilon' \biggl[- \Delta_{\mathrm{Q}} + \frac{1}{2} ( 1 - P_{2}(\mu)) S_{\mathrm{Q}}\biggr],
\label{BR2}\\
&& \Delta_{\mathrm{U}}' + i k \mu \Delta_{\mathrm{U}} = - \epsilon'  \Delta_{\mathrm{U}},
\label{BR3}\\
&& v_{\mathrm{b}}' + {\mathcal H} v_{\mathrm{b}} + \frac{\epsilon'}{R_{\mathrm{b}}} ( 3 i \Delta_{\mathrm{I}1} + v_{\mathrm{b}}) + 
i k \frac{\Omega_{\mathrm{B}} - 4 \sigma_{\mathrm{B}}}{4 R_{\mathrm{b}}}=0,
\label{BR4}
\end{eqnarray}
where $R_{\mathrm{b}}$ has been defined previously and where we defined $v_{\mathrm{b}} = \theta_{\mathrm{b}}/(i k)$. Moreover, in Eqs. (\ref{BR1}) and (\ref{BR2}):
\begin{equation}
S_{\mathrm{Q}} = \Delta_{\mathrm{I}2} + \Delta_{\mathrm{Q}0} + \Delta_{\mathrm{Q}2}.
\end{equation}
The notations $\Delta_{\mathrm{I}\ell}$ and $\Delta_{\mathrm{Q}\ell}$ denote the $\ell$-th multipole
of $\Delta_{\mathrm{I}}$ and $\Delta_{\mathrm{Q}}$. In Eqs. (\ref{BR1}) and (\ref{BR2}) 
$P_{2}(\mu) = (3\mu^2 -1)/2$ is the second Legendre polynomial. 

From Eqs. (\ref{rot3}) and (\ref{varphi}) we can take the total time derivative supposing that the initial polarization is independent on time. From the Faraday rotation 
rate we then get: 
\begin{eqnarray}
&&\Delta_{\mathrm{Q}}' +  n^{i} \partial_{i} \Delta_{\mathrm{Q}} = 2 \epsilon'F(\hat{n}) \Delta_{\mathrm{U}},
 \label{mix1}\\
 &&\Delta_{\mathrm{U}}' + n^{i} \partial_{i} \Delta_{\mathrm{U}} = -2 \epsilon' F(\hat{n}) \Delta_{\mathrm{Q}}.
 \label{mix2}
\end{eqnarray}
Equations (\ref{mix1}) and (\ref{mix2}) describe the Faraday rotation mixing. Previous
approaches to Faraday rotation, consisted in neglecting the effects of the magnetic fields 
both on the initial conditions of the Einstein-Boltzmann hierarchy and on the the dynamical equations.
The calculation of the Faraday rotation assumed then, as initial condition, just the conventional 
adiabatic initial condition of the $\Lambda$CDM mode. The EE polarization correlations 
had then no trace of the magnetized contribution.  It will be shown that 
the present numerical results, as already suggested in \cite{gk2}, do not 
support the latter conclusion.  

Before plunging into the presentation of the numerical results obtained within
the improved version of MAGcmb, two comments are in order. The first 
comment is merely technical and it has to do with the fact that, as it is 
well known, the transport equations including Faraday terms as well as other 
dispersive effects have been extensively studied in second half of the past century 
(see, for instance, \cite{saz,swi,odell} and \cite{lai} and references therein).
These studies were motivated by highly relativistic astrophysical 
plasmas leading to synchrotron emission which could be influenced, under some 
circumstances, by the Faraday effect. Consequently, the 
magnetoactive plasma of \cite{saz,swi} is relativistic. Furthermore, the 
transport equations do not include the effect of the metric inhomogeneities
and are always defined in flat space-times, as in the case of the 
conventional (polarized) heat transfer equations. Finally, with few exceptions, 
the magnetic field is always taken to be uniform or, at most, non-uniform 
along a specific Cartesian direction (see second paper in \cite{swi}).

The second issue we want to discuss has to do with the 
recent 5-year  data release of the WMAP collaboration \cite{WMAP54}. 
The WMAP 5-year data have been analyzed 
to look for possible birefringent effects in the polarization observables. We will here 
scrutinize the parametrization of birefringent effects employed 
in \cite{WMAP54}  and we will contrast it with the birefringent effects 
typical of Faraday rotation by a stochastic magnetic field. This discussion 
will also be useful to set up precisely the relation 
of the fluctuations of brightness perturbations in terms of the 
well known E-modes and B-modes.  
Two linear combinations 
of the brightness perturbations can be usefully introduced, namely:
\begin{equation}
\tilde{{\mathcal M}}_{\pm}(\hat{n}) = \Delta_{\mathrm{Q}}(\hat{n}) \pm i \Delta_{\mathrm{U}}(\hat{n}).
\label{EB1}
\end{equation}
Under a rotation of the coordinate system such as the one introduced in Eq. (\ref{rot4}) 
we clearly have that 
\begin{equation}
\tilde{{\mathcal M}}_{\pm}(\hat{n}) = e^{ \mp 2 i \varphi} {\mathcal M}_{\pm}(\hat{n}),
\label{EB2}
\end{equation}
where the tilded quantity denotes the transformed combination. Since ${\mathcal M}_{\pm}$
transforms as a spin-2 field,  we can expand it in spin-2 spherical harmonics \cite{B1,tot}, i.e. 
\begin{equation}
{\mathcal M}_{\pm}(\hat{n}) = \sum_{\ell\,m} a_{\pm 2,\,\,\ell\,m} \,\, _{\pm 2}Y_{\ell\,m}(\hat{n}),
\label{EB3}
\end{equation}
where $_{\pm 2}Y_{\ell\,m}(\hat{n})$ are the spin-2 spherical harmonics which can be 
introduced, formally, as Wigner matrix elements \cite{sud}. 
A typical 
quantum mechanical problem is to look for the representations
of the operator specifying three-dimensional rotations, i.e. $\hat{R}$; 
this problem  is usually approached within the so-called Wigner matrix elements, i.e. 
${\cal D}_{m \, m'}^{(j)}(R) = \langle j, \, m'| \hat{R} | j,\,m\rangle$ where 
$j$ denotes the eigenvalue of $J^2$ and $m$ denotes the eigenvalue of $J_{z}$. 
Now, if we replace $m' \to - {\rm s}$, $j\to \ell$, we have the definition 
of spin-s spherical harmonics in terms of the 
${\cal D}_{ -{\rm s},\, m}^{(\ell)}(\alpha,\beta, 0)$, i.e.  
\begin{equation}
_{\rm s}{\cal Y}_{\ell\, m}(\alpha,\beta) = \sqrt{\frac{2\ell +1}{4\pi}} {\cal D}_{-{\rm s},\, m}^{(\ell)}(\alpha,\beta, 0),
\label{EB4}
\end{equation}
where $\alpha$, $\beta$ and $\gamma$ (set to zero in the above definition) 
are the Euler angles defined as in \cite{sakurai}. If $s=0$, ${\cal D}_{0,\, m}^{(\ell)}(\alpha,\beta, 0) = \sqrt{(2\ell +1)/4\pi} Y_{\ell \, m}(\alpha,\beta)$ where 
$Y_{\ell \, m}(\alpha,\beta)$  are the ordinary spherical harmonics.  As 
discussed in \cite{sud} the spin-s spherical harmonics can be 
obtained from the spin-0 spherical harmonics thanks to the action of certain 
ladder operators. These ladder operators can be connected to the ladder 
operators of a putative $O(4)$ group \cite{sud}. The reason for the appearance 
of $O(4)$ stems from the fact that we are here composing two rotations: the rotations 
in the three-dimensional space and the rotations around a given point on the tangent 
plane of the celestial sphere.  A rather productive approach to obtain explicit 
expressions for the spin-2 spherical harmonics consists in studying directly the polarization 
on the 2-sphere \cite{B2}. Either with the ladder operator of \cite{sud} or with 
the more intrinsic approach of \cite{B2} it is possible to obtain more explicit 
expressions of the spin-2 spherical harmonics in terms of the spin-0 spherical harmonics and the 
result is:
\begin{equation}
_{\pm 2} Y_{\ell\, m}(\vartheta,\phi)= 2 \sqrt{\frac{(\ell -2)!}{(\ell + 2)!}} \biggl[ \partial_{\vartheta}^2 + \ell (\ell +1) 
\mp \frac{2 m}{\sin{\vartheta}}(\partial_{\vartheta} - \cot{\vartheta}) \biggr] Y_{\ell\,m}(\vartheta,\phi).
\label{EB5}
\end{equation}
The ``electric" and ``magnetic" components of polarization are 
eigenstates of parity and may be defined, from $a_{\pm,\,\,\ell\,m}$ of Eq. (\ref{EB3}), as 
\begin{equation}
a_{\ell m}^{(\rm E)}=-\frac{1}{2}(a_{2, \ell m}+a_{-2,\ell m}),\qquad
a_{\ell m}^{(\rm B)}= \frac{i}{2}(a_{2,\ell m}-a_{-2,\ell m}).
\label{EB6}
\end{equation}
Under parity inversion,  the components appearing in Eqs. (\ref{EB6}) and 
\begin{equation}
 a_{\ell m}^{\rm E} \to (-1)^{\ell }\,\,a_{\ell m}^{\rm E},\qquad
 a_{\ell m}^{\rm B} \to (-1)^{\ell +1 }\,\,a_{\ell m}^{\rm B}.
\label{EB7}
\end{equation}
Therefore, the E-modes have the same parity of the temperature correlations  
which have, in  turn, the same parity of conventional spherical harmonics, i.e. 
$(-1)^{\ell}$. On the contrary, the B-modes have $(-1)^{\ell + 1}$ parity.
The existence of linear polarization allows for 6 different cross power 
spectra to be determined, in principle, from data that measure 
the full temperature and polarization anisotropy information.
The cross power spectra can be defined in terms of the spectral functions 
$C_{\ell}^{(XY)}$ where $X$ and $Y$ stand for E, B or T depending 
on the cross-correlation one is interested in:
\begin{equation}
C_{\ell}^{(XY)} = \frac{1}{2 \pi^2} \int k^2 \, dk \sum_{m = -\ell}^{\ell} \frac{( a_{\ell m}^{X})^{\ast} a_{\ell m}^{Y}}{( 2 \ell + 1)}.
\label{defcross}
\end{equation}
Therefore, if we are interested in the TT correlations we just have to set 
$X= {\rm T}$ and $Y ={\rm T}$ and use the relevant expansions given above. 
In the present paper, following a consolidated convention, 
the correlations will be denoted, with shorthand notation, 
 as TT, EE, BB, TB and so on. Suppose 
we are interested in the TT correlations, i.e. the usual and well 
known temperature correlations. From Eq. (\ref{defcross}) we will have 
\begin{equation}
C_{\ell}^{({\rm TT})} = \frac{1}{2 \pi^2} \int k^2 \, dk \sum_{m = -\ell}^{\ell} 
\frac{[a_{\ell m}^{{\rm T}}(k)]^{\ast} a_{\ell m}^{{\rm T}}(k)}{( 2 \ell + 1)}.
\label{defcross2}
\end{equation}
Now,  using the orthogonality of spherical harmonics, we have 
that 
\begin{equation}
a_{\ell m}^{\rm T}(k) = \int d \hat{n} Y_{\ell m}^{*}(\hat{n}) \Delta_{\rm I}(\vec{k}, \hat{n}).
\label{inverse}
\end{equation}
Inserting Eq. (\ref{inverse}) into Eq. (\ref{defcross2}) and recalling the Rayleigh expansion 
of the fluctuations of the intensity:
\begin{equation}
\Delta_{\mathrm{I}}(\vec{k}, \hat{n},\tau) = \sum_{\ell} (- i)^{\ell} (2 \ell + 1) \Delta_{\mathrm{I}\ell}(k,\tau) P_{\ell}(\mu)
\label{ray1}
\end{equation}
we obtain  
\begin{equation}
C_{\ell}^{(\mathrm{TT})} = \frac{2}{\pi} \int dk \, k^2 |\Delta_{{\rm I}\,\ell}|^2.
\label{Cell}
\end{equation}
To get to Eq. (\ref{Cell}) the following two identities 
have been used, i.e. 
\begin{eqnarray}
&& \int d\hat{n} P_{\ell'}(\hat{k}\cdot\hat{n}) Y_{\ell m}^{\ast}(\hat{n}) = 
\frac{4\pi}{(2 \ell + 1)}Y_{\ell m}^{\ast}(\hat{k})  \delta_{\ell\ell'},
\nonumber\\
&& \sum_{m  = -\ell}^{\ell} Y_{\ell m}^{\ast}(\hat{k}) Y_{\ell m}(\hat{n}) =
\frac{2\ell + 1}{4\pi} P_{\ell}(\hat{k}\cdot\hat{n}).
\label{addition}
\end{eqnarray}

It will now be shown how the angular power spectrum of the B-mode can be 
related, with semi-analytical methods, to the spectrum of the Faraday rotation rate.
At the same time it is rather relevant to mention that the birefringence induced by 
stochastic magnetic fields differs from   a particular kind of birefringence which has been considered 
in \cite{WMAP54} . 
When a pseudoscalar field is coupled to $\tilde{F}^{\alpha\beta} F_{\alpha\beta}$ birefringent 
effects are expected (see, for instance, \cite{birefringence}). Furthermore, the 
presence of pseudo-scalar interactions was invoked in \cite{hk1,hk2} (see also \cite{hk3,hk4})  as 
a mechanism for the generation of magnetic helicity from a stochastic background of hypercharge 
fields with vanishing helicity. 
Suppose, therefore, a coupling in the Lagrangian density of the form $\beta \sigma \tilde{F}^{\alpha\beta} F_{\alpha\beta}/M$
where $M$ is a typical mass scale, $\beta$ is a dimensionless coupling and $\sigma$ is the dynamical 
pseudo-scalar field.  Let us suppose also, for simplicity, that the background magnetic field is absent. This 
is, in fact, the situation contemplated in \cite{WMAP54}. In this set-up the evolution equation for the 
magnetic field of the wave is given by:
\begin{equation}
\vec{b}\,'' - \nabla^2 \vec{b}  + \frac{\beta}{M} \sigma' \vec{\nabla} \times \vec{b} =0,
\label{psscal}
\end{equation}
Recalling that the prime denotes a derivation with respect to the conformal time coordinate, the 
shift in the polarization plane will be $\Delta \chi = \beta \Delta\sigma/(2\, M)$ where we denoted  
by $\chi$ the angle to avoid confusion with the other angular variables 
previously introduced. Now, it is clear that while $\Delta\chi$ is frequency independent,
$\Delta \varphi^{(\mathrm{F})}$  depends upon the frequency. Furthermore, the inhomogeneity 
scale of the magnetic field plays enters directly $\Delta \varphi^{(\mathrm{F})}$. On the contrary 
$\Delta \chi = \beta \Delta\sigma/(2\, M)$ is a fully homogeneous quantity.  Now, in \cite{WMAP54}
$\Delta\chi$ is observationally constrained by adding one single parameter to the $\Lambda$CDM model.

The rotation by $\Delta\chi$ implies, at the level of 
$a^{(\mathrm{E})}_{\ell\,m}$ and of $a^{(\mathrm{B})}_{\ell\,m}$
\begin{eqnarray}
&& \tilde{a}^{(\mathrm{E})}_{\ell\,m} = \cos{(2 \Delta \chi)} a^{(\mathrm{E})}_{\ell\,m} + \sin{(2 \Delta\chi)} a^{(\mathrm{B})}_{\ell\,m},
\label{EB8}\\
&& \tilde{a}^{(\mathrm{B})}_{\ell\,m} = - \sin{(2 \Delta \chi)} a^{(\mathrm{E})}_{\ell\,m} + \cos{(2 \Delta\chi)} a^{(\mathrm{B})}_{\ell\,m}.
\label{EB9}
\end{eqnarray}
where the tilded quantity denotes, as in Eq. (\ref{EB2}) the rotated variable.
Using now Eq. (\ref{defcross}) and assuming that $\Delta\chi$ does not depend 
upon the Fourier wavenumber, the transformed EE and BB  autocorrelations can be swiftly 
obtained
\begin{eqnarray}
&& \tilde{C}^{(\mathrm{EE})}_{\ell} = \cos^2{(2 \Delta \chi)} C^{(\mathrm{EE})}_{\ell} + \sin^2{(2 \Delta \chi)} C^{(\mathrm{BB})}_{\ell},
\label{EB10}\\
&& \tilde{C}^{(\mathrm{BB})}_{\ell} = \cos^2{(2 \Delta \chi)} C^{(\mathrm{EE})}_{\ell} + \sin^2{(2 \Delta \chi)} C^{(\mathrm{BB})}_{\ell},
\label{EB11}\\
&& \tilde{C}^{(\mathrm{EB})}_{\ell} = \frac{1}{2}(C_{\ell}^{(\mathrm{EE})} - C_{\ell}^{(\mathrm{BB})}) \sin{(4 \Delta\chi)}.
\label{EB12}
\end{eqnarray}
According to the analysis of \cite{WMAP54} 
the tilded quantities denote the observed power spectra. In deriving Eqs. (\ref{EB10}), (\ref{EB11}) and (\ref{EB12}) ,
the initial $C^{(\mathrm{EB})}_{\ell}$ is assumed to be vanishing for consistency with \cite{WMAP54}.
Furthermore, if we would take literally the $\Lambda$CDM model with no tensors, then $C_{\ell}^{(\mathrm{BB})} =0$ (as also 
assumed in \cite{WMAP54}). In the latter case the induced BB angular power spectrum will be 
simply given by Eq. (\ref{EB11}) with $C^{(\mathrm{BB})}_{\ell}=0$. 

In the case of birefringence induced by a stochastic magnetic field the calculation is more involved and it is 
reported, in full detail, in Appendix \ref{APPC}. 
Using the shorthand notation 
$F(\hat{n}) = {\mathcal A} \vec{B}\cdot \hat{n}$, $F(\hat{n})$ can be expandes in series of (scalar) spherical
harmonics. Thus, since the magnetic field does not break spatial isotropy, the angular power 
spectrum of Faraday rotation can be written as:
\begin{equation}
\langle F(\hat{n}_{1}) F(\hat{n}_{2})\rangle = \frac{1}{4\pi}\sum_{\ell} (2 \ell + 1)C_{\ell}^{(\mathrm{F})} P_{\ell}(\hat{n}_{1} \cdot
\hat{n}_{2})
\label{APSfar}
\end{equation}
where
\begin{equation}
C_{\ell}^{(\mathrm{F})} = 4\pi {\mathcal A}^2 \ell (\ell +1) \int \frac{d k}{k} {\mathcal P}_{B}(k) \frac{j_{\ell}^2(k\tau_{0})}{k^2 \tau_{0}^2}.
\label{F13a}
\end{equation}
As it will be shown in a moment, the magnetic field affects directly 
the EE angular power spectra and, therefore, the equation for the BB angular power spectrum 
which follows from Eq. (\ref{F13}) can be written, formally, as
\begin{equation}
{\mathcal C}_{\ell}^{\mathrm{BB}} = \sum_{\ell_{1},\,\,\ell_{2}}  {\mathcal G}(\ell, \ell_1, \ell_{2}) C_{\ell_{2}}^{(\mathrm{EE})} C_{\ell_{1}}^{(\mathrm{F})}
\label{F14a}
\end{equation}
where ${\mathcal G}(\ell, \ell_1, \ell_{2})$ is a rather cumbersome function of the multipole moments which contains 
also a Clebsch-Gordon coefficient. The expression for ${\mathcal G}(\ell, \ell_{1},\ell_{2})$ is reported in Appendix \ref{APPC} 
and has been derived in \cite{far5}. 

Even if the cases described by Eqs. (\ref{EB10})--(\ref{EB12})  and Eqs. (\ref{F13a})--(\ref{F14a}) 
 are conceptually very different, it is tempting to interpret the bounds on $\Delta \chi$ as constraints on the 
 averaged rotation angle, i.e. 
 \begin{equation}
\langle |\Delta \varphi^{(\mathrm{F})}|^2\rangle  = \sum_{\ell} \frac{2\ell + 1}{4\pi} C_{\ell}^{(\mathrm{F})} \simeq \int \frac{\ell (\ell+1)}{ 2\pi } 
 C_{\ell}^{(\mathrm{F})} \frac{d\ell}{\ell}. 
\label{FRR15}
\end{equation}
In spite of the fact that this identification is  rather gross and, as a consequence, the derived 
bounds are not so significant and inherently ambiguous. One of the sources of ambiguity stems from the frequency dependence 
of $C_{\ell}^{(\mathrm{F})}$ which scales, according to our conventions, as $(\nu_{\mathrm{max}}/\nu)^4$ where 
$\nu_{\mathrm{max}} = 222.617$ GHz is the maximum of the CMB blackbody \footnote{It is practical to refer 
the frequency dependence to $\nu_{\mathrm{max}}$ defined as in Eq. (\ref{F20}) and (\ref{F21}). Consequently, the power 
spectrum of Faraday rotation will scale as $(\nu_{\mathrm{max}}/\nu)^4$ (see also Eqs. (\ref{F26}) and (\ref{F27}) of Appendix 
\ref{APPA}).}. 

The WMAP experiment observes the microwave sky in five frequency channels ranging from \cite{WMAP32} $23$ GHz to $94$ GHz.
The bandwidth, correspondingly, increases from small to high frequencies signaling that probably the best 
sensitivity to polarization comes the high frequency channels\footnote{More precisely the five frequency channels 
of the WMAP experiment are centered at $23$, $33$, $41$, $61$ and $94$ in units of GHz.}. 
On the contrary, the Faraday rotation signal is larger at low frequencies and, therefore, a possibility would be 
to compare the bounds on $\Delta\chi$ derived in \cite{WMAP54} with Faraday rotation signal evaluated at the 
lowest frequency of observation.  The lowest frequency of observation \cite{WMAP32} is, however, used as a
foreground template. Therefore the lowest available effective frequency would be indeed around 
$30$ GHz which is, approximately, the lowest frequency channel of the Planck experiment \cite{planck} (see also 
later, in the first part of section \ref{sec7}).  
The results reported in \cite{WMAP54} do not allow to determine how the 
various frequency channels have been combined in the analysis. 
The authors of \cite{WMAP54} simply assumed that the rotation 
angle was frequency independent, which is not our case, as it will also 
be discussed later on. 

\renewcommand{\theequation}{6.\arabic{equation}}
\setcounter{equation}{0}
\section{Magnetized TE and EE angular power spectra}
\label{sec6}
Three fiducial sets of parameters will be used for the illustration 
of the numerical results. The first set of parameters is obtained from 
the WMAP 5-year data alone \cite{WMAP51,WMAP52,WMAP53,WMAP54,WMAP55}  analyzed in the light of the 
conventional $\Lambda$CDM model\footnote{In the present script the differential optical depth 
has been denoted by $\epsilon'$ while a more standard notation is $\tau'$. Here
the variable $\tau$ denotes instead the conformal time coordinate. To avoid possible confusions 
in the identifications of the m$\Lambda$CDM parameters the conventional notation 
will be here restored.}:
\begin{equation}
(\Omega_{\mathrm{b}0},\, \Omega_{\mathrm{c}0}, \,\Omega_{\mathrm{\Lambda}}, \,h_{0}, \,n_{\mathrm{s}},\,\tau)= (0.0441,\, 0.214,\, 0.742,\, 0.719,\, 0.963,\,0.087).
\label{best1}
\end{equation}
If compared to the best fit parameters of the WMAP 3-year data alone \cite{WMAP31,WMAP32}, we can notice: a slight increase of the scalar spectral index (from $0.958$ to $0.963$); a slight increase of $\Omega_{\mathrm{b}0}$; a slight increase of $\Omega_{\mathrm{c}0}$ and a consequent decrease of $\Omega_{\Lambda}$. The features discussed with one set of data 
have a corresponding qualitative counterpart within different data sets. 
The WMAP 5-year data have been also combined with the ACBAR 
\footnote{The Arcminute Cosmology Bolometer Array 
Receiver (ACBAR) operates at three frequencies, i.e. $150$, $219$ and $274$ GHz.} data set \cite{rein}. In the latter case, 
the derived values of the cosmological parameters are:
\begin{equation}
(\Omega_{\mathrm{b}0},\, \Omega_{\mathrm{c}0}, \,\Omega_{\mathrm{\Lambda}}, \,h_{0}, \,n_{\mathrm{s}},\,\tau)= (0.0441,\, 0.215,\, 0.741,\, 0.720,\, 0.964,\,0.088).
\label{acb}
\end{equation}
Finally, a third set of cosmological parameters will be the one 
suggested by the analysis  of the WMAP 5-year data in combination with the large-scale structure data (in particular 2df and SDDS data) \cite{perci}, in turn 
combined with all the available supernova data\footnote{In the plots we will denote this 
class of data by WMAP 5-year + BAO + SNall.  The acronym BAO refers 
to the baryon acoustic oscillations which could be deduced from the large-scale structure data.} \cite{SN1,SN2}:
\begin{equation}
(\Omega_{\mathrm{b}0},\, \Omega_{\mathrm{c}0}, \,\Omega_{\mathrm{\Lambda}}, \,h_{0}, \,n_{\mathrm{s}},\,\tau)= (0.0462,\, 0.233,\, 0.721,\, 0.701,\, 0.960,\,0.084).
\label{baosn}
\end{equation}
\begin{figure}[!ht]
\centering
\includegraphics[height=6.5cm]{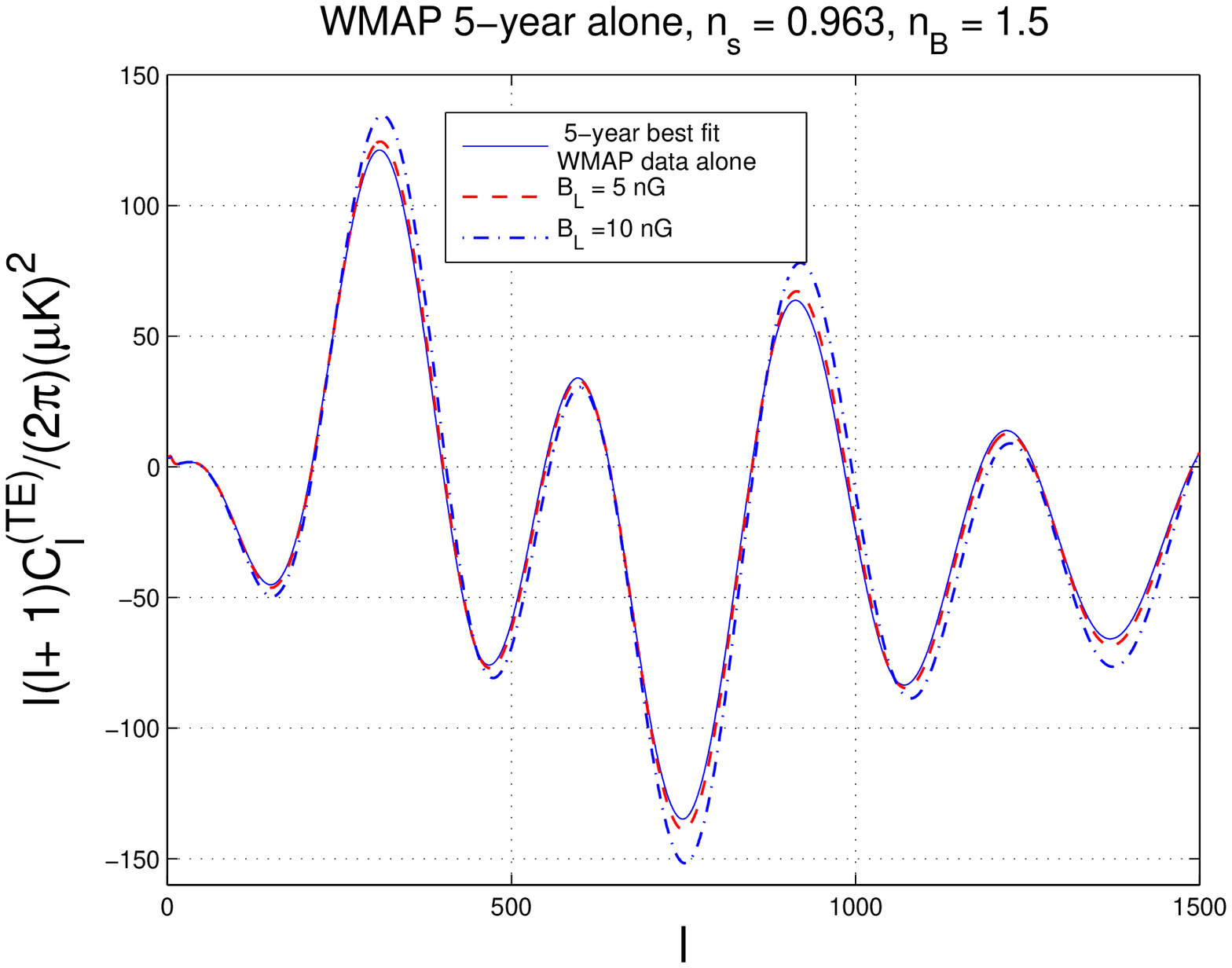}
\includegraphics[height=6.5cm]{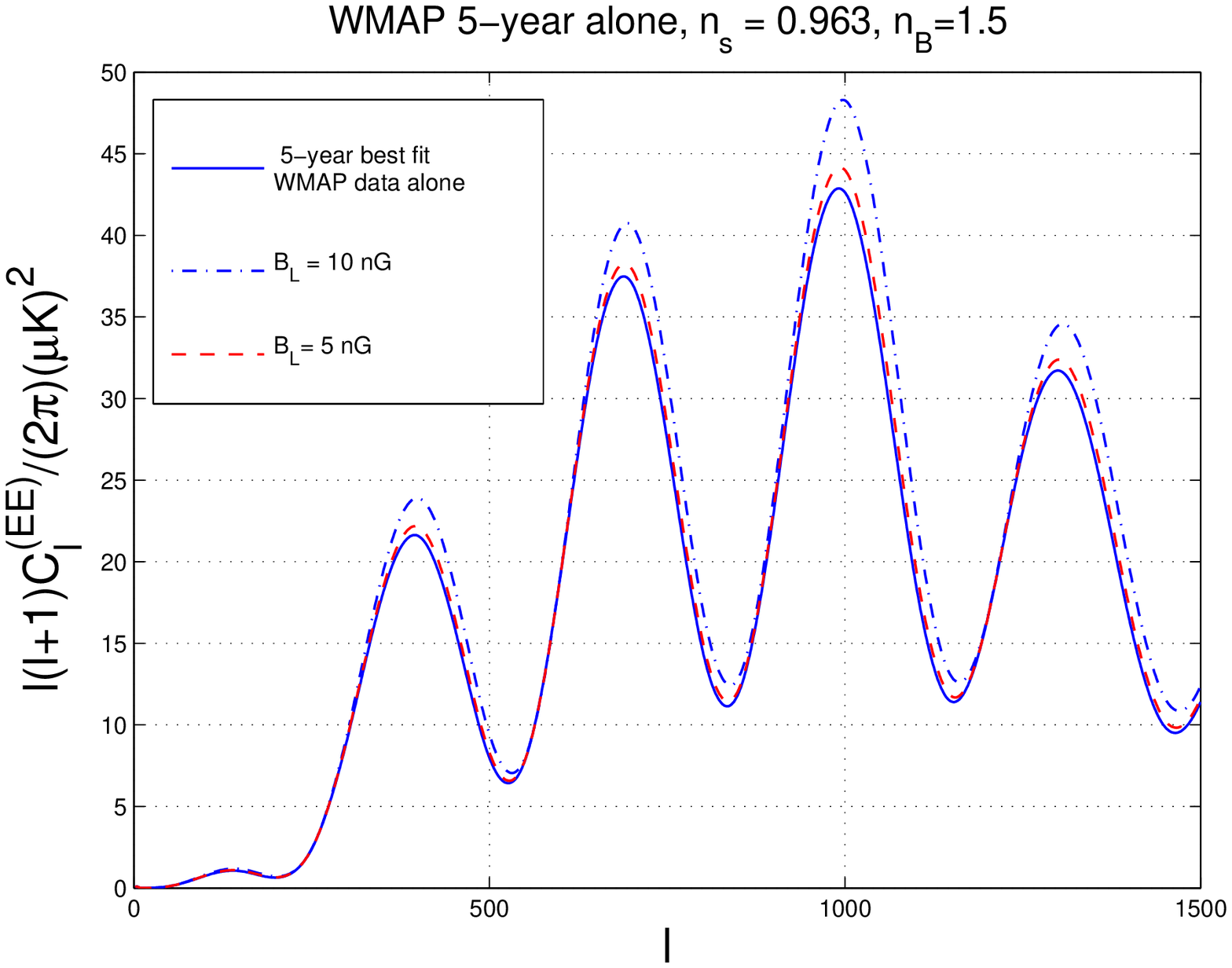}
\caption[a]{The TE and EE angular power spectra in the m$\Lambda$CDM model. The full 
line corresponds to the best fit of the WMAP 5-year alone data analyzed in the light 
of the conventional $\Lambda$CDM paradigm.}
\label{Figure1}      
\end{figure}
In Fig. \ref{Figure1} the results of the numerical integration of Eqs. (\ref{BR1})--(\ref{BR4}) are reported  for the choice 
of parameters given in Eq. (\ref{best1}). The magnetic spectral index has been fixed, in both plots, to 
$n_{\mathrm{B}}=1.5$.  The purpose of Fig. \ref{Figure1} is twofold. In the first place we point out, as 
already discussed in \cite{gk2},  that we are sensitive to nG magnetic fields. The dot-dashed curve in 
Fig. \ref{Figure1} indeed corresponds to a magnetic field\footnote{The regularized amplitude 
of the magnetic field, $B_{\mathrm{L}}$, is defined as in \cite{gk2}. For blue magnetic spectral 
index (i.e. $n_{\mathrm{B}} > 1$) the window function is taken to be Gaussian. For red spectral index (i.e. 
$n_{\mathrm{B}} < 1$) the window 
function is a step with smooth edges. The magnetic pivot scale, denoted by $k_{\mathrm{L}}$ is always 
taken, for illustrative purposes, to be $1\, \mathrm{Mpc}^{-1}$. In Appendix \ref{APPC} the precise definition 
of $B_{\mathrm{L}}$ is reviewed since it is needed for a consistent evaluation 
of the Faraday rotation power spectrum.}
of  $10$ nG while the dashed curve corresponds to a magnetic field of $5$ nG. 

The second occurrence which is evident from Fig \ref{Figure1} is that the large-scale magnetic 
fields do affect the TE and EE angular power spectra. This cannot happen, by construction, 
if the magnetic field is uniform as in the case discussed, for instance, in \cite{far1,far3} and \cite{far4}.
Furthermore, for a stochastic magnetic field  the TB and EB correlations will be vanishing.
\begin{figure}[!ht]
\centering
\includegraphics[height=6.5cm]{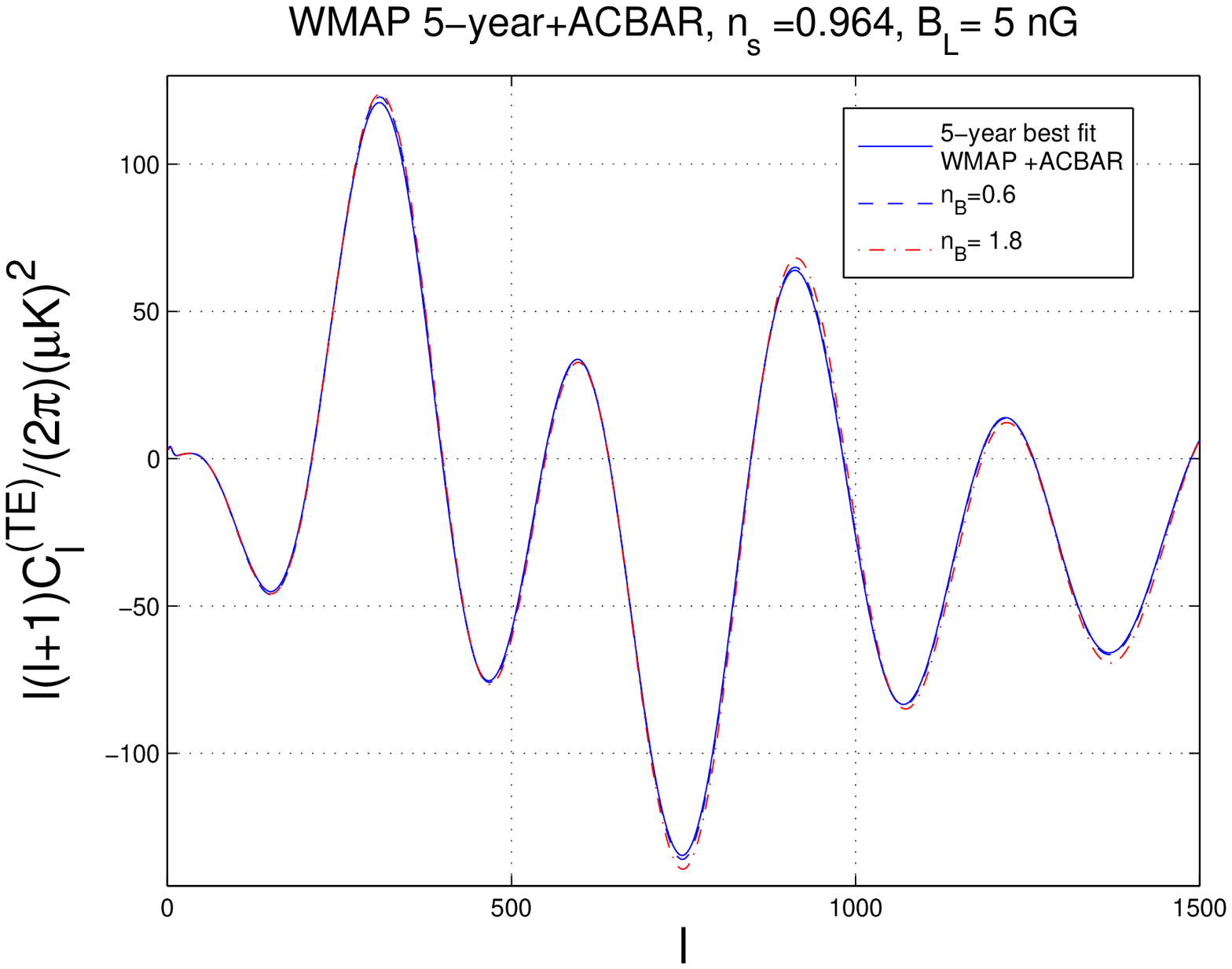}
\includegraphics[height=6.5cm]{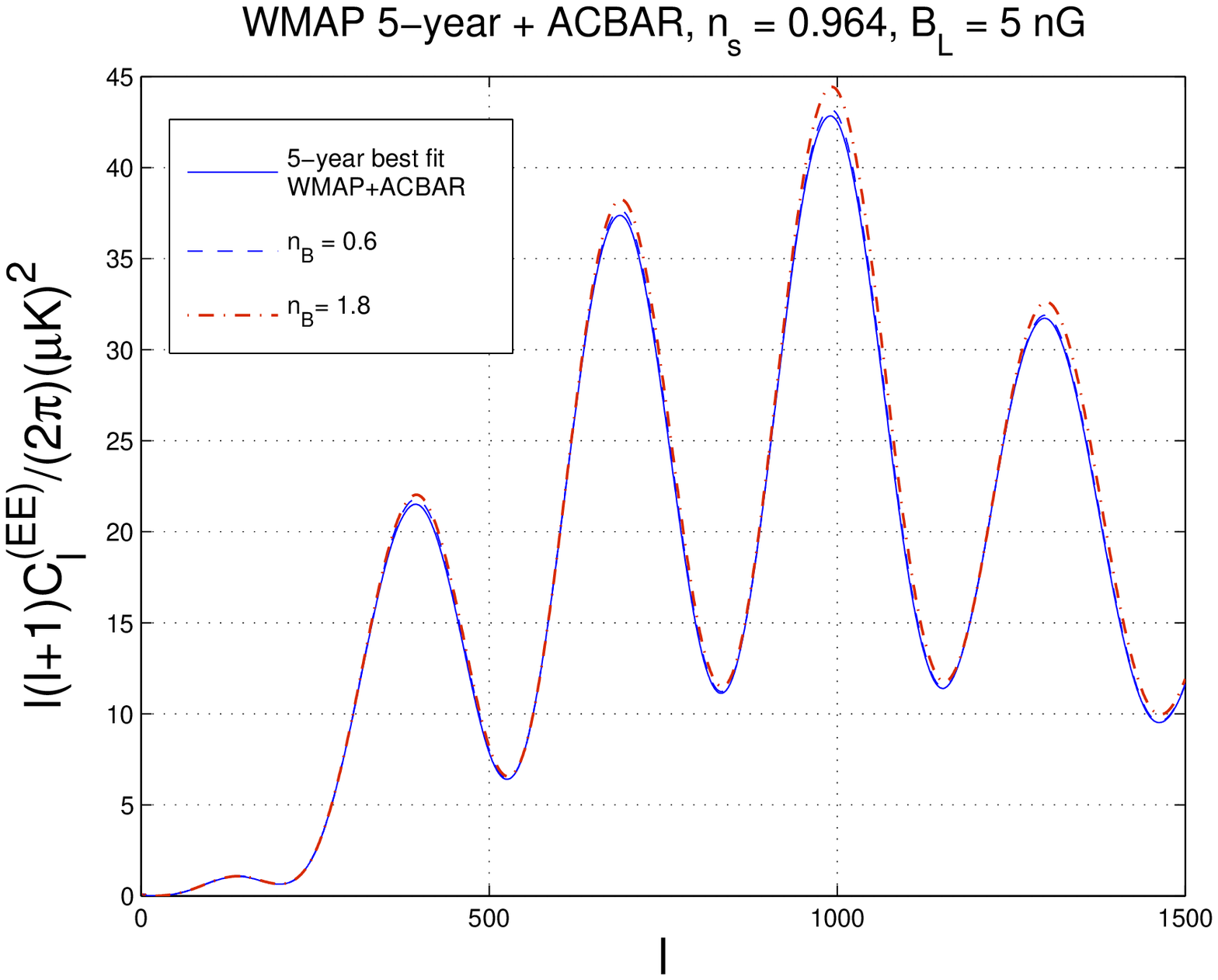}
\caption[a]{As in Fig. \ref{Figure1} the TE and EE angular power spectra for a different choice of magnetic field 
parameters. With the full line the best fit to the conventional $\Lambda$CDM paradigm is illustrated and it corresponds 
to the parameters reported in Eq. (\ref{acb}). }
\label{Figure2}      
\end{figure}
The same situation illustrated in Fig. \ref{Figure1} is also illustrated in Figs. \ref{Figure2} and \ref{Figure3} 
but with two different sets of parameters. In Fig. \ref{Figure2} the WMAP data are combined 
with the ACBAR data and the full line is then the best-fit with parameters fixed as in Eq. (\ref{acb}).
\begin{figure}[!ht]
\centering
\includegraphics[height=6.5cm]{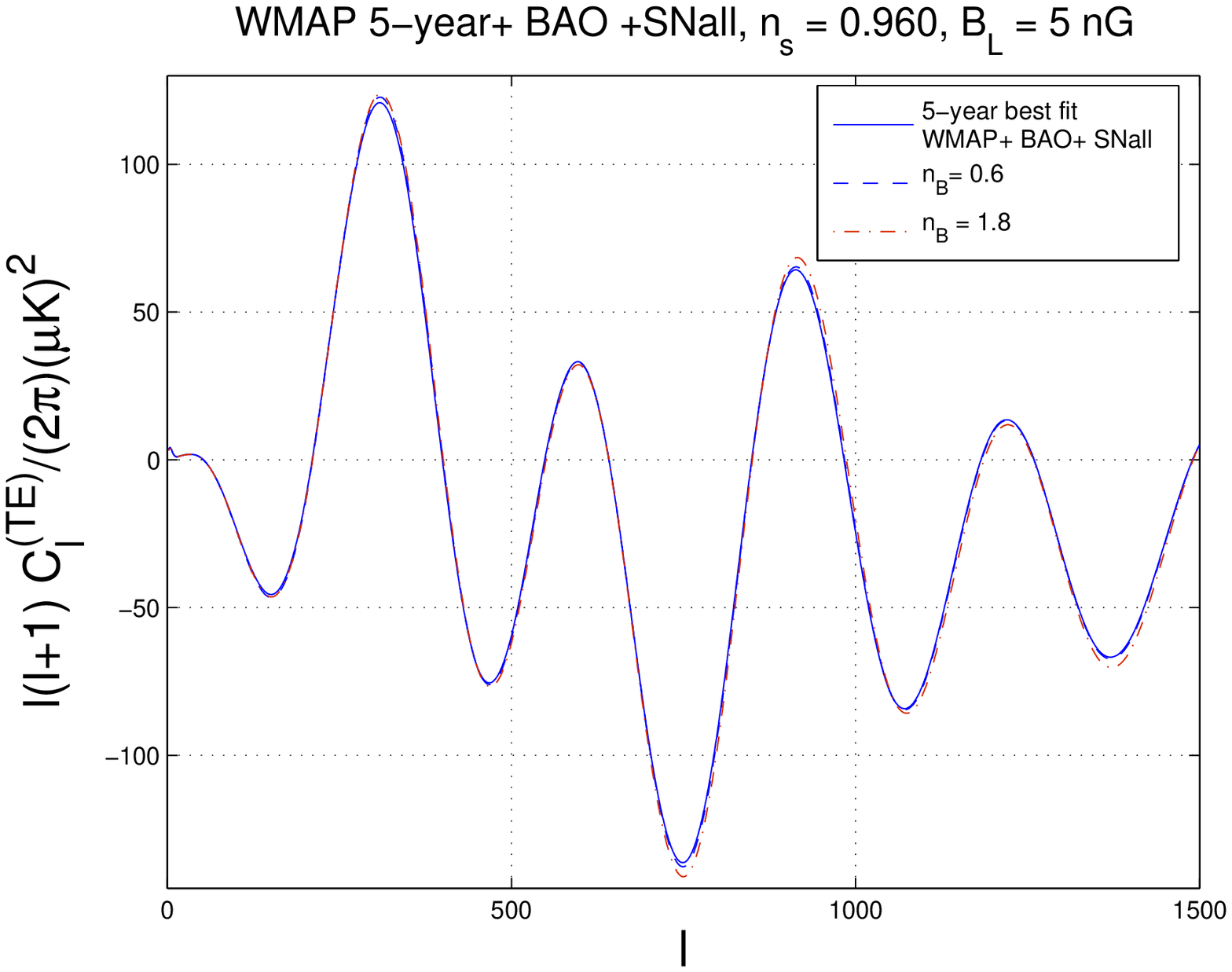}
\includegraphics[height=6.5cm]{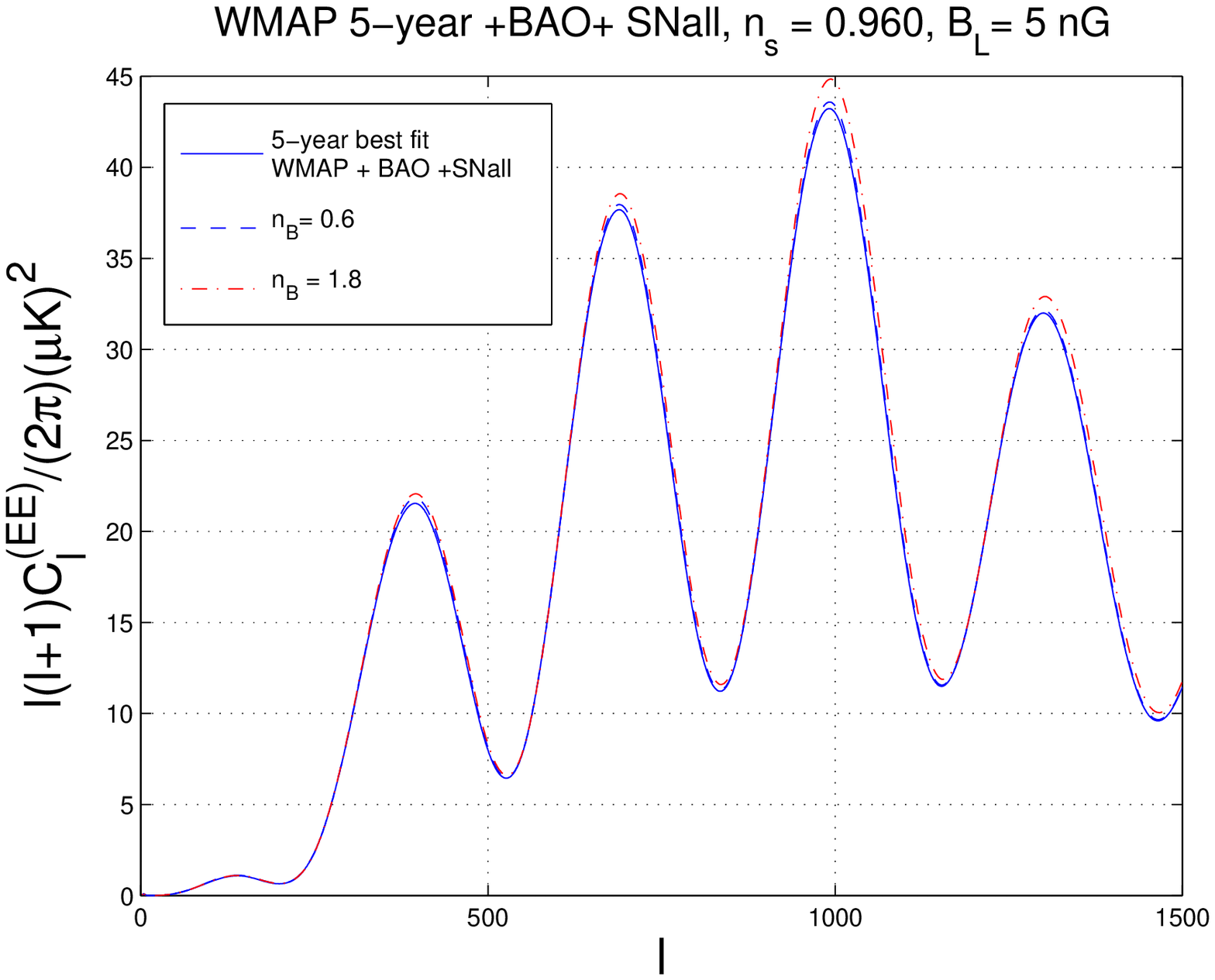}
\caption[a]{The same as in Fig. \ref{Figure2} but with the $\Lambda$CDM parameters reported in Eq. (\ref{Figure3}). }
\label{Figure3}      
\end{figure}
The range of the m$\Lambda$CDM parameters illustrated in Figs. \ref{Figure2} and \ref{Figure3} is narrower than in the case of  
Fig. \ref{Figure1}. Indeed, as it can be argued by comparing the legends of the three figures, in Fig. \ref{Figure1} a magnetic field 
of $10$ nG has been allowed. On the contrary in Figs. \ref{Figure2} and \ref{Figure3} a sizably smaller magnetic field 
has been assumed (i.e. $B_{\mathrm{L}} = 5$ nG) but the magnetic spectral index $n_{\mathrm{B}}$ has been illustrated 
for two different blue values. 
Different sets of parameters have been studied within the m$\Lambda$CDM model and the qualitative conclusions are the same.
 Contrary to what normally assumed large-scale magnetic fields do affect both the TE and EE 
angular power spectra and nG magnetic fields lead to clearly observable features.

To conclude the present section it is relevant to point out, very swiftly, that the TT angular power spectra 
have been studied, both, analytically and numerically in \cite{mg2,mg3} and in \cite{gk1,gk2}.
The typical patterns of correlated distortions of the  acoustic peaks 
have been thoroughly investigated (in the light of the minimal m$\Lambda$CDM model)
in \cite{gk1,gk2}. 
\begin{figure}[!ht]
\centering
\includegraphics[height=6.5cm]{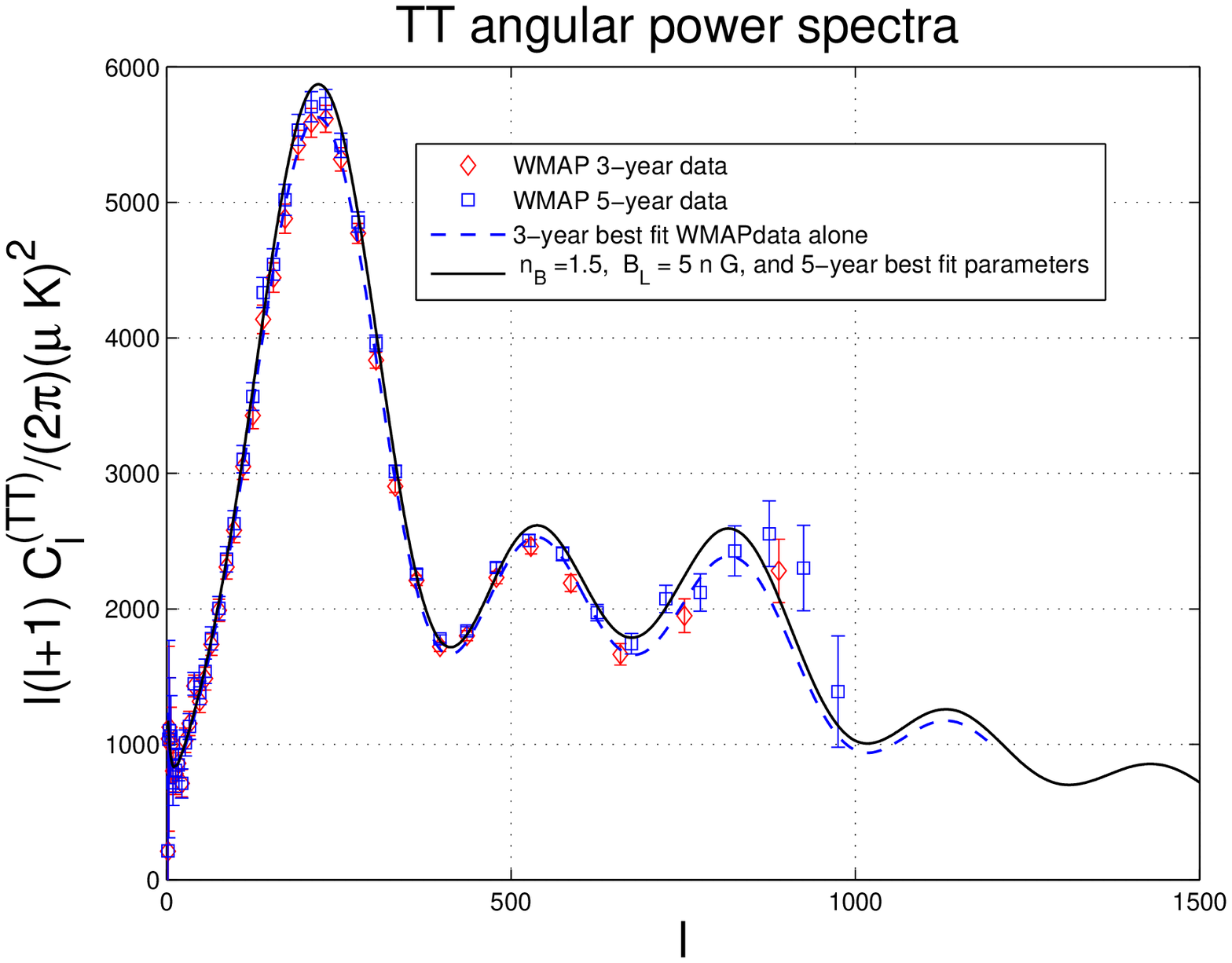}
\includegraphics[height=6.5cm]{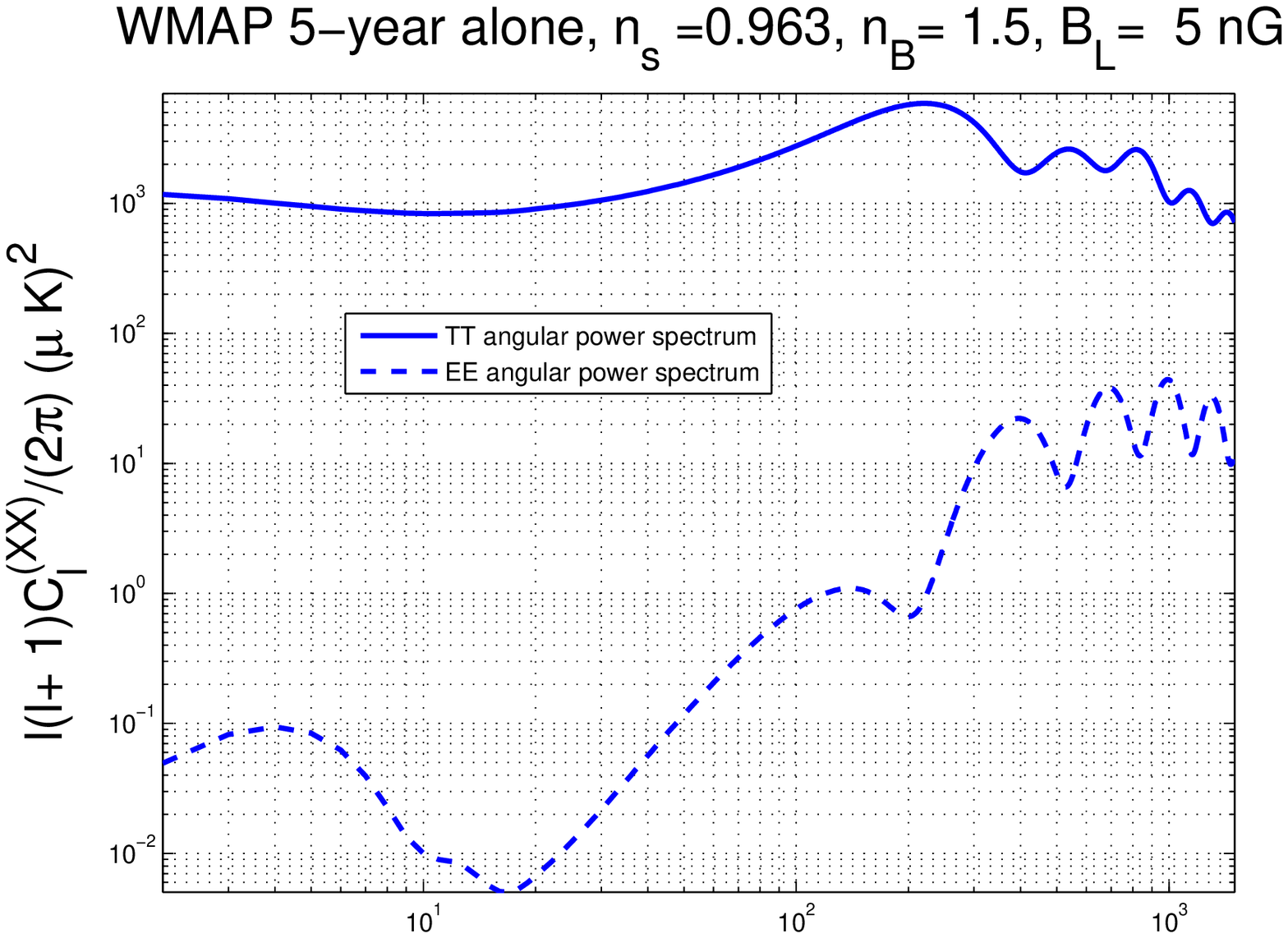}
\caption[a]{The magnetized TT angular power spectra in linear scale (plot at the left) and in double 
logarithmic scale (plot at the right).}
\label{Figure4}      
\end{figure}
In Fig. \ref{Figure4} the plots for the TT angular power spectra are reported. In the plot at the left
the diamonds are the WMAP 3-year data 
the boxes the WMAP 5-year data. With the dashed line the best fit to the WMAP 3-year data 
alone is illustrated. With the full line the magnetic filed is included with the other
 parameters fixed to the 5-year best fit. In  the plot at the right the TT and EE angular power spectra 
 are reported in a double logarithmic scale to show the relative magnitude of the two angular power spectra.

\renewcommand{\theequation}{7.\arabic{equation}}
\setcounter{equation}{0}
\section{Faraday screened BB angular power spectra}
\label{sec7}
The $C_{\ell}^{(\mathrm{BB})}$ depend not only upon the parameters of the magnetized 
background but also upon the comoving frequency. This important occurrence, as explained 
in the previous sections, is a direct consequence of the dispersion relations. 
\begin{figure}[!ht]
\centering
\includegraphics[height=6.5cm]{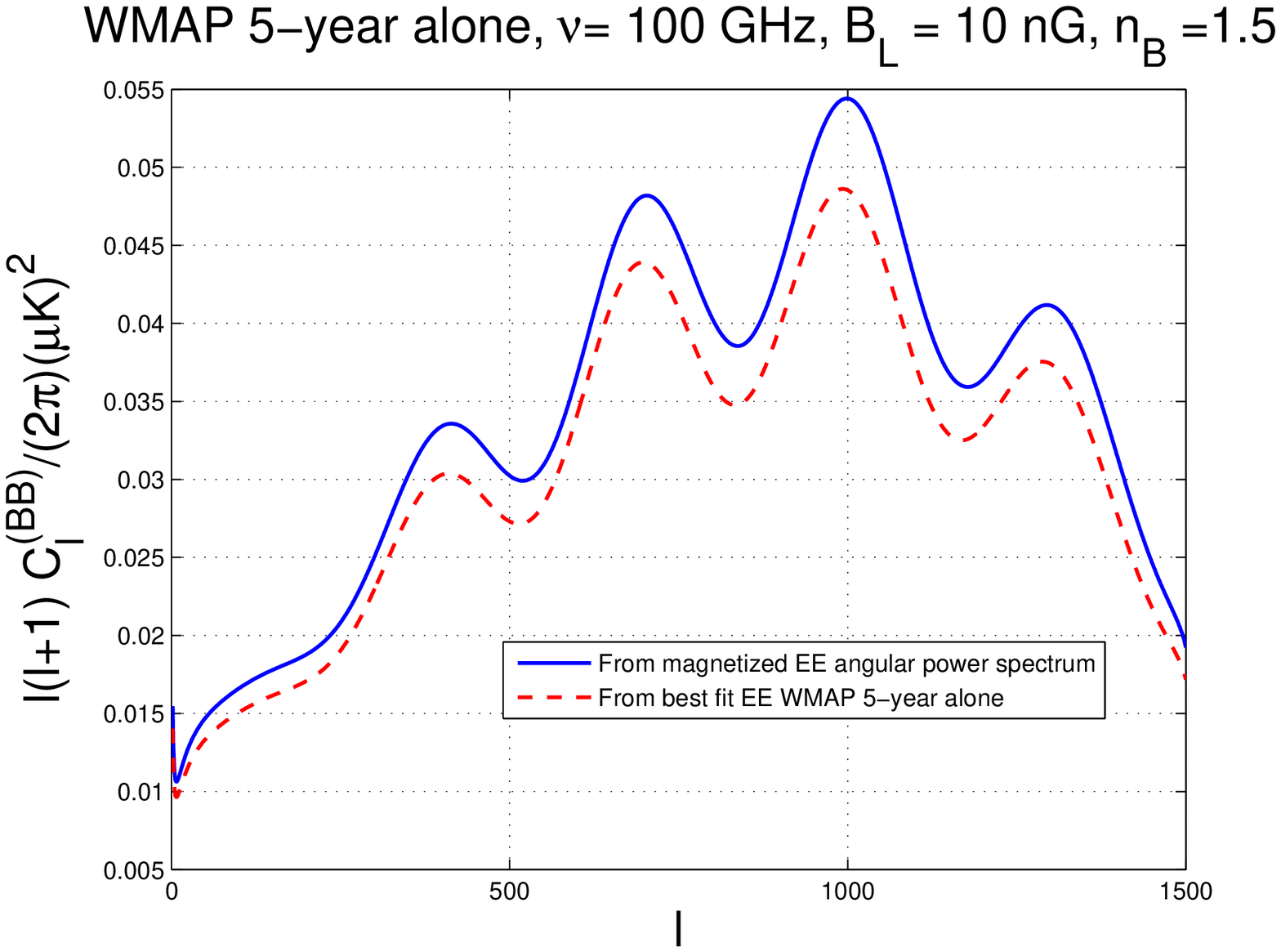}
\includegraphics[height=6.5cm]{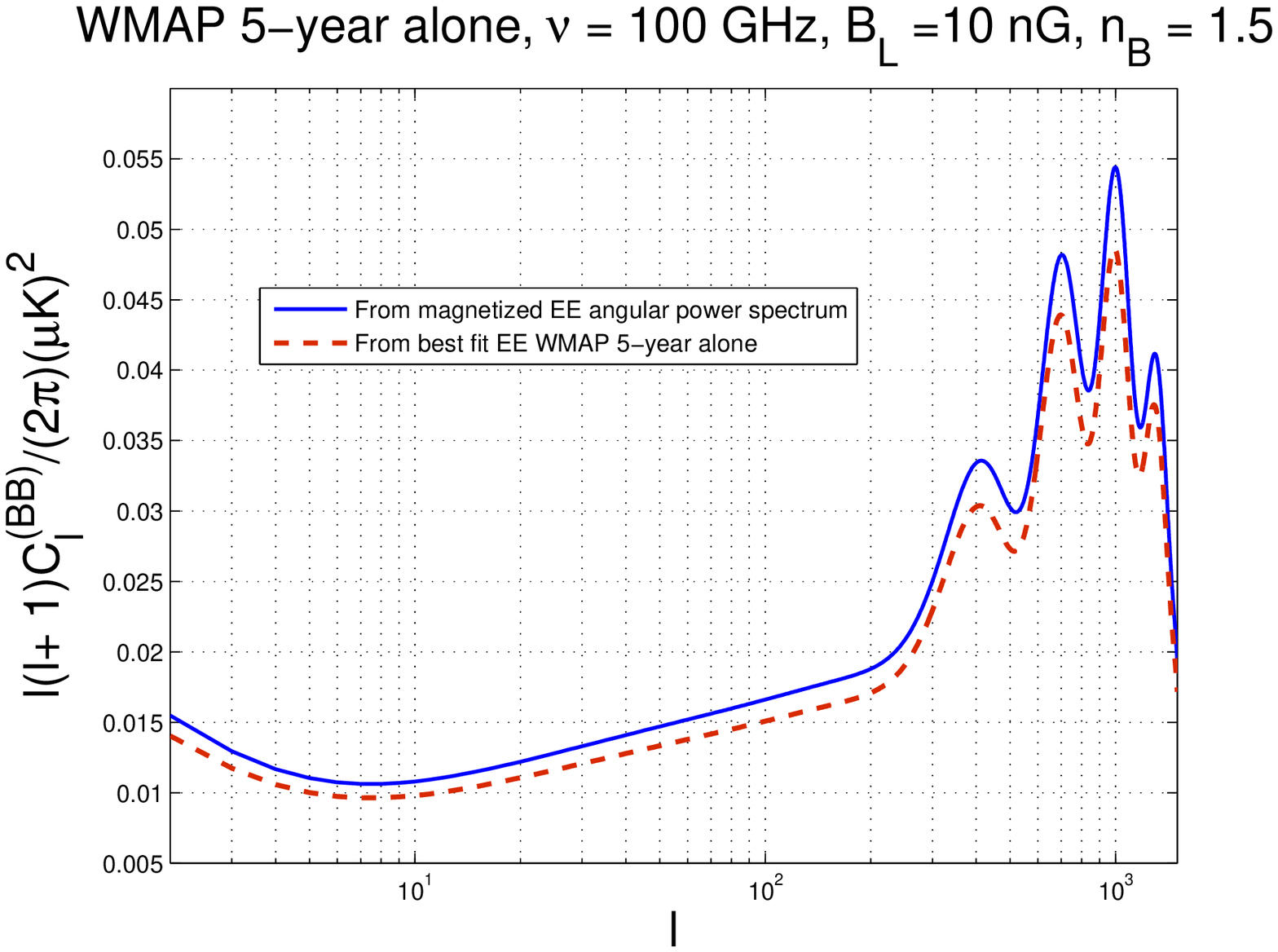}
\caption[a]{The BB angular power spectrum in linear scale (plot at the left) and in logarithmic scale (plot at the right).}
\label{Figure5}      
\end{figure}
The Planck experiment \cite{planck} will observe the microwave sky in nine frequency channels:
three frequency channels 
(i.e. $\nu= 30,\,44,\,70$ GHz) belong to the low frequency instrument (LFI);  six 
channels (i.e. $\nu= 100,\,143,\,217,\,353,\,545,\,857$ GHz) belong to the high 
frequency instrument (HFI). The BB power spectra for all the relevant 
frequency channels. There are reasons to expect that the sensitivity to polarization 
will be larger at high frequency \cite{planck}. At the same time the expected signal will be larger at small frequencies.
Since our source of information is primarily the Planck bluebook\footnote{We are aware 
of larger sensitivities possibly achievable in the context of the LFI. However, since these 
results have not been published, we will stick to the informations contained in the bluebook.} we will illustrate our results 
for a putative frequency of $100$ GHz. Needless to say that all the relevant magnetized polarization 
observables can be computed, with MAGcmb, for any relevant frequency and explicit examples 
will be given in what follows.
\begin{figure}[!ht]
\centering
\includegraphics[height=6.5cm]{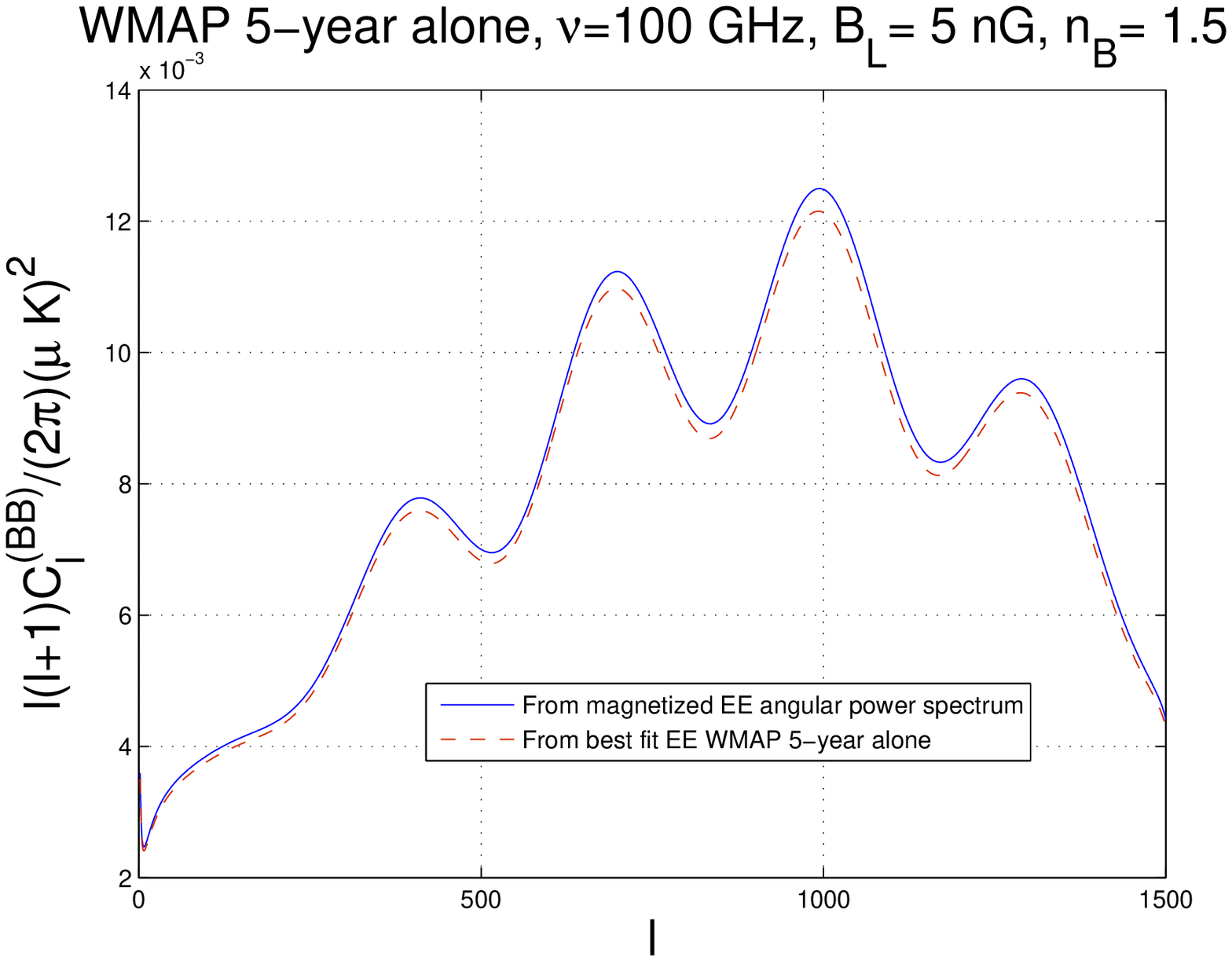}
\includegraphics[height=6.5cm]{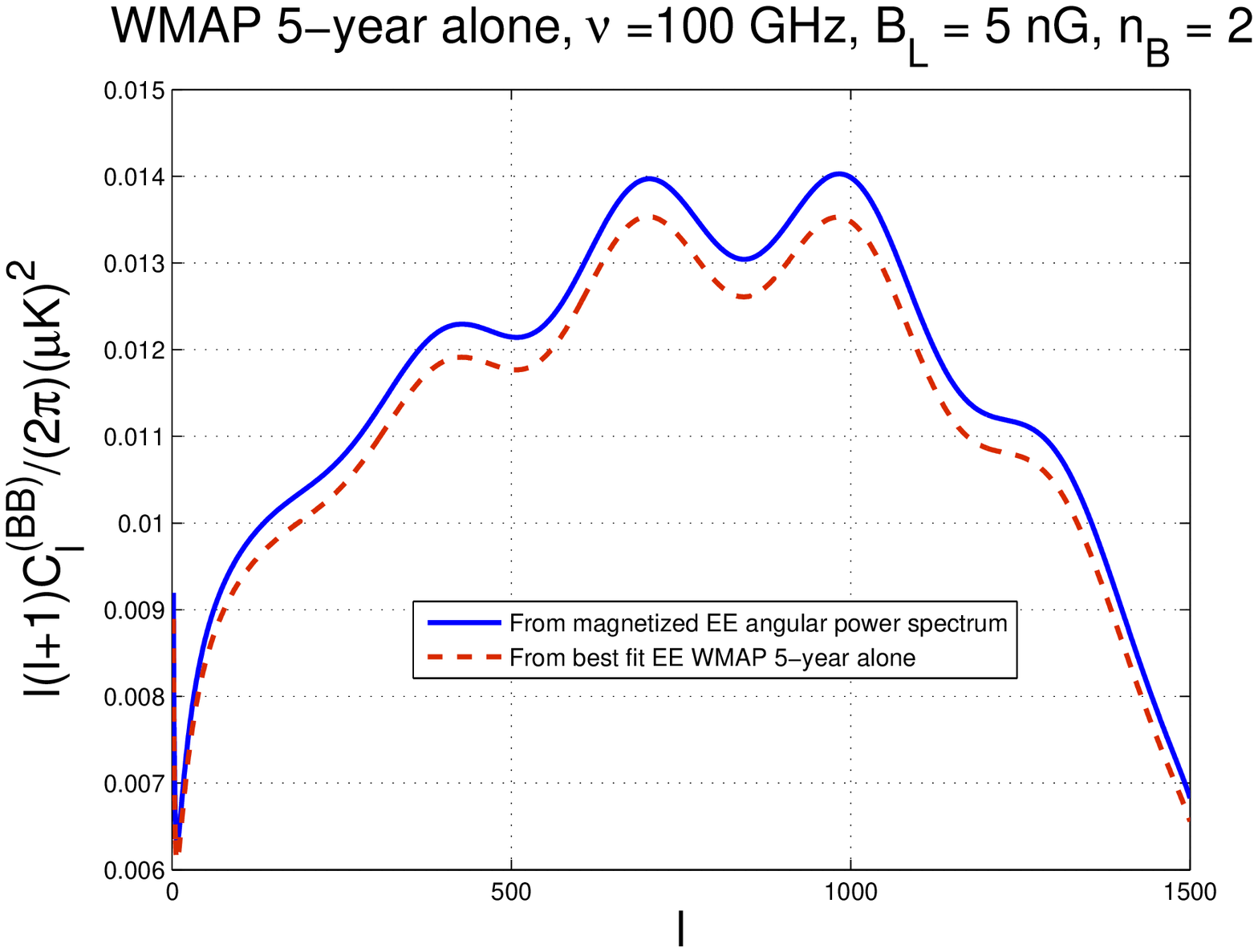}
\caption[a]{The effect of the increase in the spectral index in the region of the peaks.}
\label{Figure6}      
\end{figure}
In Fig. \ref{Figure5} the BB angular power spectrum is reported for the parameters listed in the title of the figure. 
The difference between the left and the right plot is just the scale which is linear (at the left) and which 
is logarithmic (at the right). While the linear scale is rather good in capturing the region of the peaks, the logarithmic 
scale accounts better for the low multipoles.
In Fig. \ref{Figure5}, with the full line, the BB angular power spectrum is reported in the case when 
the the EE and the TE spectra are computed taking into account the magnetized contribution. The dashed 
line, on the contrary, is obtained when the EE power spectrum is just taken to be the one
implied by the 5-year WMAP data analyzed in the light of the conventional $\Lambda$CDM model. Figure \ref{Figure5}  demonstrates that if the E-mode is computed consistently, the resulting B-mode 
is not the Faraday screened E-mode produced by the conventional adiabatic mode.

By increasing the magnetic spectral index $n_{\mathrm{B}}$ the peaks become less pronounced while the corresponding 
amplitude of the BB angular power spectrum decreases. This aspect is captured by Fig. \ref{Figure6}. At the 
left the region of the peaks is illustrated for $B_{\mathrm{L}} = 5$ nG and $n_{\mathrm{B}}= 1.5$. At the right, for the same 
value of the regularized amplitude, the spectral index is increased from $1.5$ to $2$.  
\begin{figure}[!ht]
\centering
\includegraphics[height=6.5cm]{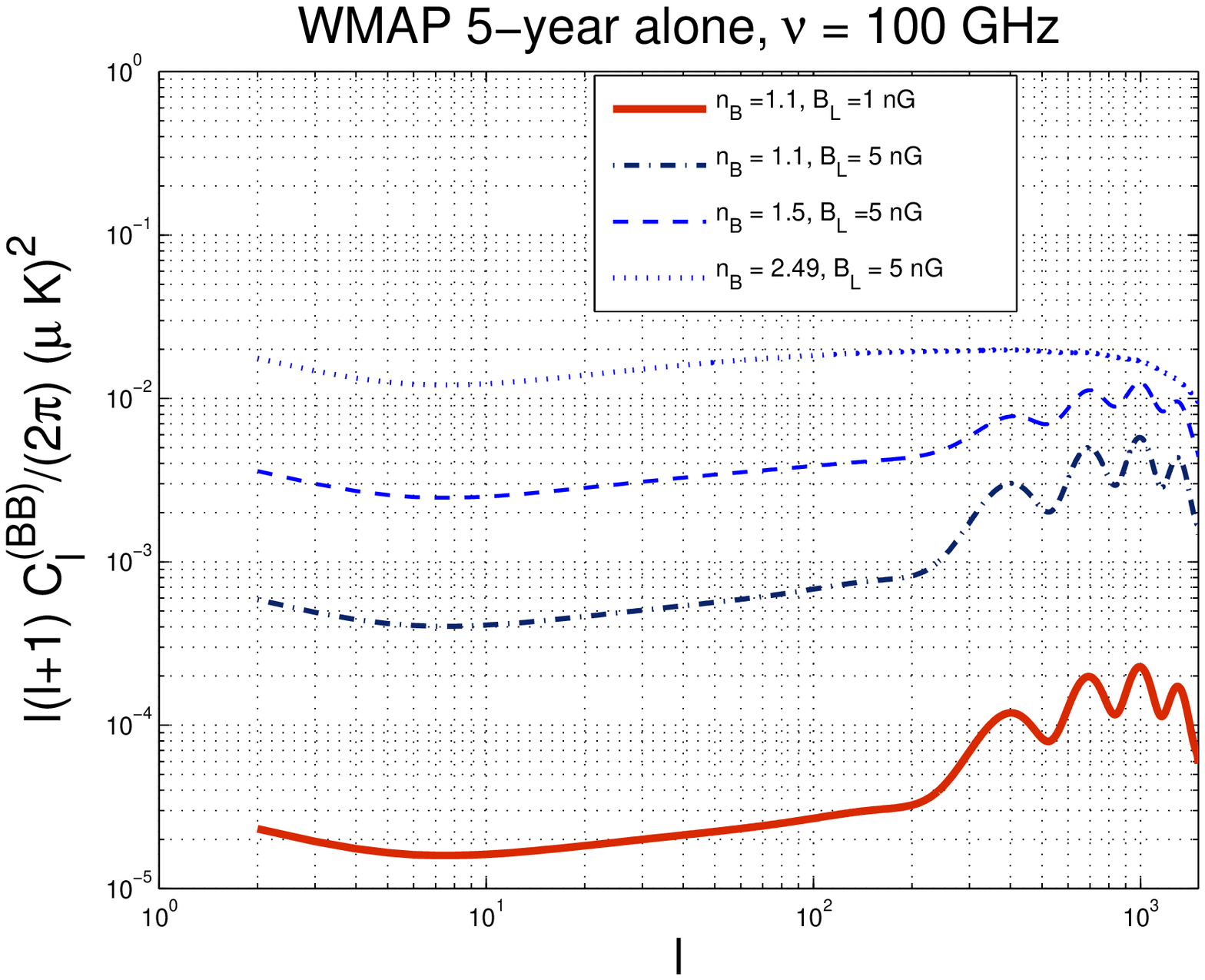}
\includegraphics[height=6.5cm]{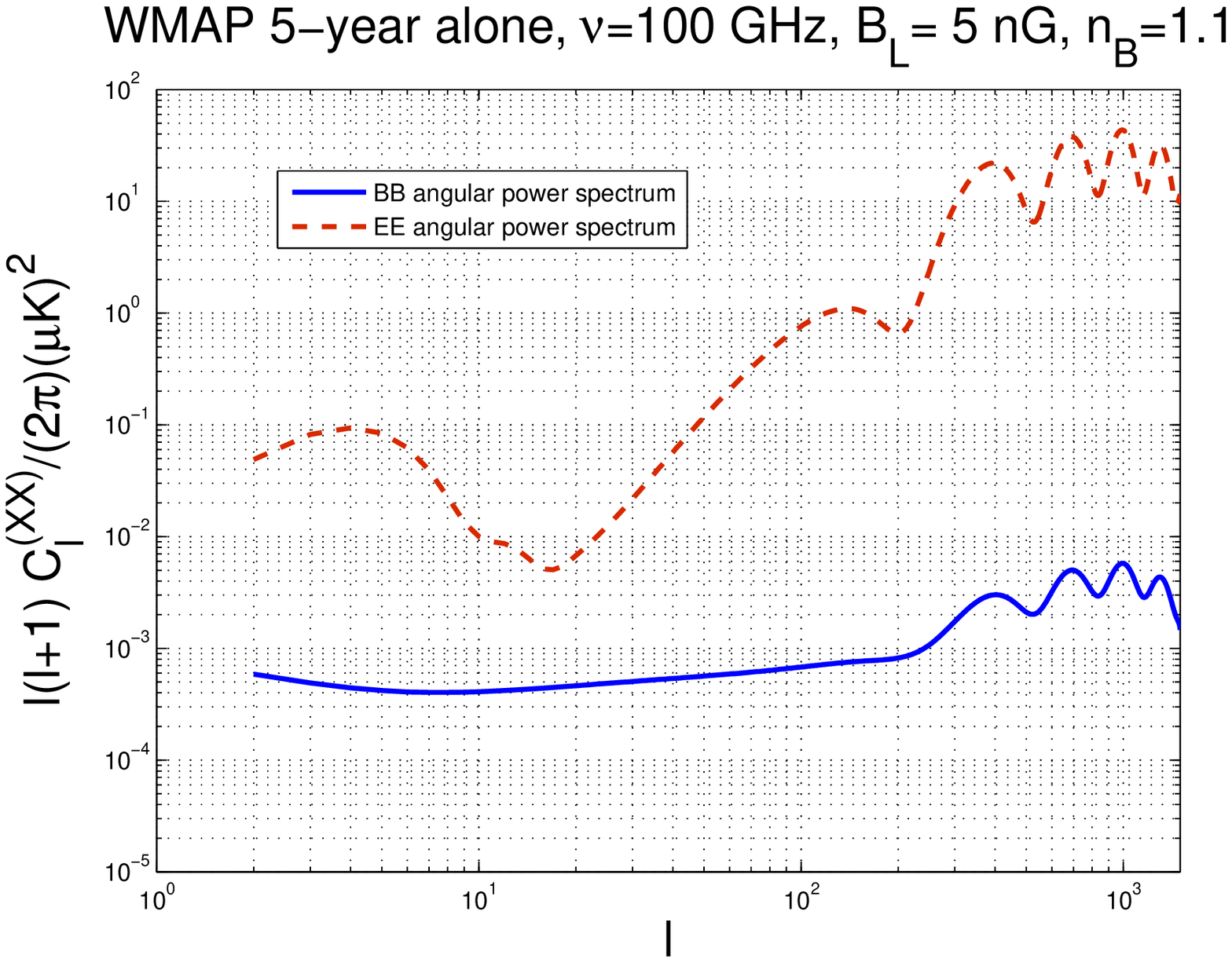}
\caption[a]{BB power spectra for different values of the parameters of the magnetized 
background (plot at the left). EE and BB power spectra are compared (plot at the right) with a
double logarithmic scale.}
\label{Figure7}      
\end{figure}
This trend is even sharper in Fig. \ref{Figure7}. In the plot at the left, with the full line, the case of $B_{\mathrm{L}}= 1$ nG 
is illustrated. The other three curves (i.e. dot-dashed, dashed and dotted) all refer to the case of $B_{\mathrm{L}} = 5$nG  but
with progressively increasing spectral indices. 
Always in Fig. \ref{Figure7} (plot at the right) the relative amplitude of the EE and BB power spectra is illustrated.

By looking at the relative amplitude of the EE and BB angular power spectra (see, for instance, right plot 
in Fig. \ref{Figure7}), it could be argued that, in principle, the full line (i.e. the BB angular power spectrum) 
and the dashed line (i.e. the EE angular power spectrum) might intersect for frequencies 
$\nu< \nu_{\mathrm{max}}$. Indeed, while the EE spectrum does not depend 
upon the frequency, the BB spectrum increases as the frequency decreases well below $\nu_{\mathrm{max}}$.
\begin{figure}[!ht]
\centering
\includegraphics[height=6.5cm]{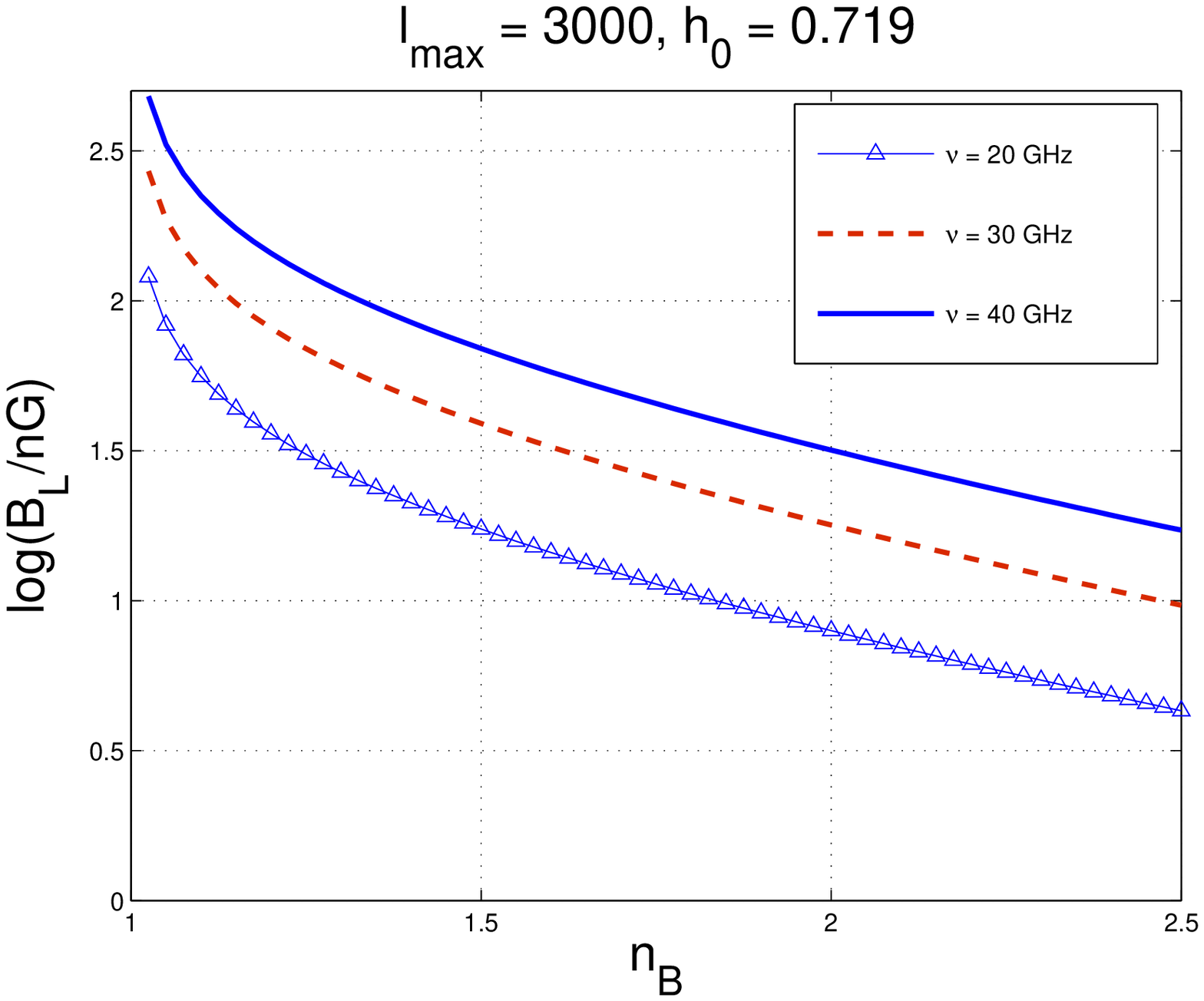}
\includegraphics[height=6.5cm]{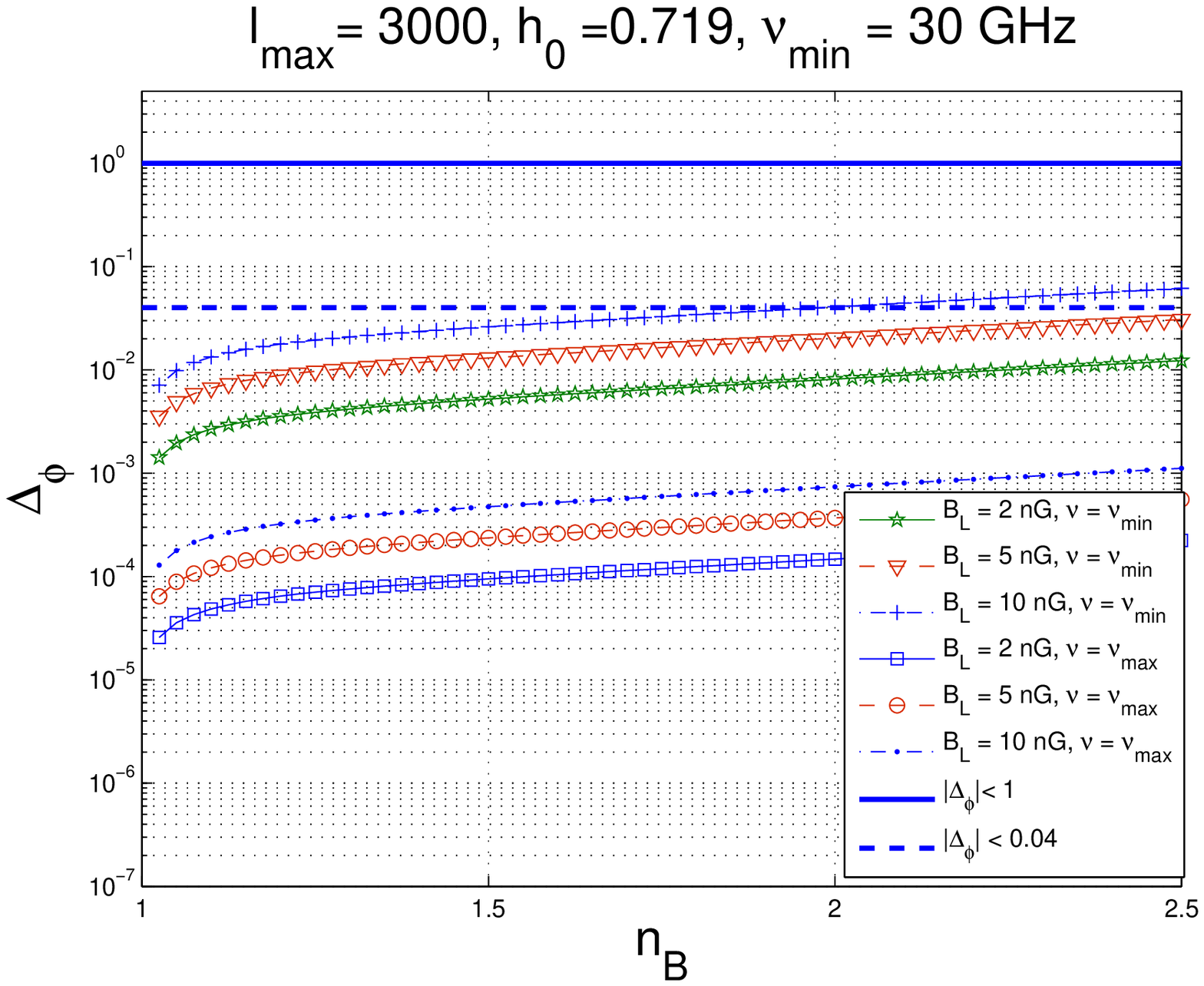}
\caption[a]{Bounds stemming from the consistency of the calculation.}
\label{Figure8}      
\end{figure}
The analytical and numerical considerations pursued so far hold under the assumption of sufficiently small 
Faraday rotation rate, i.e. $F(\hat{n})<1$. This is not a severe limitation and does not help to exclude 
the mutual cross-over of the EE and BB power spectra. Indeed, the requirement $F(\hat{n})<1$ implies a condition 
on Eq. (\ref{FRR15}). In particular, to keep the estimate semi-analytical, we can observe 
that the integrand appearing at the right hand side always increases (as a function of $\ell$) if we consider, just 
for illustrative purposes, the case $n_{\mathrm{B}}> 1$. Thus the integral will be dominated, in the 
first approximation, by $\ell \simeq \ell_{\mathrm{max}}$ and this observation leads to a (not so constraining) 
bound which is reported in Fig. \ref{Figure8} (plot at the left).  In the $(B_{\mathrm{L}}, n_{\mathrm{B}})$ 
plane the constraints will be different depending upon the specific frequency. Always in Fig. \ref{Figure8} (see the plot at the right) the mean square rotation angle (i.e. 
$|\Delta_{\phi}| = \sqrt{\langle |\Delta\varphi^{(F)}|^2 \rangle}$, see also Eq. (\ref{FRR15})) is reported for 
different values of the regularized magnetic field intensity $B_{\mathrm{L}}$ and for different 
frequencies.  The maximal frequency will be taken to correspond to the frequency of the CMB maximum (i.e. 
$\nu_{\mathrm{max}} = 222.617$). The minimal frequency will be taken to be $\nu_{\mathrm{min}} = 30$GHz.
Now, the WMAP 5-year data give a series of bounds on the rotation 
angle due to the birefringent nature of the primeval plasma. As already 
discussed the analysis of \cite{WMAP54} does not apply to the present case for, at least, two reasons. 
In the case studied in the present paper 
the rotation angle has a well defined power spectrum (i.e. an explicit $\ell$ dependence).
Secondly, the rotation angle also depend upon the frequency. 
In Fig. \ref{Figure8} with the full thick line we report the bound $|\Delta_{\phi}|< 1$, as implied by the 
internal consistency of the calculation.
With the dashed (thick) line the bound $|\Delta_{\phi}|<0.04$ is reported.
The latter figure would read, in degrees,  $|\Delta_{\phi}|<2.4$ deg and it is one of the most constraining 
bounds reported in  \cite{WMAP54}.  

The WMAP 5-year data (see, in particular, \cite{WMAP55}) imply that, when averaged over $\ell= 2-6$,
$\ell (\ell +1) C_{\ell}^{(\mathrm{BB})}/(2\pi) < 0.15 (\mu\mathrm{K})^2$ ( $95\,\%$ C.L.). Slightly less 
stringent bounds were obtained by using the 3-year data \cite{WMAP32}. 
\begin{figure}[!ht]
\centering
\includegraphics[height=6.5cm]{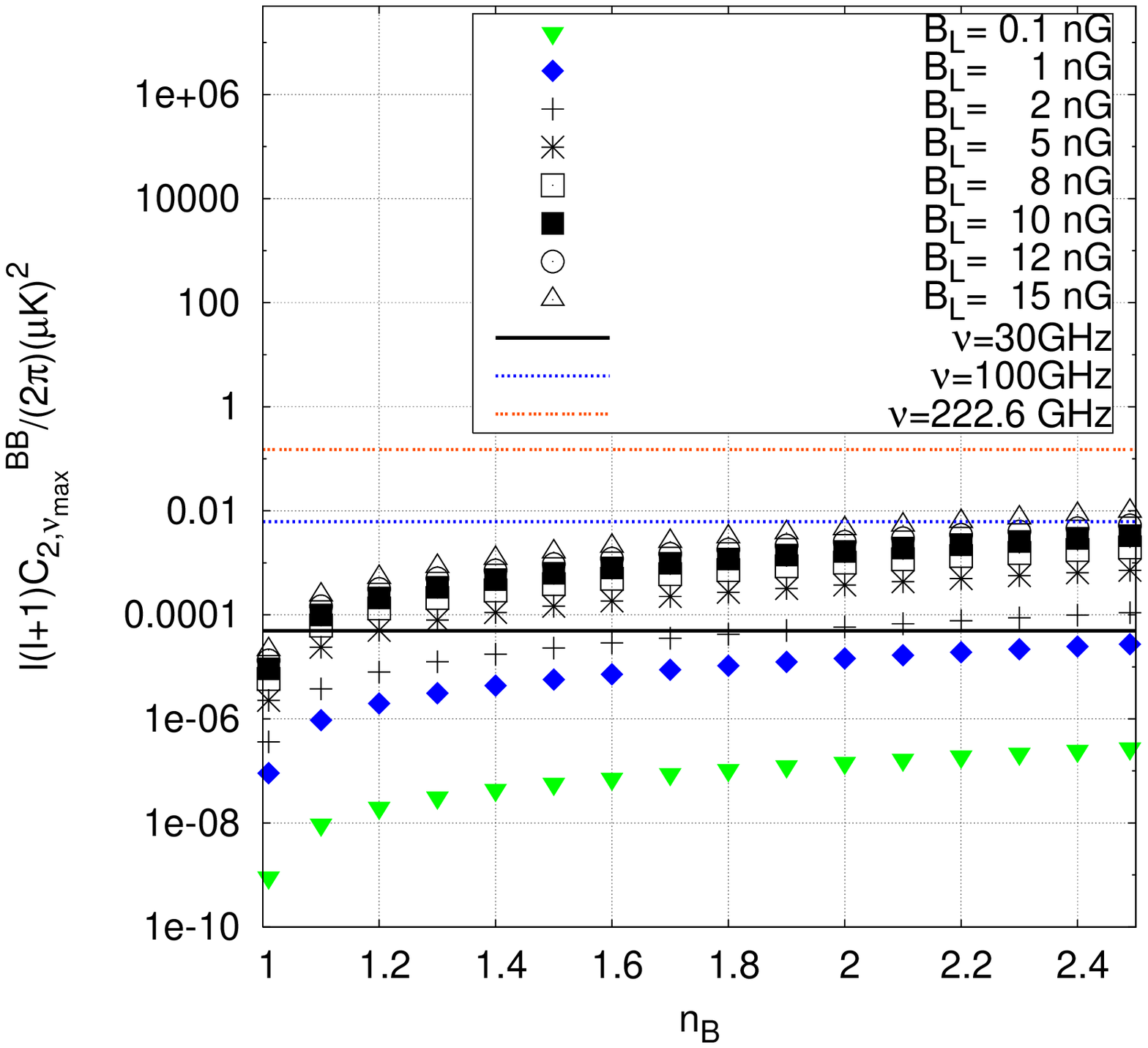}
\includegraphics[height=6.5cm]{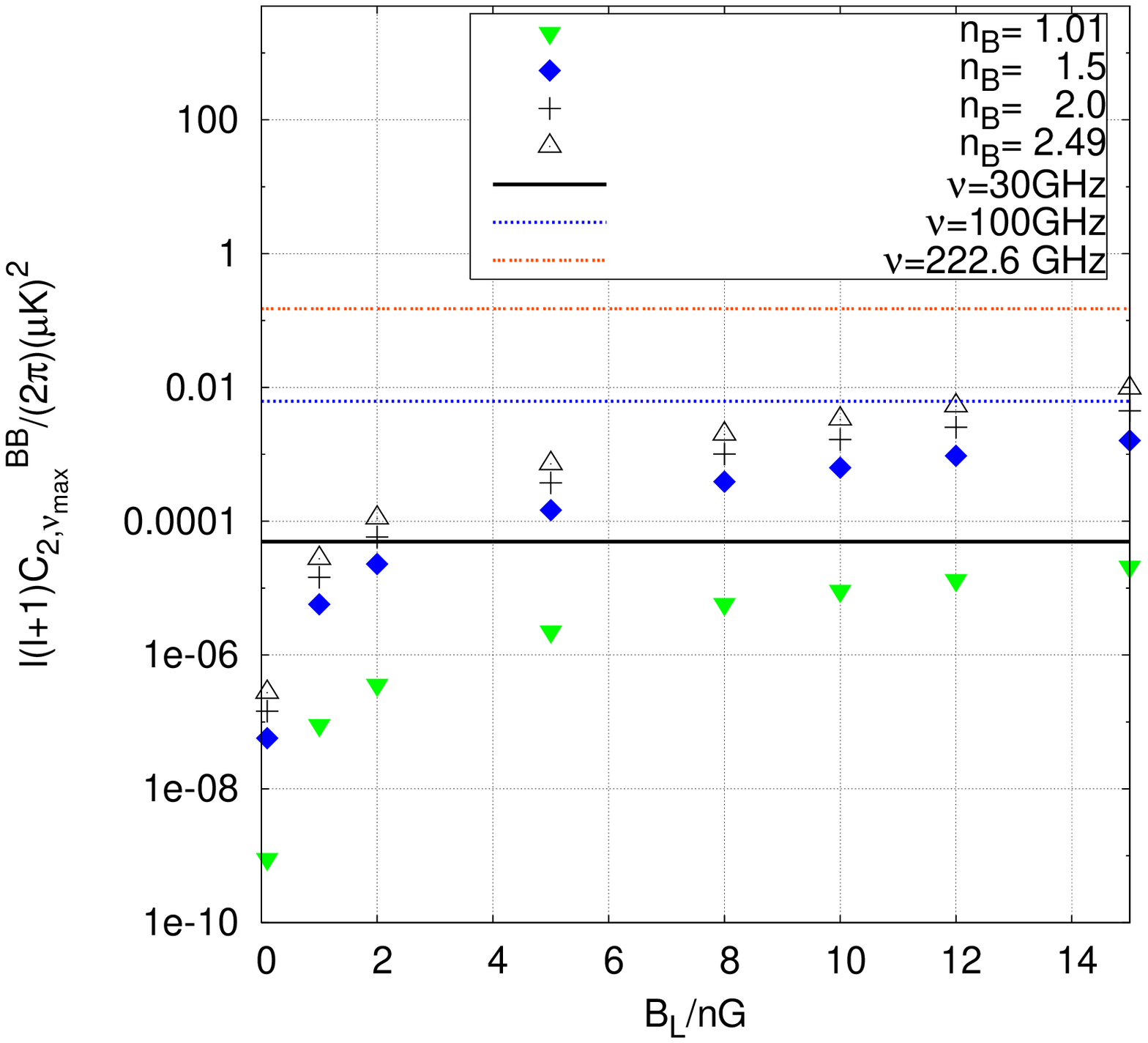}
\caption[a]{Direct constraints on the BB angular power spectrum are illustrated, for different 
frequencies, with the full, dashed and dot-dashed horizontal (thick) lines.}
\label{Figure9}      
\end{figure}
As already mentioned the putative constraint  of \cite{WMAP55} does not make reference 
to a specific frequency. So we have to assume that it holds for all the frequencies.
Then, in our case it should be imposed at the lowest frequency which is 
the most constraining one. The lowest available 
frequency for this purpose would be for $27$ GHz. The preceding frequency (i.e. $23$ GHz) 
has been used as a foreground template \cite{WMAP32} and, consequently, the EE and BB
power spectra have not been freed from the foreground contamination.
We therefore choose to set the bound for a minimal frequency of $30$ GHz since this 
not only intermediate between the KKa and KQ bands of the WMAP experiment 
but it is also the putative (lowest) frequency of the Planck experiment.
\begin{figure}[!ht]
\centering
\includegraphics[height=6.5cm]{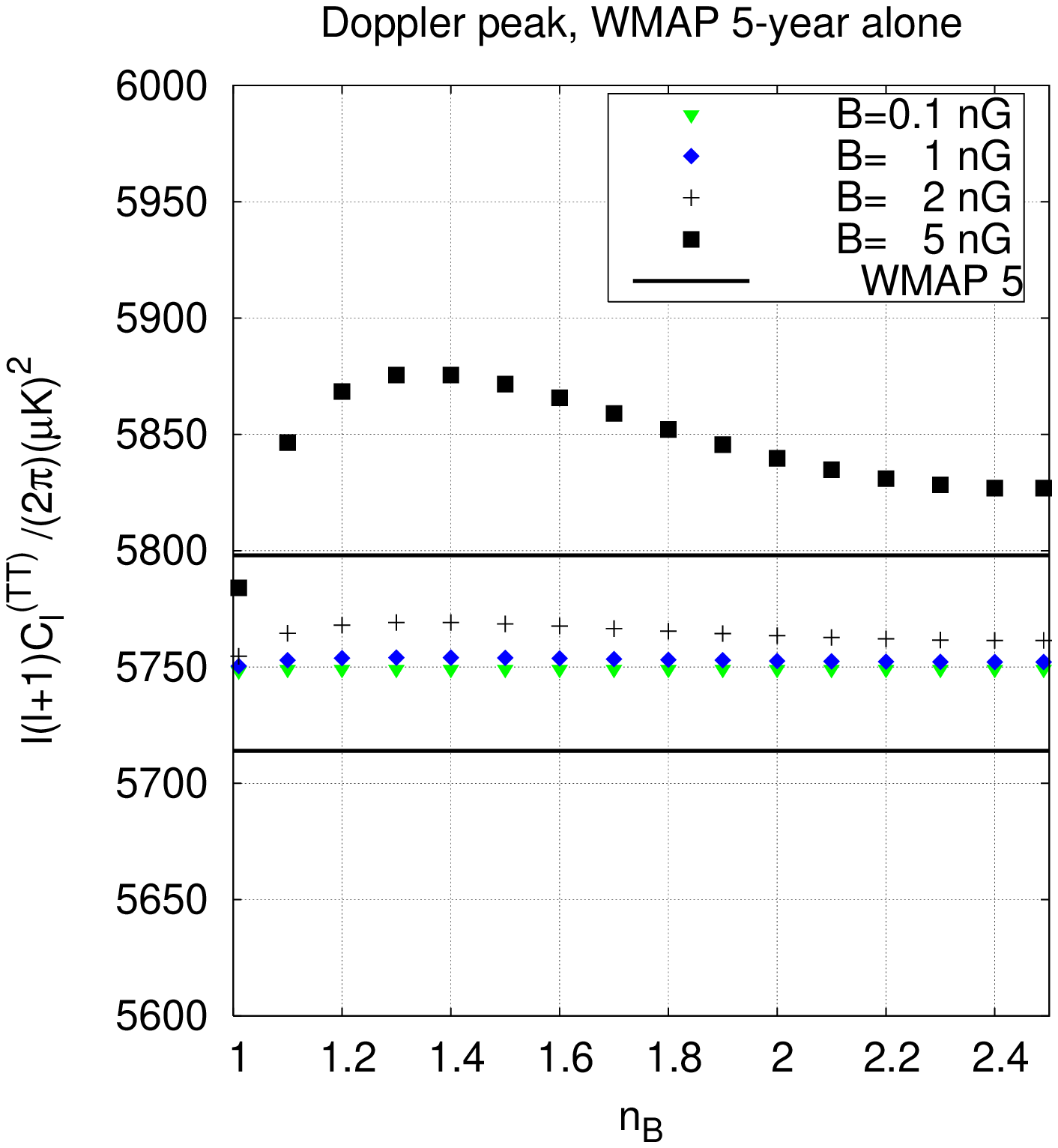}
\includegraphics[height=6.5cm]{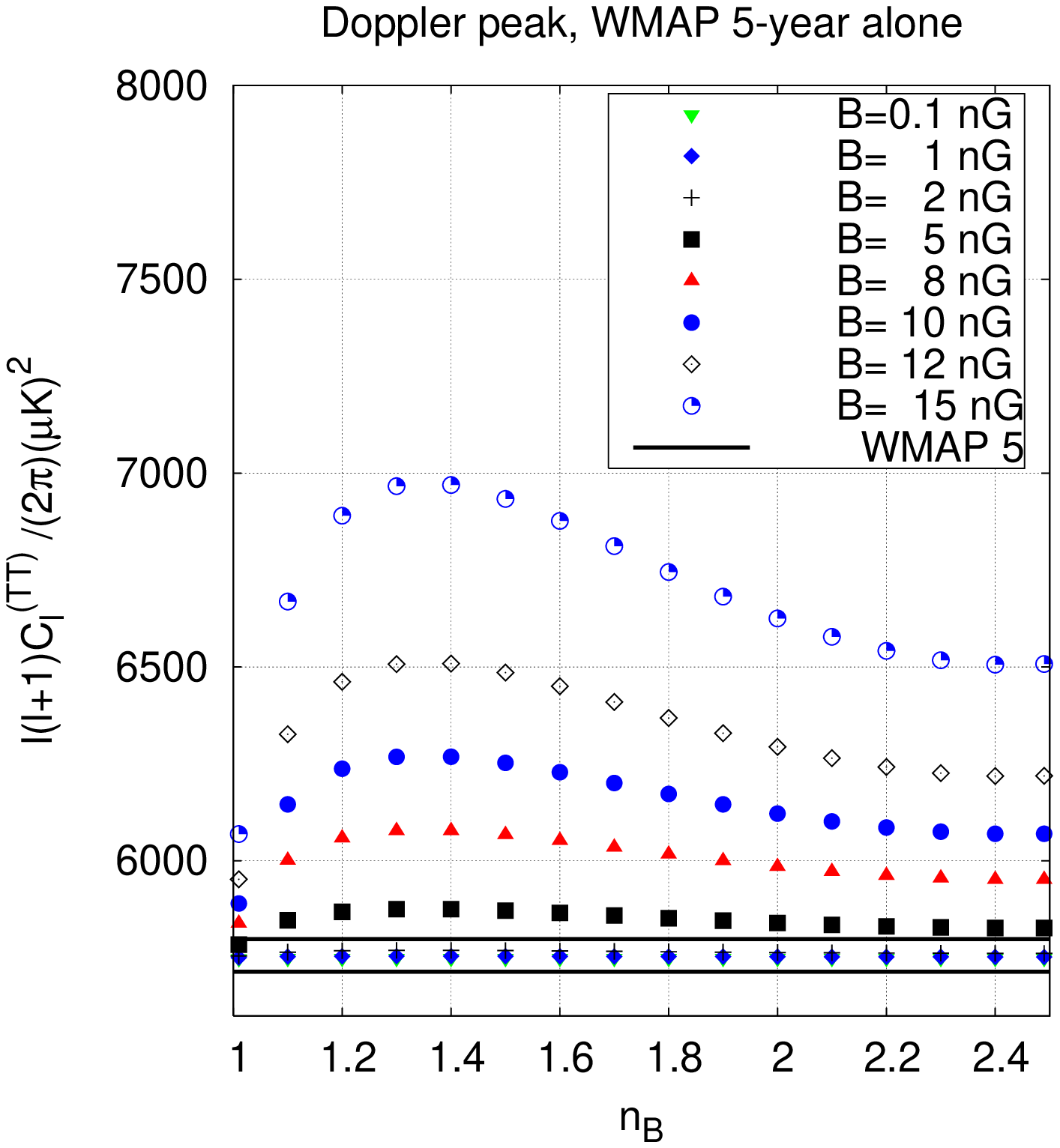}
\caption[a]{Direct constraints on the distortion and increase of the first acoustic peak 
of the TT angular power spectrum are reported.}
\label{Figure10}      
\end{figure}
As expected, the direct bound on the $BB$ angular power spectrum is not so constraining. This 
aspect is  visible in Fig. \ref{Figure9} where, for illustration, the case of blue 
spectrum has been considered. The different bounds (horizontal lines) refer to different 
fiducial frequencies. The full, dashed and dot dashed lines refer, respectively, to the cases 
of $\nu =30$ GHz, $\nu = 100$ GHz and $\nu= \nu_{\mathrm{max}}$. 

In \cite{gk1,gk2} it was argued that an idea on the range of variation of the parameters 
of the magnetized background can be obtained by monitoring the height and the position 
of the acoustic peaks. In particular, the WMAP collaboration measures  that the Doppler 
multipole is given by \cite{WMAP54}
\begin{equation}
\frac{\ell (\ell +1)}{2\pi} C_{\mathrm{Doppler}}^{(\mathrm{TT})} = 5756 \pm 42 \,\, (\mu K)^2.
\label{acous}
\end{equation}
In Fig. \ref{Figure10} the Doppler multipole is computed for different values of the parameters of the magnetized 
background. The full (thick) lines give the region allowed by error bar quoted in Eq. (\ref{acous}).

It is finally useful to recall that in \cite{gk2} it has been suggested that the highest peaks in the acoustic oscillations 
might be sensitive to large-scale magnetic fields. Since the 5-year WMAP data have been combined with ACBAR 
data we might suspect that the bounds stemming from the acoustic peaks of higher order (i.e. compatible with the ACBAR
range) would put more severe constraints. 
\begin{figure}[!ht]
\centering
\includegraphics[height=6.5cm]{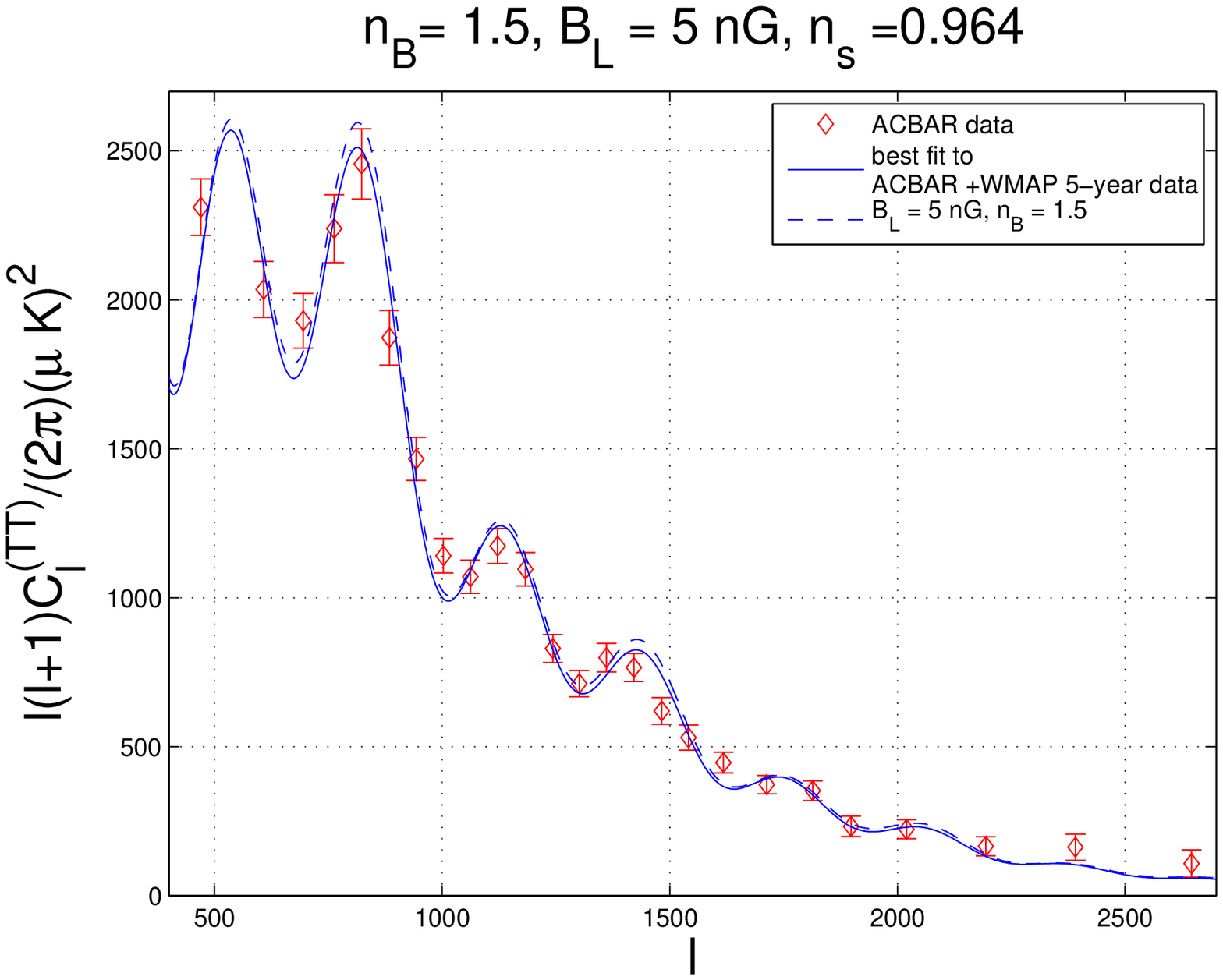}
\includegraphics[height=6.5cm]{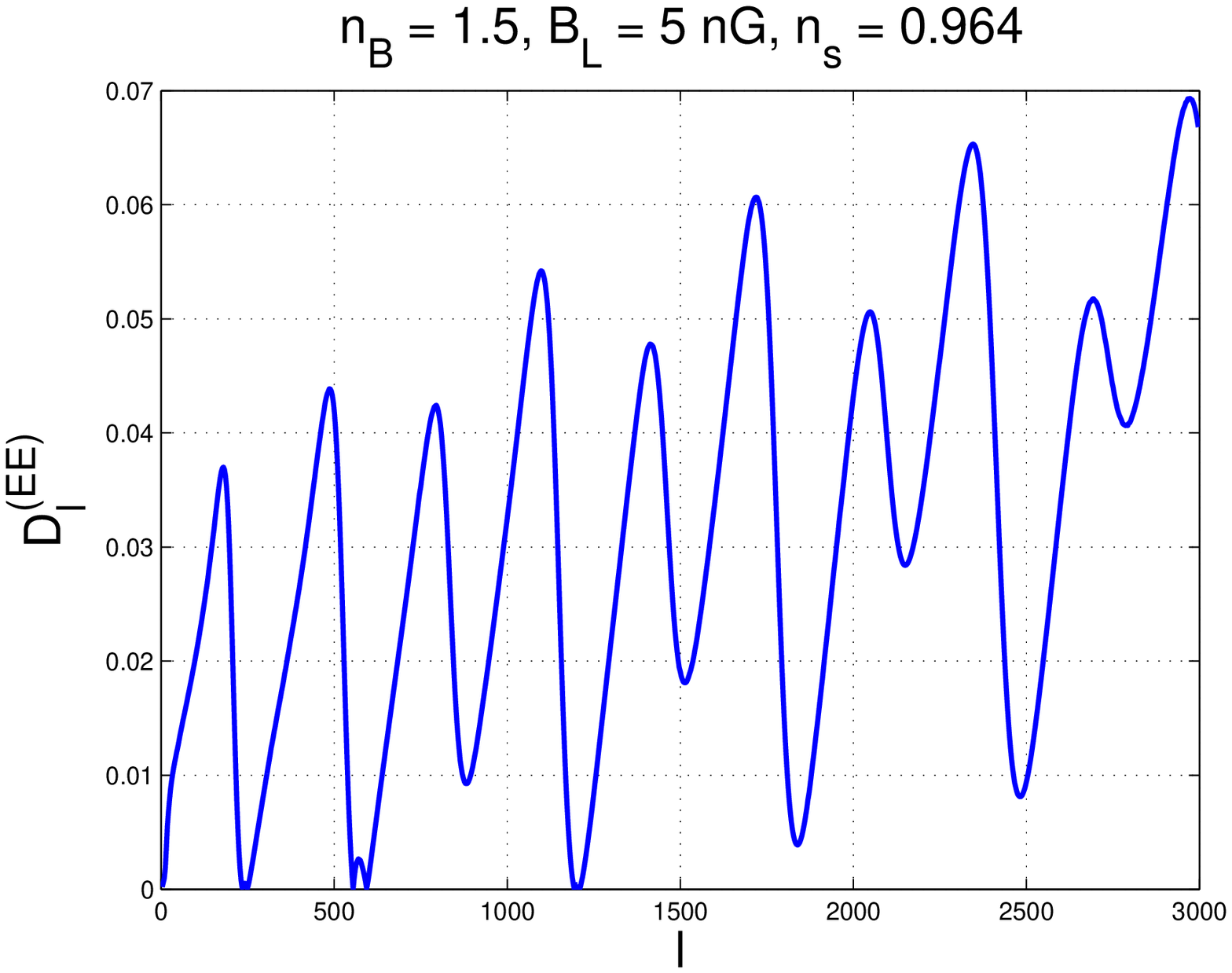}
\caption[a]{Angular power spectra at high $\ell$.}
\label{Figure11}      
\end{figure}
This is certainly a perspective which should be investigated. From our 
preliminary investigation it seems that the height of the first acoustic peak is 
still one of the best indicators of the typical patterns of correlated distortions induced by a large-scale 
magnetic field. This conclusion can be drawn by comparing, for instance, the values of the parameters 
in Fig. \ref{Figure10} with the plot at the left in Fig. \ref{Figure11}.  A magnetic field of $5$ nG and relatively 
high spectral index is barely tolerable by the ACBAR data but it is definitely ruled out by the accuracy on the 
first peak. Already with a magnetic field of $1$ nG the situation much less constrained. This 
also shows that our calculation is sensitive to magnetic fields of nG strength and smaller. This 
was never achieved in the past.  Needless to say that the high-$\ell$ region is expected to be 
accurately measured by the Planck explorer satellite \cite{planck} and this was also 
one of the motivations of our analysis. We want to stress that it would be very interesting 
to achieve a reasonable accuracy also on the EE power spectra at high multipoles. 
In Fig. \ref{Figure11} (plot at the right) we report $D_{\ell}^{(\mathrm{EE})} = |\overline{C}_{\ell}^{(\mathrm{EE})}
- C_{\ell}^{(\mathrm{EE})}|/\overline{C}_{\ell}^{(\mathrm{EE})}$, where $\overline{C}_{\ell}^{(\mathrm{EE})}$ is computed 
without any magnetized contribution and for the best -fit to the 5-year WMAP data supplemented by the ACBAR data.
On the contrary $C_{\ell}^{(\mathrm{EE})}$ is computed in the presence of a magnetized background 
with parameters specified in the title of the figure. This gives an idea of the required accuracy of the observations at high multipoles.

Up to now it has been shown how the magnetized polarization observables 
can be accurately computed. This analysis concludes the first part of the program 
commenced in \cite{gk1} and further developed in \cite{gk2,gk3}. The results 
obtained in the present paper allow to compute not only the TT  but also the TE, the EE and the 
BB power spectra which are the ones expected in the case of a fully inhomogeneous 
magnetic field\footnote{As already stressed in the introduction, the uniform field case, breaking 
explicitly spatial isotropy, can only be considered as useful toy model which is however unrealistic 
to begin with.}. The next step is to feed MAGcmb in one of the standard strategies 
of parameter estimation and this analysis is already in progress. 

In the case of birefringent effects due to pseudo-scalar interactions the strategy adopted in 
\cite{WMAP54} has been to check for a potential reparametrization 
of the BB and EE angular power spectra.  Such a reparametrization would be directly 
induced by birefringent effects and has been also discussed in Eqs. (\ref{EB10}) and (\ref{EB11}).
The advantage of such a strategy is that, of course, only one extra parameter is 
introduced on top of the parameter content of the conventional $\Lambda$CDM paradigm.

In what follows we would like to suggest a complementary strategy which would 
allow, in  principle, to determine if the birefringent effects are really due to pseudo-scalar 
particles or rather to a stochastically distributed large-scale magnetic fields. The key 
observation, in this respect, is the following. In spite of the fact that the magnetic field 
intensity enters, in a rather simple way, the Faraday rotation rate, the BB angular 
power spectrum does not have a simple scaling property as a function 
of $B_{\mathrm{L}}$ the reason is that, as explicitly demonstrated, the magnetic field 
modifies also the EE and TE spectra which are needed, in turn, to compute the BB 
power spectrum.  Conversely, a simple scaling law of the signal can be 
deduced by looking at the frequency dependence. In short the logic 
is that while the EE and the TT autocorrelations do not depend upon the 
frequency, the BB autocorrelations do depend upon $\nu$ as a consequence 
of the dispersion relations. Denoting by 
\begin{equation}
{\mathcal G}_{\ell}^{(\mathrm{T})} = \frac{\ell (\ell +1)}{2\pi} C_{\ell}^{(\mathrm{TT})}, \qquad 
{\mathcal G}_{\ell}^{(\mathrm{E})} = \frac{\ell (\ell +1)}{2\pi} C_{\ell}^{(\mathrm{EE})}, \qquad
{\mathcal G}_{\ell}^{(\mathrm{B})} = \frac{\ell (\ell +1)}{2\pi} C_{\ell}^{(\mathrm{BB})},
\label{scal1}
\end{equation}
the results obtained in the present paper  suggest that when the frequency is rescaled 
as $\nu \to \tilde{\nu}$, the three relevant autocorrelations of Eq. (\ref{scal1}) change as: 
\begin{equation}
{\mathcal G}_{\ell}^{(\mathrm{T})}(\nu) = \tilde{{\mathcal G}}_{\ell}^{(\mathrm{T})}(\tilde{\nu}), \qquad 
{\mathcal G}_{\ell}^{(\mathrm{E})}(\nu) = \tilde{{\mathcal G}}_{\ell}^{(\mathrm{E})}(\tilde{\nu}),\qquad  \nu^4  {\mathcal G}_{\ell}^{(\mathrm{B})}(\nu) = \tilde{\nu}^4 \tilde{{\mathcal G}}_{\ell}^{(\mathrm{B})}(\tilde{\nu}).
\label{scal2}
\end{equation}
There are two possible objections to this argument. The first one is that the  frequency 
channels should be sufficiently numerous. The second objection is instead that the scaling law 
expressed by Eq. (\ref{scal2}) could be contaminated by spurious scalings induced by 
the know (and yet unknown) foregrounds. 

It seems that nine frequency channels (such as the 
ones of the Planck satellite) are sufficient for testing the scaling law of Eq. (\ref{scal2}). There is, however, a proviso.
It seems that for an effective determination of the scaling law (\ref{scal2}) the polarization 
sensitivity of the three low frequency channels is an essential requirement.

Usually, in CMB studies, the frequency dependence is not used, as we suggest, to deduce the nature 
of the signal but rather in order to disentangle the possible foreground contaminations. Indeed 
both the synchrotron and the free-free emissions lead to specific frequency slopes \footnote{The
frequency slopes, directly known in limited regions of the spectrum, are often extrapolated over the 
whole physiccal range.}.  So the question becomes: are the known foregrounds able 
to simulate the scaling law of Eq. (\ref{scal2}). The answer to the latter question is negative.

To demonstrate the previous statement let us adopt the parametrization of 
the foregrounds adopted in \cite{WMAP32}, namely, in the notations of Eq. (\ref{scal1})
\begin{eqnarray}
&& \overline{{\mathcal G}}_{\ell}^{(\mathrm{EE})} = \biggl[ Q_{\mathrm{sE}} \biggl(\frac{\nu}{65\mathrm{GHz}}\biggr)^{2 \beta_{\mathrm{sE}}} + 
Q_{\mathrm{dE}} \biggl(\frac{\nu}{65\mathrm{GHz}}\biggr)^{2 \beta_{\mathrm{dE}}} \biggr] \ell^{m},
\label{scal3}\\
&& \overline{{\mathcal G}}_{\ell}^{(\mathrm{BB})} = \biggl[ Q_{\mathrm{sB}} \biggl(\frac{\nu}{65\mathrm{GHz}}\biggr)^{2 \beta_{\mathrm{sB}}} + 
Q_{\mathrm{dB}} \biggl(\frac{\nu}{65\mathrm{GHz}}\biggr)^{2 \beta_{\mathrm{dB}}} \biggr] \ell^{m}.
\label{scal4}
\end{eqnarray}
Equation (\ref{scal3}) is a simplified parametrization of the EE foreground while Eq. (\ref{scal4}) is a parametrization 
of the BB foreground. Both parametrization hold outside the galactic mask (conventionally called P06). 
In Eqs. (\ref{scal3}) and (\ref{scal4}) the two components of the foreground are given, respectively, by the dust 
(characterized by the subscript d) and by the synchrotron (characterized by the subscript s). The values 
of the various parameters appearing in Eqs. (\ref{scal3}) and (\ref{scal4}) are: 
\begin{eqnarray}
&& Q_{\mathrm{sE}} = 0.36 (\mu \mathrm{K})^2,\qquad \beta_{\mathrm{sE}}= - 3.0,\qquad 
Q_{\mathrm{dE}} = 1.0 (\mu \mathrm{K})^2,\qquad \beta_{\mathrm{dE}}= 1.5,
\label{scal5}\\
&&  Q_{\mathrm{sB}} = 0.30 (\mu \mathrm{K})^2,\qquad \beta_{\mathrm{sB}}= - 2.8,\qquad 
Q_{\mathrm{dE}} = 1.0 (\mu \mathrm{K})^2,\qquad \beta_{\mathrm{dB}}= 1.5.
\label{scal6}
\end{eqnarray}
In both cases $m= -0.6$.  Already from this rather naive argument (which can be applied, stricto sensu, only for sufficiently 
low multipoles, i.e. $\ell < 100$) interesting conclusions can be drawn.  Consider, indeed, the ratio 
of the foregrounds:
\begin{eqnarray}
&&\frac{\overline{{\mathcal G}}_{\ell}^{(\mathrm{BB})}}{\overline{{\mathcal G}}_{\ell}^{(\mathrm{EE})}} \simeq 0.83
 \biggl(\frac{\nu}{65\mathrm{GHz}}\biggr)^{0.4},\qquad \nu \ll  65\,\,\mathrm{GHz}, 
\nonumber\\
&& \frac{\overline{{\mathcal G}}_{\ell}^{(\mathrm{BB})}}{\overline{{\mathcal G}}_{\ell}^{(\mathrm{EE})}}\simeq 0.5, \qquad 
 \nu \gg  65\,\,\mathrm{GHz}.
\label{scal7}
\end{eqnarray}
The frequency dependence of Eq. (\ref{scal7}) (i.e. of the foregrounds) is much more shallow than the one expressed 
by Eq. (\ref{scal2}) which would imply, in terms of the same ratio of Eq. (\ref{scal7}) 
\begin{equation}
\frac{{\mathcal G}_{\ell}^{(\mathrm{BB})}}{{\mathcal G}_{\ell}^{(\mathrm{EE})}} \propto \biggl(\frac{\nu}{\nu_{\mathrm{p}}}\biggr)^{-4} 
\label{scal8}
\end{equation}
where $\nu_{\mathrm{p}}$ is an appropriate pivot frequency which can be chosen either to coincide 
with $\nu_{\mathrm{max}}$ or to be fixed experimentally on the basis of the specific features 
of the analysis. For instance, it could be useful to fix $\nu_{\mathrm{p}}$ to the value of the lowest 
(or the highest) frequency channel sensitive to polarization.

\renewcommand{\theequation}{8.\arabic{equation}}
\setcounter{equation}{0}
\section{Magnetized CMB maps}
\label{sec8}
Satellite experiments like WMAP observe the cosmic microwave background in several frequency channels. The observations are encoded in temperature and polarization maps. 
These have to be cleaned from foreground emissions, mainly coming from our own galaxy but also from extra-galactic sources.
The resulting maps are used to deduce the angular power spectra 
 $C_{\ell}^{(XY)}$ introduced in Eq. (\ref{defcross}).

Some strategies of parameter estimation 
are based on the comparison between the observed and the 
computed $C_{\ell}^{(XY)}$. In a complementary perspective, 
the temperature maps already contain valuable information which can be used to distinguish between different cosmological paradigms such as the one 
suggested by the m$\Lambda$CDM model.  
The local extrema in the temperature maps provide  an interesting diagnostic that can be used to constrain the underlying cosmological scenario. 
Heeding observations, local extrema (i.e. hot and cold spots in the cosmic microwave background), have a large signal-to-noise ratio and are therefore easily detectable. On the theoretical side the statistics of  local extrema assuming Gaussian fluctuations has been already studied in detail 
\cite{hotcold1,hotcold2,hotcold3,hotcold4,hotcold5}. 

To simulate CMB maps it is mandatory to have an efficient way of estimating power spectra in terms of a finite number of cosmological parameters. 
This step can be achieved via a Boltzmann solver 
such as COSMICS, CMBFAST and, among their descendants, MAGcmb. 
The calculated $C_{\ell}^{(XY)}$  can be fed back into the HEALPix  package \cite{healp1,healp2} which provides a pixelization scheme for the data on the sphere. This is essential in order to pass from the angular power spectra $C_{\ell}^{(XY)}$ to a full sky map of the temperature fluctuations and polarization as observed by different satellite experiments.  
 
Examples of sky maps will now be provided for different values of the pre-decopuling magnetic field and its spectral index\footnote{For the calculations reported in this section the use of the HEALPix package \cite{healp1,healp2} is warmly acknowledged.}. In particular, the attention will be focussed on three particular choices of parameters:
\begin{itemize}
\item{} the best fit of the WMAP 5-year data alone (see Eq. (\ref{best1}));
 \item{}  the best fit of the WMAP 5-year data alone supplemented by a magnetic field with $B_{\mathrm{L}}=5$nG and spectral index $n_{\mathrm{B}}=1.5$;
 \item{} the best fit to the WMAP 5-year data alone supplemented by a 
 magnetic field  $B_{\mathrm{L}}=20$ nG and spectral index $n_{\mathrm{B}}=2.49$. 
 \end{itemize}
The last choice of parameters is already ruled out by the observational bounds on the first acoustic peak, as discussed in Section \ref{sec6}. Nonetheless, it is instructive 
to consider also the last (extreme) model to emphasize a trend which is still there 
also for lower values of the magnetized background but which would be 
more difficult to visualize.
The sky maps are simulated using the 
improved version of MAGcmb which has been 
described and illustrated in the previous sections.  
The magnetized angular power spectra have been calculated for a maximal 
multipole $\ell_{\mathrm{max}}=2500$ and at a frequency 217 GHz. This frequency 
is the {\em lowest Planck frequency} with the {\em highest} sensitivity,
at least according to the Planck bluebook \cite{planck}. Again, if lower frequency channels 
will be more sensitive than what claimed at the moment, this will be 
very interesting for our study, as the skilled reader can appreciate. 

\begin{figure}[!ht]
\centering
\includegraphics[height=5.2cm]{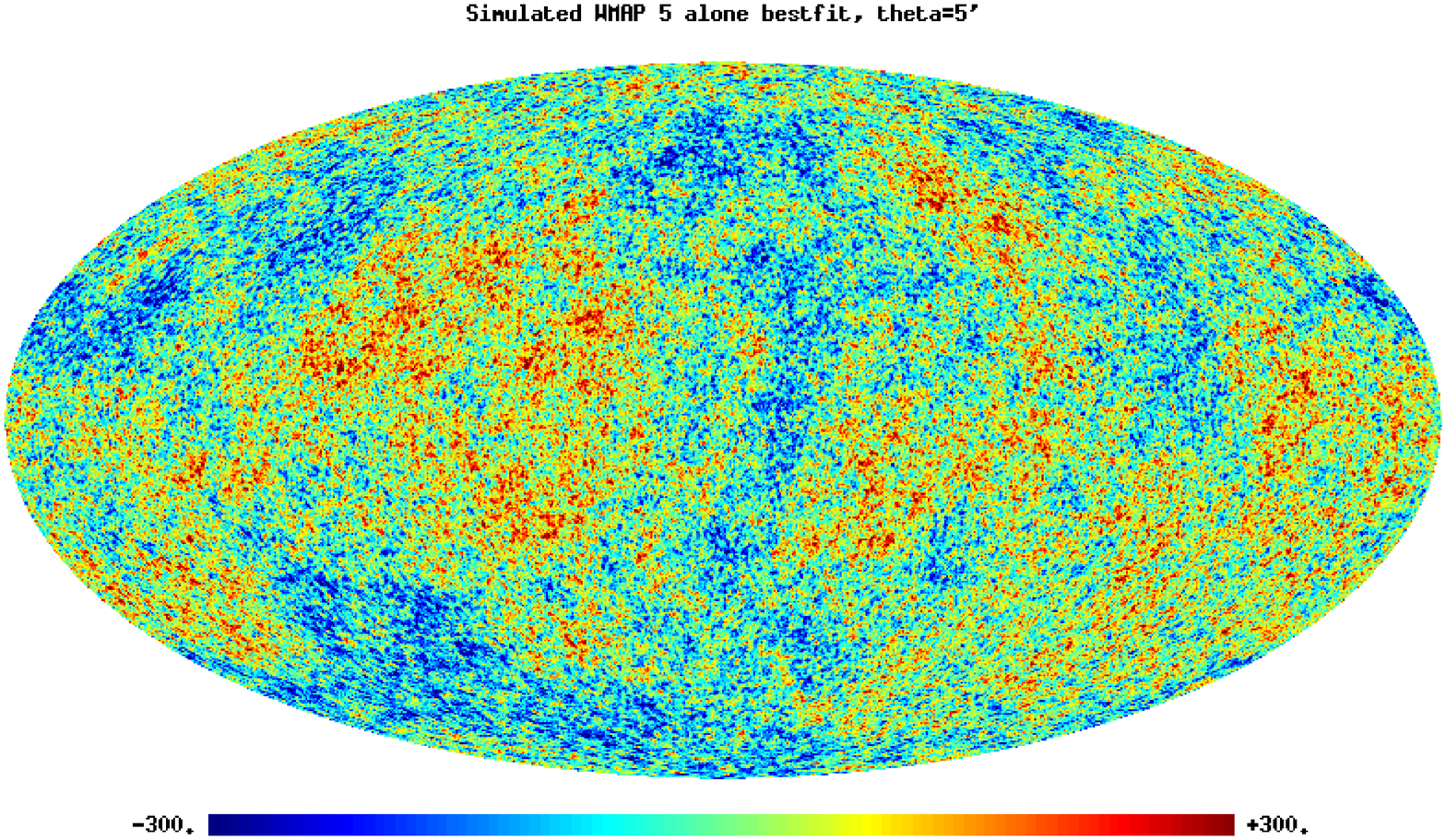}\hspace{0.7cm}
\includegraphics[height=5.2cm]{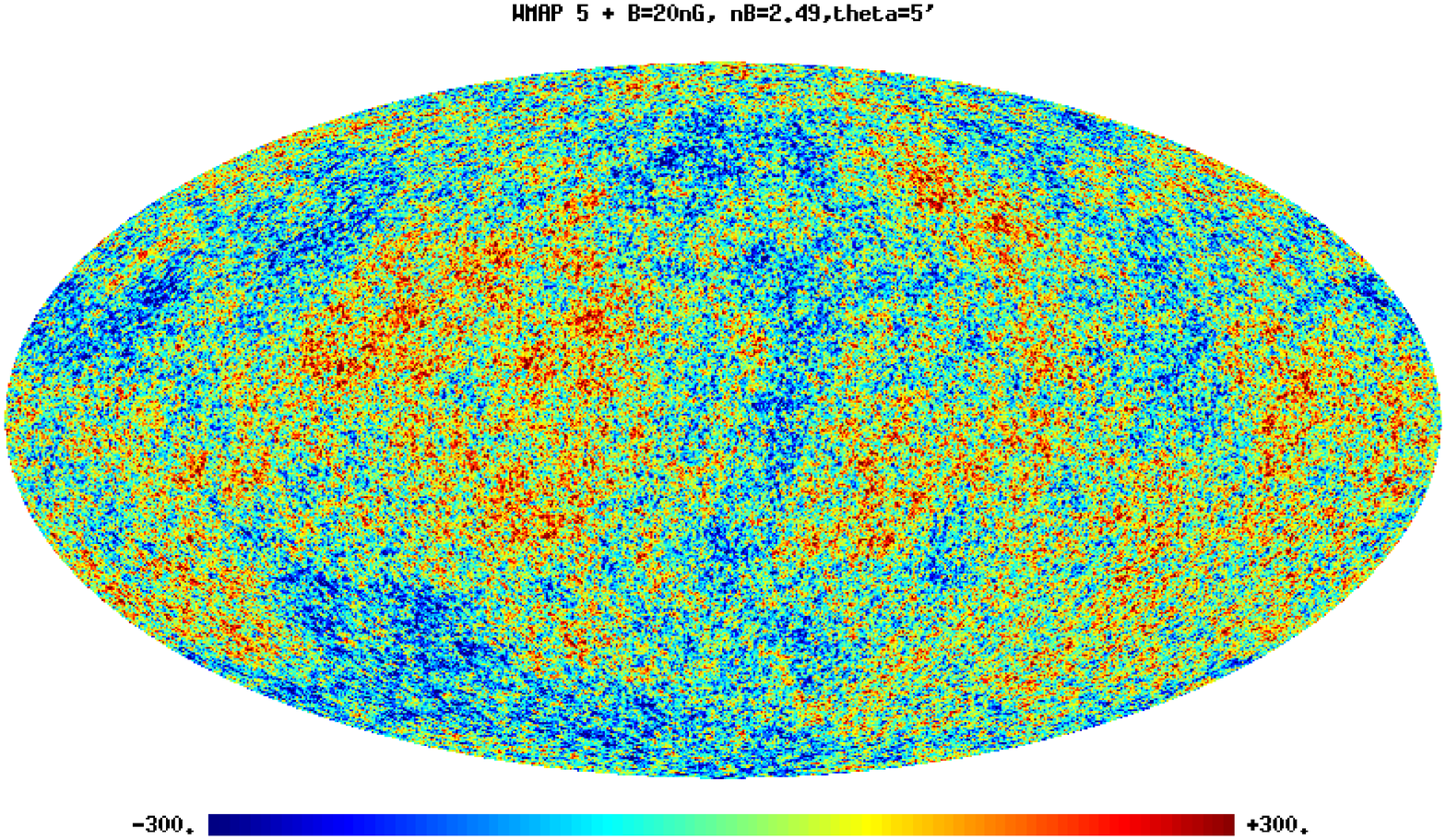}
\caption[a]{Temperature maps (in $\mu$K) for the best fit WMAP 5 model without a magnetic field (left) and with a magnetic field $B_{\mathrm{L}}=20$ nG and $n_{\mathrm{B}}=2.49$ (right) simulated assuming a Gaussian beam size $\theta_{\mathrm{FWHM}}=5'$. }
\label{figmap1}      
\end{figure}

\begin{figure}[!ht]
\centering
\includegraphics[height=6.2cm]{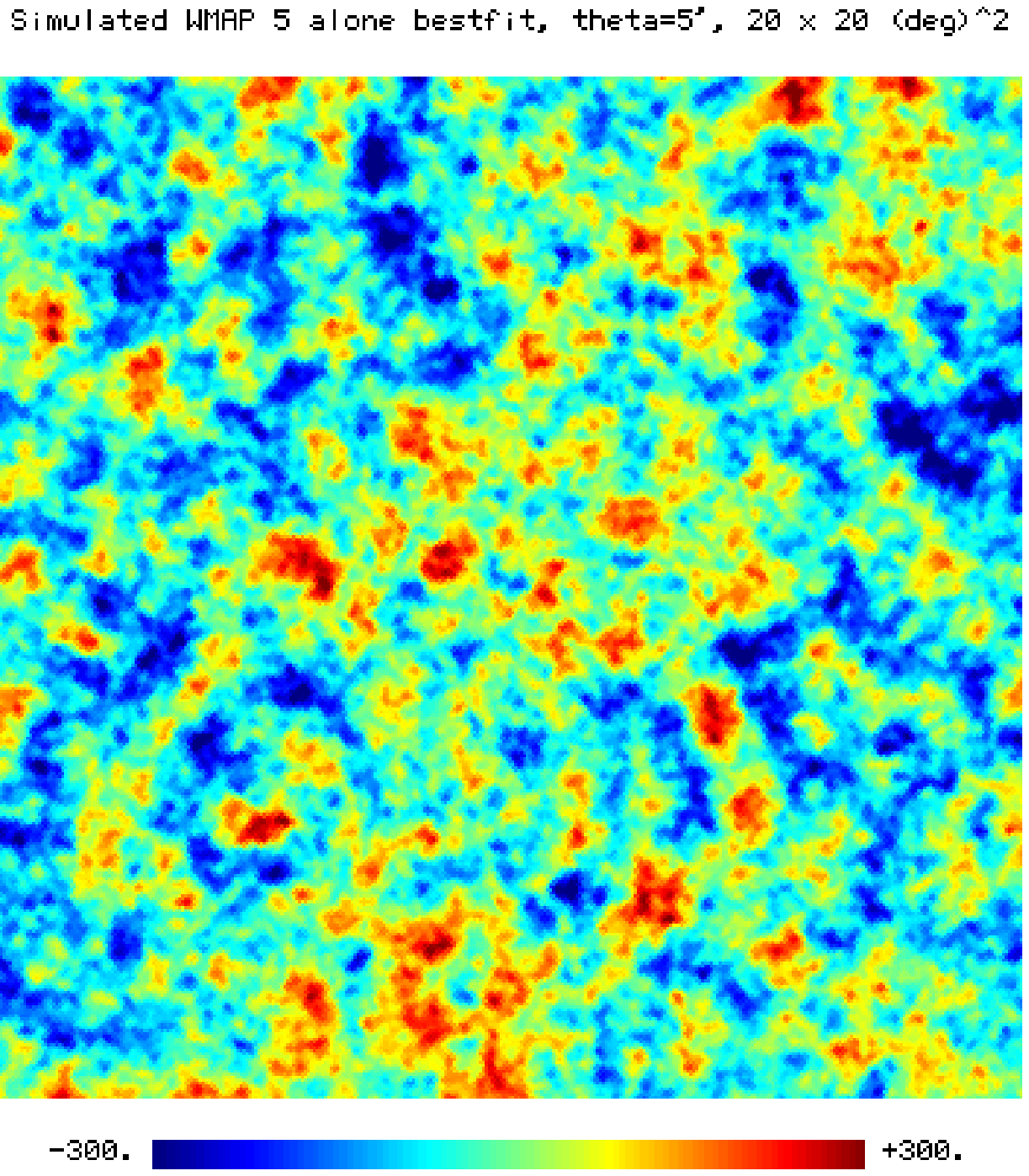}\hspace{0.7cm}
\includegraphics[height=6.2cm]{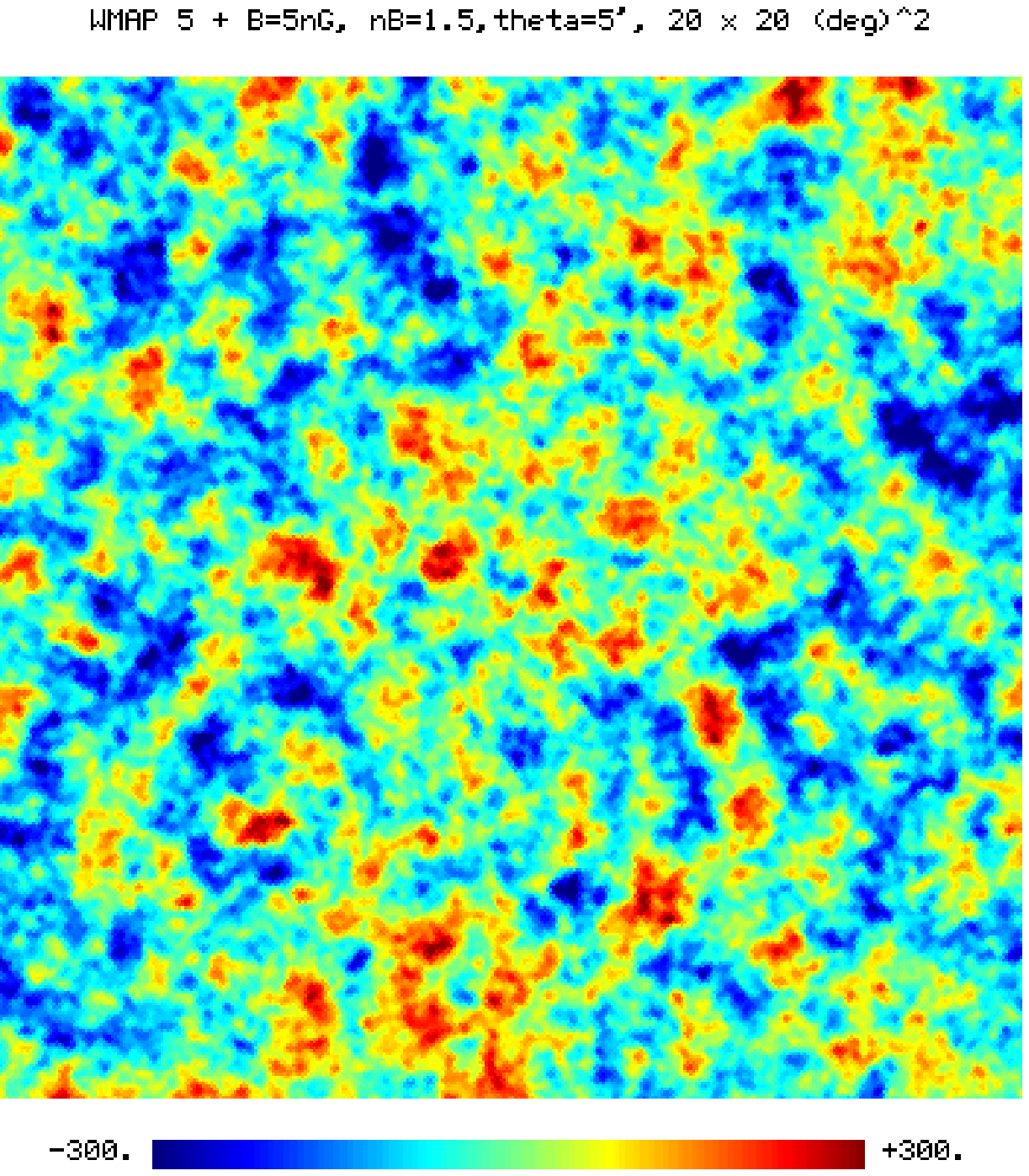}
\hspace{0.7cm}
\includegraphics[height=6.2cm]{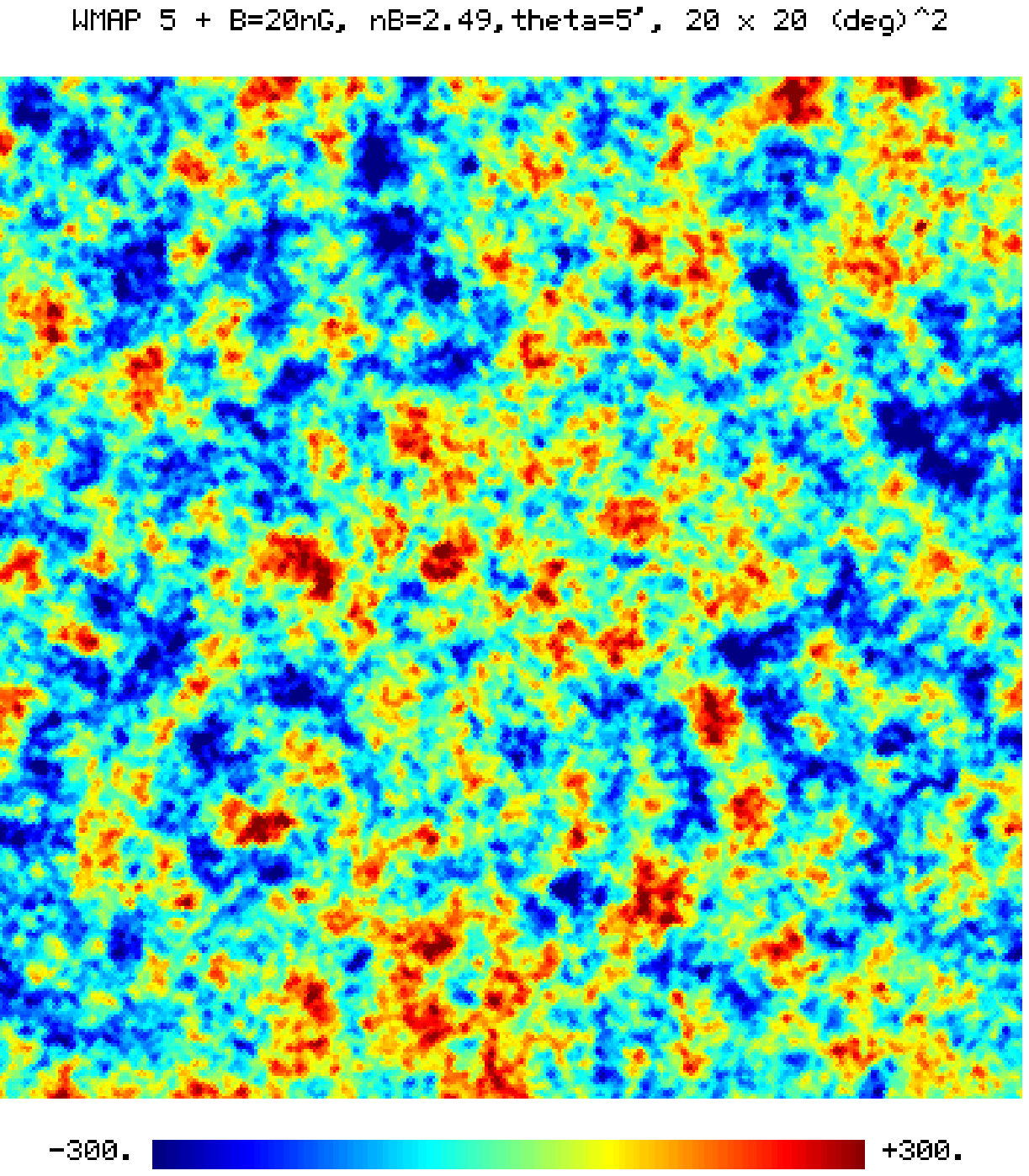}
\caption[a]{$20^{\circ}\times 20^{\circ}$ fields of the temperature maps (in $\mu$K) in gnomonic projection for the best fit WMAP 5 model without a magnetic field (left), with a magnetic field $B_{\mathrm{L}}=5$ nG and $n_{\mathrm{B}}=1.5$ (center) and with a magnetic field $B_{\mathrm{L}}=20$ nG and $n_{\mathrm{B}}=2.49$ (right) simulated assuming a Gaussian beam size $\theta_{\mathrm{FWHM}}=5'$. }
\label{figmap2}      
\end{figure}

\begin{figure}[!ht]
\centering
\includegraphics[height=5.2cm]{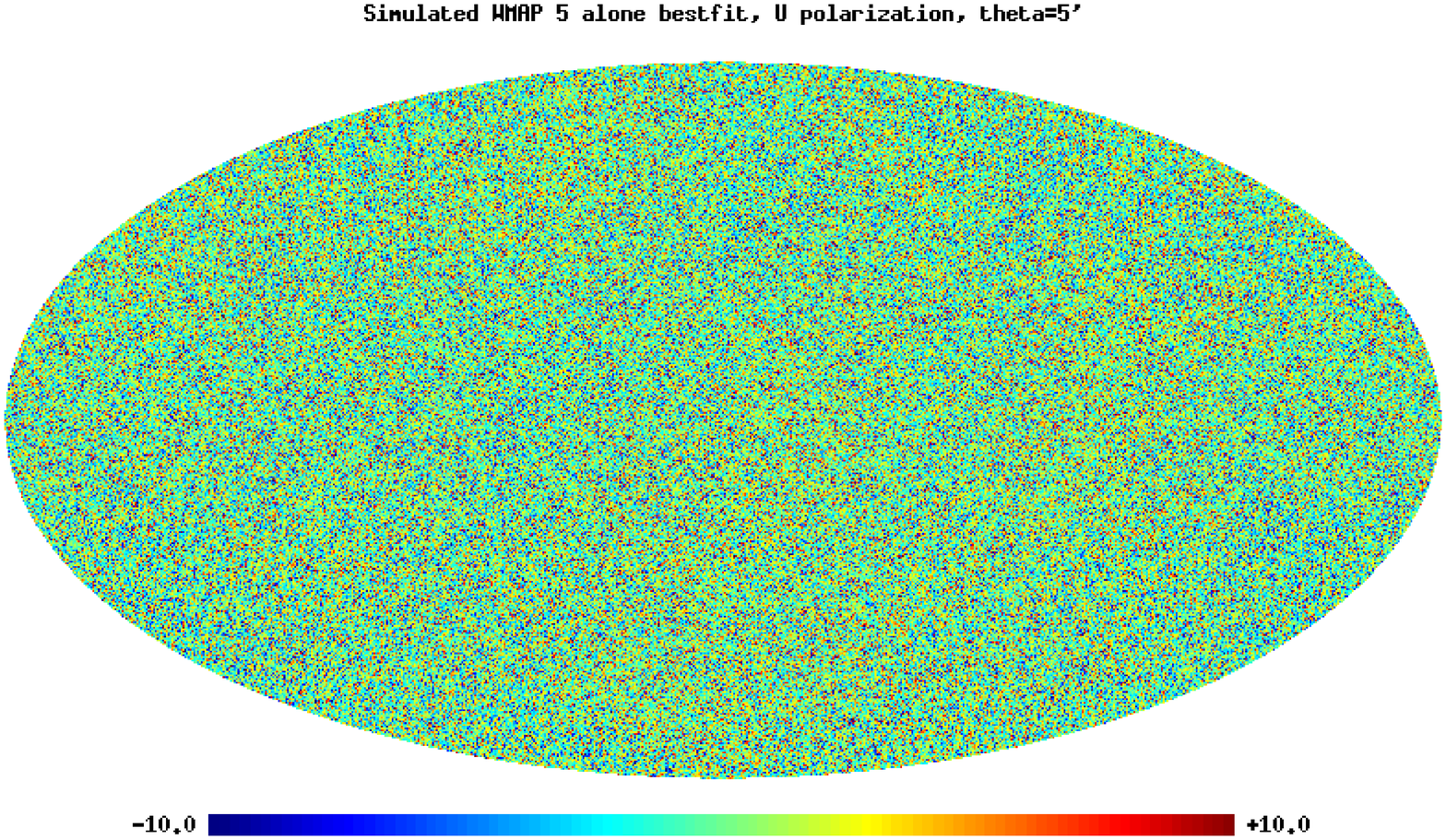}\hspace{0.7cm}
\includegraphics[height=5.2cm]{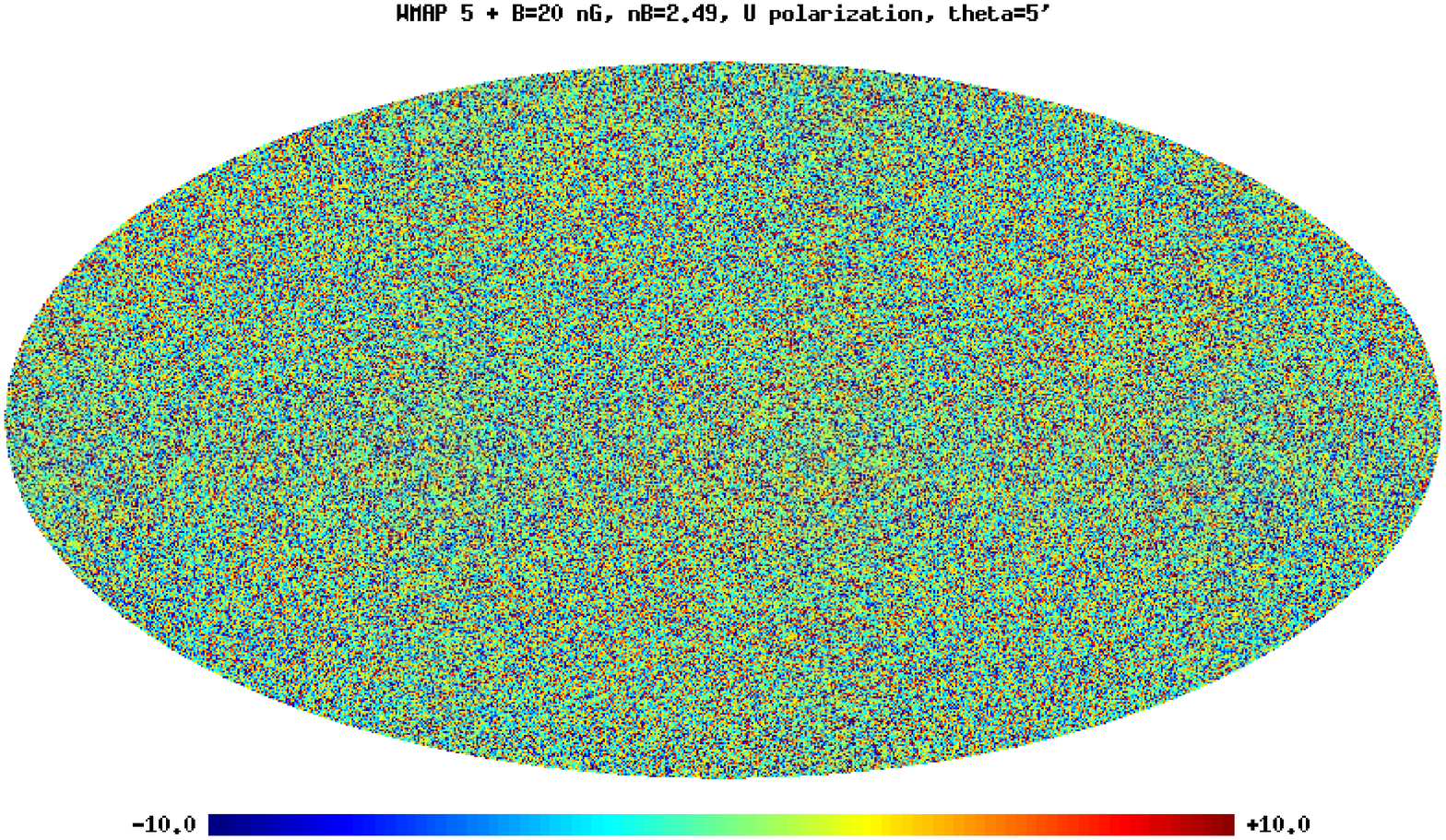}
\caption[a]{U polarization maps (in $\mu$K) for the best fit WMAP 5 model without a magnetic field (left) and with a magnetic field $B_{\mathrm{L}}=20$ nG and $n_{\mathrm{B}}=2.49$ (right) simulated assuming a Gaussian beam size $\theta_{\mathrm{FWHM}}=5'$. }
\label{figmap3}      
\end{figure}

For the simulations the value  $N_{\mathrm{side}}=1024$ has been adopted. This
figure corresponds to a resolution parameter 10 used by the WMAP team. Furthermore,  the first simulation was done using a Gaussian beam with beam size of 5 arcmin, corresponding to the Planck  channel at 217 GHz. In a second step the resulting maps were smoothed to  a Gaussian beam size $\theta_{\mathrm{FWHM}}=1^{\circ}$ i.e.  of the order of the  resolution of the five frequency channels of WMAP, which are between $0.22^{\circ}$ in the highest frequency channel at 94 GHz and $0.88^{\circ}$ in the lowest frequncy channel at 23 GHz.
In Fig. \ref{figmap1} examples of  temperature maps at Planck resolution are shown.
The effect of the presence of the magnetic field is an augmentation of the hot spots as well as a general raise of their mean temperature as will be discussed below in more detail.  The number of cold spots is increasing with increasing magnetic field strength and spectral index while their average temperature is decreasing. 
It can then be argued that the local extrema are becoming more pronounced as the magnetic field becomes more intense. 
A relatively strong magnetic field of $B_{\mathrm{L}}=20$ nG does change the appearance of the simulated temperature map. Conversely, a magnetic field of 5 nG does not have an effect simply visible in the temperature maps. 
The latter statement can be corroborated by looking at a
 $20^{\circ}\times 20^{\circ}$ patch of the sky. Indeed,  Fig. 
 \ref{figmap2} shows, visually, that it is difficult to distinguish the imprint of a magnetic field strength $B_{\mathrm{L}}= 5$ nG. By comparison,  a 20 nG field can instead be 
 distinguished (see Fig. \ref{figmap2}, plot at the far right).
The effects of a magnetic field can be also seen in the simulated polarization maps.
In Fig. \ref{figmap3} the U polarization map is reported, just for purposes 
of  illustration. The general visual result is that the U (and to a lesser extent the Q) 
maps are modified only for sufficiently intense values of the magnetic fields.
What has been reported in Figs. \ref{figmap1}, \ref{figmap2} and \ref{figmap3}
represent clearly a feasibility proof of the strategy pursued in this paper. 
It also represents a novelty since, to the best of our knowledge, magnetized CMB
maps have never been computed before within a well defined numerical approach, such as the one described here and initiated in \cite{gk1,gk2,gk3}. 

In spite of the fact that CMB maps contain a lot of precious informations, 
they cannot be just analyzed with ``visual" methods. We are going here 
to point out less anthropic ways of looking at magnetized CMB maps.
A primordial magnetic field increases the number of extrema in the temperature map. Furthermore, it increases the mean temperature of the hot spots and lowers the mean temperature of the cold spots. To make this statement more prescise the HOTSPOT routine which is part of the HEALPix package has been used to find the extrema in the three cases at hand. 
This values of the local extrema and their localization can then be obtained.
Suppose, for simplicity, we are only interested in the one-point statistics (hence the localization of the extrema is not taken into account). A similar approach has been taken, for instance,  in \cite{hotcold5}. 
The statistics of the temperature values  (number, mean, variance, skewness and kurtosis) of the minima and maxima is illustrated in Table 1.
For comparison, the simulations were done for $N_{\mathrm{side}}=1024$ which results in $N_{\mathrm{pix}}=12N_{\mathrm{side}}^2$ given by
$N_{\mathrm{pix}}=12582912$. This means that about  4\% of the total data points are local extrema.
\begin{table}[!ht]
\begin{center}
\begin{tabular}{||l|c|c|c|c|c|c||}
\hline
Model&type& number&mean($\mu$K)&var($\mu$K)$^2$&skew& curt\\
\hline
$B_{\mathrm{L}}=0$ & Max& 202174 & 76.951& 5909.1 & -0.9012 & -0.0329\\
$B_{\mathrm{L}}=5$ nG, $n_{\mathrm{B}}=1.5$& Max & 203083 &  77.778 & 6000.0& -0.9032 & -0.0340\\
$B_{\mathrm{L}}=20$ nG, $n_{\mathrm{B}}=2.49$&Max&303089& 96.355& 7542.7&-0.9728&-0.0507\\
\hline\hline
$B_{\mathrm{L}}=0$ &  Min & 201601& -76.175&5776.5 & 0.9305& -0.1108\\
$B_{\mathrm{L}}=5$ nG, $n_{\mathrm{B}}=1.5$&Min & 202495 & -76.915 & 5829.3 & 0.9360& -0.1093\\
$B_{\mathrm{L}}=20$ nG, $n_{\mathrm{B}}=2.49$&Min&302607&-92.029&6951.7&1.0603& -0.0949\\
\hline\hline
\end{tabular}
\caption{Statistics of the temperature values (number, mean, variance, skewness and curtosis) of the local extrema (maxima/minima) of the simulated temperature maps at $\theta_{\mathrm{FWHM}}=5'$ for the WMAP 5-year  bestfit model  alone (i.e. $B_{\mathrm{L}}=0$) and for the WMAP 5-year  bestfit supplemented by a magnetic field with strength $B_{\mathrm{L}}$ and spectral index $n_{\mathrm{B}}$.}
\end{center}
\label{tmap1}
\end{table}
There are further quantities characterizing the statistics of the local extrema of the temperature map, like for example the two point correlation function. This has been used to investigate the detectability of weak lensing in the distribution of hot spots \cite{hotcold4}. We leave this as well as other developments to forthcoming 
studies \cite{mk}.

\begin{figure}[!ht]
\centering
\includegraphics[height=6.2cm]{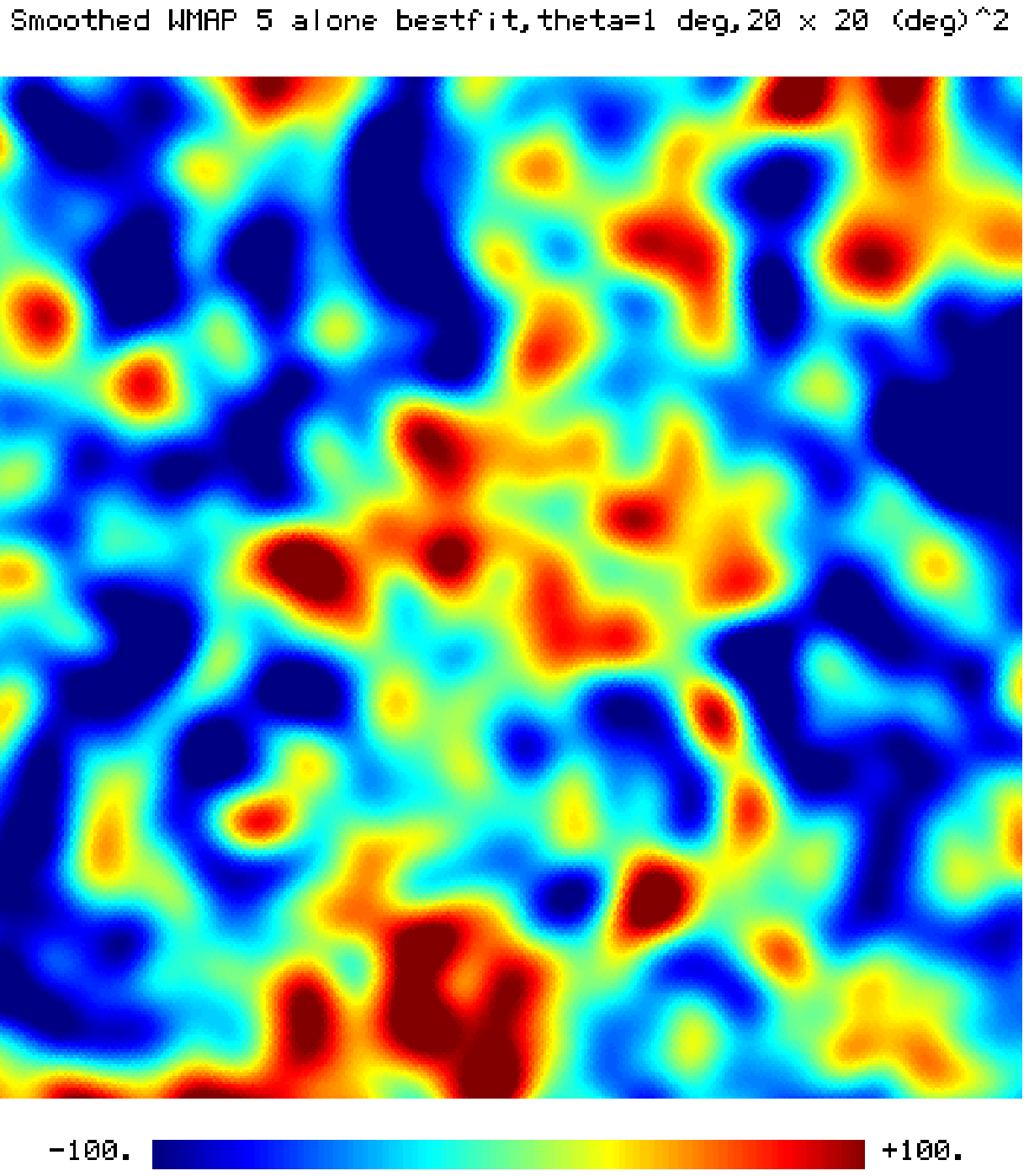}\hspace{0.7cm}
\includegraphics[height=6.2cm]{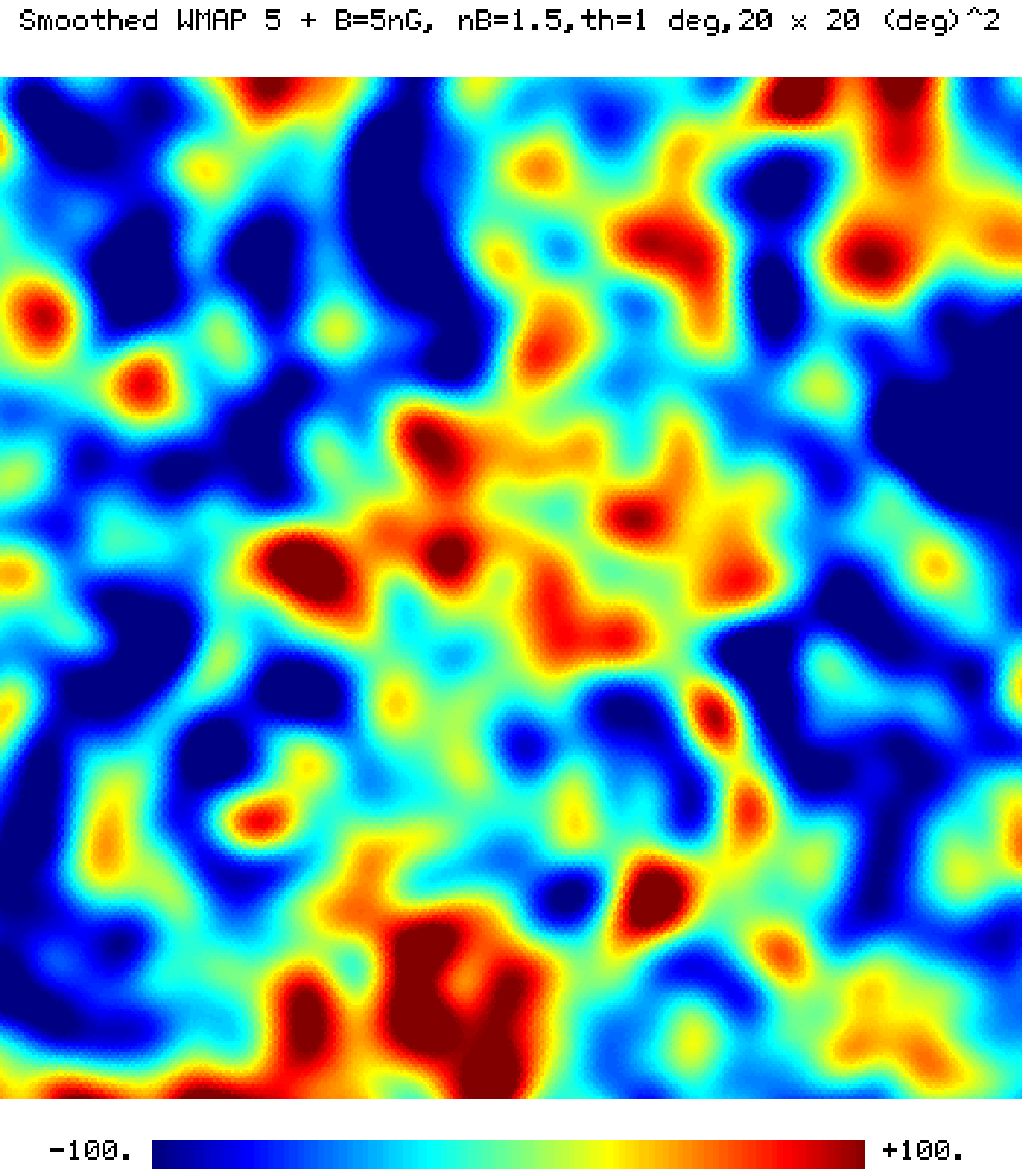}
\hspace{0.7cm}
\includegraphics[height=6.2cm]{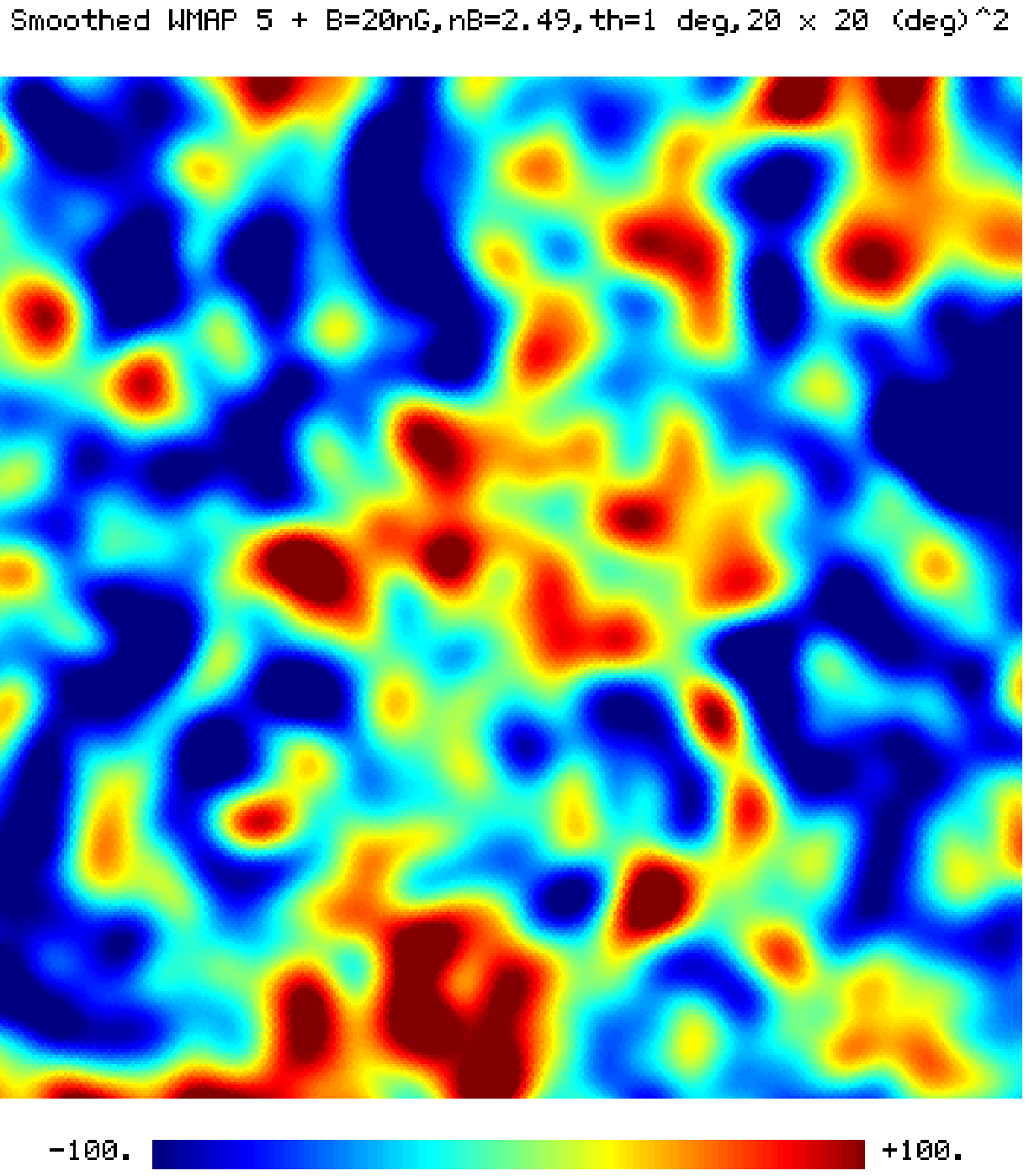}
\caption[a]{$20^{\circ}\times 20^{\circ}$ fields of the temperature maps (in $\mu$K) in gnomonic projection for the best fit WMAP 5 model without a magnetic field (left), with a magnetic field $B_{\mathrm{L}}=5$ nG and $n_{\mathrm{B}}=1.5$ (center) and with a magnetic field $B_{\mathrm{L}}=20$ nG and $n_{\mathrm{B}}=2.49$ (right) smoothed assuming a Gaussian beam size $\theta_{\mathrm{FWHM}}=1^{\circ}$. }
\label{figmap4}      
\end{figure}

Smoothing the maps with a Gaussian beam of size $\theta_{\mathrm{FWHM}}=1^{\circ}$ results in maps of lower resolution. This is equivalent to going from the resolution of the Planck experiment to WMAP.
 The resulting temperature maps are shown in Fig. \ref{figmap4}.
As it can be appreciated from Fig. \ref{figmap4}, this resolution the effect of even a considerably strong magnetic field is hardly visible.

Applying the  HOTSPOT routine to the  smoothed maps results in a considerably smaller number of local extrema. The trend of the magnetic field increasing the number of extrema is still observable. In this case, however,  the 
increase is less pronounced, as can it be argued from Table 2.
\begin{table}[!ht]
\begin{center}
\begin{tabular}{||l|c|c|c|c|c|c||}
\hline
Model&type& number&mean($\mu$K)&var($\mu$K)$^2$&skew& curt\\
\hline
$B_{\mathrm{L}}=0$ & Max& 7274 & 63.005& 2715.5& -1.0426 & -0.1742\\
$B_{\mathrm{L}}=5$ nG, $n_{\mathrm{B}}=1.5$& Max & 7298 &  63.555 & 2731.2& -1.0484 & -0.1676\\
$B_{\mathrm{L}}=20$ nG, $n_{\mathrm{B}}=2.49$&Max&7606&66.198 & 2868.1&-0.1051 &-0.1843\\
\hline
WMAP 5 ILC & Max& 8066& 65.597& 2691.1&-1.1370&-1.1617\\
\hline\hline
$B_{\mathrm{L}}=0$ &  Min & 7252& -63.482&2778.3 & 0.9763& -0.3041\\
$B_{\mathrm{L}}=5$ nG , $n_{\mathrm{B}}=1.5$&Min & 7277 & -63.809& 2796.3& 0.9770& -0.3039\\
$B_{\mathrm{L}}=20$ nG, $n_{\mathrm{B}}=2.49$&Min&7600& -66.509&2913.2 & 0.9955& -0.2846\\
\hline
WMAP 5-year ILC &Min&8120&-64.903&2647.6&1.1306&-1.6289\\
\hline\hline
\end{tabular}
\caption{Statistics of the temperature values (number, mean, variance, skewness and curtosis) of the local extrema (maxima/minima) of the simulated temperature maps smoothed with a gaussian beam with  $\theta_{\mathrm{FWHM}}=1^{\circ}$ for the WMAP 5 bestfit model with and without a magnetic field with field strength $B_{\mathrm{L}}$ and spectral index $n_{\mathrm{B}}$.  For comparison the results of the Internal Linear Combination (ILC) map of WMAP 5 has been included.}
\end{center}
\label{tabmap2}
\end{table}
For comparison the statistics of the temperature values of the local extrema of 
the 5-year Internal Linear Combination (ILC) map provided by the WMAP collaboration has been included. This was obtained applying the HOTSPOT routine to the data file provided at the LAMBDA web site  \cite{LAMBDA} .
The ILC map is a weighted linear combination over the smoothed temperature maps obtained from each of the 5 frequency channels.  It has minimal Galactic foreground contribution and is assumed to give a reliable signal of the cosmic microwave background at angular scales greater than $10^{\circ}$. Its  resolution parameter is 9, which is equivalent to $N_{\mathrm{side}}=512$ and it has a resolution of $1^{\circ}$.
The number of local extrema in the simulated maps and the ILC map are of the same order.
Comparing naively the mean temperature of the hot spots of the simulated maps with the ILC map seems to indicate that actually models with a non vanishing magnetic field  are  a better fit to the data.
Likewise the models with a non vanishing magnetic field have a lower mean temperature of the cold spots which is in better accordance with ILC data then the WMAP 5-year best fit model without a magnetic field.
To draw such  conclusions one should use the WMAP 5-year temperature maps in the different frequency channels and perform a more thorough exploration of  the 
parameter space of the models using Monte Carlo simulations \cite{hotcold5}. 
This is beyond the scope of the current article \cite{mk}.
The comparison with the ILC map was used for illustrative purposes showing that the local extrema of the simulated maps are good indicators of the one point statistics of the observed temperature map.

Thus in conclusion, we see that primordial magnetic fields do have an effect on the structure of the temperature and polarization maps of the cosmic microwave background. These effects are clearly visible for rather large values of the primordial magnetic field at high resolution as expected to be provided by the Planck experiment. To quantify the effect of a magnetic field on the temperature map the statistics of the local extrema has been used. On the basis of the reported 
results, the Planck experiment provides an interesting perspective of constraining magnetized models using temperature maps.

\renewcommand{\theequation}{9.\arabic{equation}}
\setcounter{equation}{0}
\section{Concluding considerations}
\label{sec9}
In the present paper a viable strategy for the complete calculation of the magnetized polarization observables has been 
presented. To achieve this goal it was mandatory to employ, within the different branches of the spectrum 
of plasma excitations, the appropriate description. In previous studies this aspect was never discussed in detail 
and never applied. The reported results are a necessary physical step for any 
consistent calculation of the magnetized polarization observables either at the semi-analytical or at 
the fully numerical level.

At a more operational level, 
the accurate calculation of the TE, EE and BB angular power spectra beyond the MHD approximation, will be of upmost importance for our program of parameter estimation which is already under way.
In front of difficult problems, such as the one we are scrutinizing, there could be
the tendency of substituting words for actions. Even if this tendency could be potentially rewarding in the short run, it will be the bane for future comparisons with observational data. Our primary objective, in this respect, is very modest: we want 
 to develop a sound approach where all the relevant physical effects (i.e. 
 dispersive and not dispersive) are realistically modeled.  Absent this necessary step, any parameter estimation would be forlorn and theoretically biased. 

The second line of developments of the present 
considerations imply a strategy for deducing the typical signatures of a Faraday-induced B-mode.
The reported results suggests that an unambiguous way of pinning down the 
Faraday-induced B-mode is to look at the scaling properties of the measured BB angular power spectrum
as a function of the frequency and to compare both with computed BB power spectrum as well as 
with the other power spectra of the cross-correlations (i.e. EE and TT angular power spectra).  

Realistic simulations of maps including the effects of large-scale magnetic fields have been derived and presented. For experiments with high (nominal) resolution (such as the one of the forthcoming Planck explorer mission) the effect of a primordial magnetic field on the temperature and polarization maps is distinguishable. Lowering a bit  the resolution changes the structure of the temperature maps but, according to our preliminary results, does not totally jeopardize the possibility of reading-off the effects of (sufficiently strong) pre-decoupling magnetic fields.
Polarization maps  have been also studied for different magnetic field parameters. For purposes 
of presentation, only few examples have been illustrated.
As a general rule of thumb, a magnetic field increases the number of hot and cold spots in the simulated temperature maps. 
This has been quantified in the determination of local extrema and their one-point statistics.
There is a potential correlation between the number of local extrema and their corresponding temperatures. The differences between the mean temperatures for the hot and cold spots, respectively,  for different values of the magnetic field strength and its spectral index are larger at a higher resolution of the experiment, or in other words for a smaller beam size. 
Our results open interesting perspectives for the 
Planck explorer mission: the m$\Lambda$CDM scenario (as well as any of its non-minimal extensions) 
can be analyzed by studying the  one- and two-point statistics of the temperature maps 

The original spirit of our endeavors has been, also in the past, to bring the treatment 
of magnetized CMB anisotropies to the same standards typical of the conventional case. At the moment, a 
number of preparatory tasks for the forthcoming Planck experiment has been completed. The 
fulfilled tasks permit, at the moment, explicit and accurate calculations of the effects of pre-decoupling magnetic fields on the various CMB observables within a faithful dynamical framework.

There are other intriguing physical problems which are linked to the ones we discussed in this paper. 
An example along this direction is already apparent from the scaling 
properties (with frequency) of the temperature and polarization autocorrelations 
which has been suggested, in this paper, as a model-independent diagnostic 
of magnetized birefringence. It is here important to stress 
that the experimenters will have to apply a mask selecting those regions with 
low contamination of Faraday rotation due not to the pre-decoupling magnetic field 
but rather to the galactic magnetic field.  It seems that here there will be, in the months 
to come, a rather interesting interplay between the latter problem and the recent findings of cosmic ray experiments 
\footnote{The latest analyses of the AUGER experiment 
demonstrated a correlation between the arrival directions of cosmic rays 
with energy above $6\times10^{19}$ eV and the positions of active galactic 
nuclei within $75$ Mpc \cite{auger1}.  At smaller energies it has been 
convincingly demonstrated \cite{auger2} that overdensities on windows of $5$ deg radius (and for energies $10^{17.9} \mathrm{eV} < E < 10^{18.5} \mathrm{eV}$) are compatible 
with an isotropic distribution. Thus,  in the highest energy domain (i.e. energies larger  than $60$ EeV),
cosmic rays are not appreciably deflected: within a cocoon of $70$ Mpc the intensity of the (uniform) component of the putative magnetic field should be smaller than the nG.  On a theoretical ground, the existence of much larger  magnetic fields (i.e. ${\mathcal O}(\mu\mathrm{G})$) cannot be justified already  if
the correlation scale is of the order of $20$ Mpc.}
as far as the properties of the large-scale magnetic fields in the local Universe 
are concerned. This problem, for the reasons mentioned above,  is also rather interesting for us \cite{mk}.

\section*{Acknowlegments}
 K.E.K.  is supported by the ``Ram\'on y Cajal'' programme 
 and acknowlegdes partial support by Spanish Science Ministry grants 
 FPA2005-04823, FIS2006-05319 and CSD2007-00042.
 \newpage
\begin{appendix}
\renewcommand{\theequation}{A.\arabic{equation}}
\setcounter{equation}{0}
\section{Vlasov-Landau equation in curved space-time}
\label{APPA}
 The Vlasov-Landau 
equation for charged species is rarely written in the presence 
of relativistic fluctuations of the geometry.  In this Appendix 
the Vlasov-Landau approach \cite{vla,lan} (see also \cite{lif})
 will be derived in curved space-time \footnote{Even dedicated discussions of kinetic theory 
 in curved space-times \cite{ber}, do not address the problem of the Vlasov-Landau description in curved backgrounds when curvature inhomogeneities 
 are simultaneously present. This is one of the aims of the present Appendix.} 
for a generic species which we will take to be positively 
charged. The extension to negatively charged species is trivial.

In a curved background, the best pivot variables for 
the Vlasov-Landau approach are the comoving three-momenta 
(very much as it happens in the case of the conventional Boltzmann 
equation which is implemented in standard Boltzmann solvers). 
In the present case the main difference is that the geodesic of a charged 
species is affected  by the electromagnetic fields.  
Consider, to begin with, the conjugate momenta $P^{\alpha}$  and their mass-shell condition for a
generic (massive) species:
\begin{equation}
P^{\alpha} = m u^{\alpha} = m \frac{d x^{\alpha}}{d\lambda},\qquad 
g_{\alpha\beta} P^{\alpha} P^{\beta} = m^2,
\label{VL1}
\end{equation}
where $u^{\alpha}$ is the four-velocity of the generic  species 
and where $\lambda$ denotes, throughout this Appendix, the affine parameter.
The mass-shell condition of Eq. (\ref{VL1}) implies, as it 
has to, $g_{\alpha\beta} u^{\alpha} u^{\beta} =1 $.
From the second relation of Eq. (\ref{VL1}) the three-momentum $p^{i}$ can be introduced in terms of the conjugate momenta as 
$g_{ij} P^{i} P^{j} = - \delta_{ij} p^{i} p^{j}$. The comoving 
three-momentum is then, by definition, $q^{i} = a p^{i}$.
According to Eq. (\ref{VL1}) the components of the conjugate momenta 
and the components of the comoving three-momenta are related as:
\begin{eqnarray}
&&P^{0} = \frac{1}{a^2} \sqrt{m^2 a^2 + q^2}, \qquad  P_{0} = \sqrt{m^2 a^2 + q^2},
\label{VL2}\\
&& q^{i} = a^2 \biggl[ \delta_{j}^{i} - \frac{h_{j}^{i}}{2}\biggr] P^{j},
\label{VL3}
\end{eqnarray}
where $h_{ij}$ is the metric fluctuation defined in Eq. (\ref{M2}).
It is useful to relate the comoving three-momentum related to the comoving 
three-velocity $\vec{v}$ which has been used in the bulk of the paper 
and which arises naturally in the kinetic treatment.  The relation between $\vec{v}$ and $\vec{u}$ is given by:
\begin{equation}
u^{i} = \frac{dx^{i}}{d\lambda} = u^{0} v^{i}, \qquad u^{0} = \frac{d\tau}{d\lambda} = \frac{\gamma(v)}{a}, \qquad \gamma(v) = \frac{1}{\sqrt{1 - v^2}}.
\label{VL4}
\end{equation}
Since, by definition, $ v^{i} = P^{i}/P^{0}$, we shall also have
\begin{equation}
\vec{v} = \frac{\vec{q}}{\sqrt{q^2 + m^2 a^2}}, \qquad 
\vec{q} = m a \gamma(v) \vec{v}.
\label{VL5}
\end{equation}
Having introduced all the relevant variables we can write 
the geodesic for charged particles, namely\footnote{We will consider, for sake of concreteness, the case of a particle with charge $+e$ and mass $m$.}:
\begin{equation}
\frac{d u^{\mu}}{d\lambda} + \Gamma_{\alpha\beta}^{\mu} u^{\alpha} u^{\beta} = \frac{e}{m} F^{\mu\alpha} u_{\alpha},
\label{VL6}
\end{equation}
which becomes 
\begin{eqnarray}
&& \frac{ d \vec{v}}{d\tau} + {\mathcal H} \frac{\vec{v}}{\gamma^2}  = \frac{e}{m a \gamma^3}( \vec{E} + \vec{v}\times \vec{B}),
\label{VL7}\\
&& \frac{d u^{0}}{d\tau} + {\mathcal H} u^{0} (1 + v^2) = \frac{e}{m a^2}\vec{E}\cdot\vec{v},
\label{VL8}
\end{eqnarray}
where $\vec{E} = a^2 \vec{{\mathcal E}}$ and $\vec{B} = a^2 \vec{{\mathcal B}}$. Concerning Eqs. (\ref{VL7}) and (\ref{VL8}) two comments 
are in order:
\begin{itemize}
\item{} even if we used rescaled electromagnetic fields, the evolution 
equations are not the ones we would have in flat space-time;
\item{} Eq. (\ref{VL8}) is implied by Eq. (\ref{VL7}) once 
we recall that, by definition, $u^{0} = \gamma(v)/a$.
\end{itemize}
The rationale for the first comment is that, of course, the particles
are massive, in our case and the very presence of the mass 
breaks Weyl invariance. Notice, however, that by using the definition 
of the comoving three-momentum $\vec{q}$ in terms of the comoving 
three-velocity $\vec{v}$ (i.e. Eq. (\ref{VL5})), Eq. (\ref{VL7}) can be 
written as 
\begin{equation}
\frac{d\vec{q}}{d\tau} = e ( \vec{E} + \vec{v} \times \vec{B}).
\label{VL9}
\end{equation}
Now, in the ultrarelativistic limit $\vec{v} = \vec{q}/|\vec{q}|$ (and 
the evolution equations would have the same flat-space-time 
form). In the non-relativistic limit $\vec{v}= \vec{q}/(m a)$ 
(and conformal invariance is now broken). Clearly, in the 
problem addressed in the present paper the electrons and 
the ions  are always non-relativistic throughout all the stages 
of the calculation.  
Recalling that, to first order in the synchronous fluctuations 
of the geometry, 
\begin{equation}
\Gamma_{ij}^{0} = {\mathcal H} \delta_{ij}  - \frac{1}{2}( h_{ij}' + 2 {\mathcal H} h_{ij}),\qquad 
\Gamma_{i0}^{j} = {\mathcal H} \delta_{i}^{j} - \frac{1}{2}{h_{i}^{j}}',
\label{VL10}
\end{equation}
Eq. (\ref{VL9}) gets modified as 
\begin{equation}
\frac{d q^{i}}{d\tau} = e ( E^{i} + \epsilon^{k j i} v_{k} B_{j}) + \frac{q^{j}}{2} {h^{i}_{j}}', \qquad 
q^{i} = m a v^{j} \gamma(v)\biggl[\delta^{i}_{j} - \frac{h_{j}^{i}}{2}\biggr]. 
\label{VL11}
\end{equation}
The Vlasov-Landau equation can then be written as:
\begin{equation}
\frac{\partial f}{\partial \tau} + v^{i} \frac{\partial f}{\partial x^{i}} + \biggl[\frac{{h_{j}^{i}}'}{2} q^{j} + 
e ( E^{i} + \epsilon^{k j i} v_{k} B_{j})\biggr]
\frac{\partial f}{\partial q^{i}} 
= {\mathcal C}_{\mathrm{coll}}.
\label{VL12}
\end{equation}
By taking the integral of the right hand and of the left hand sides of Eq. (\ref{VL12}) we do obtain the 
canonical form of the continuity equation 
\begin{equation}
\frac{\partial n}{\partial \tau} + \vec{\nabla} \cdot (n \vec{v}) - \frac{h'}{2} n =0,
\label{VL13}
\end{equation}
where we assumed that the massive species are non-relativistic. The result of Eq. (\ref{VL13}) 
can be easily obtained using integration by parts exactly as in the flat-space analog \cite{spitzer}.
Notice, indeed, that the integral of the collision term over the velocity vanishes and that the 
derivative of the Lorentz force upon the velocity also vanishes.
Equation (\ref{VL13}) can be obtained directly from the covariant conservation of the current, i.e.
\begin{equation}
\nabla_{\mu} j^{\mu} =0, \qquad j^{\mu} = e \tilde{n} u^{\mu}
\label{VL14}
\end{equation}
by recalling that $\nabla_{\mu} j^{\mu} = \partial_{\mu} j^{\mu} + \Gamma_{\mu\alpha}^{\mu} j^{\alpha}$ and 
that $n = a^3 \tilde{n}$.
Neglecting $h_{ij}$ Eq. (\ref{VL12}) can be written, in the non-relativistic limit, as:
\begin{equation}
\frac{\partial f}{\partial \tau} + \vec{v} \frac{\partial f}{\partial \vec{x}} + \frac{e}{m a} (\vec{E} + \vec{v} \times\vec{B}) 
\frac{\partial f}{\partial \vec{v}} = {\mathcal C}_{\mathrm{coll}}.
\label{VL15}
\end{equation}
Equation (\ref{VL15}) describes, for instance, the evolution of  the electrons and ions prior to equality 
and throughout decoupling.  In this case the equilibrium distribution will be Maxwellian. 
At the same time Eq.  (\ref{VL12}) has been deduced without specifying the equilibrium distribution
and it can then be applied in more general terms. Consider, indeed, the situation of massless 
neutrinos. From Eq. (\ref{VL12}) setting $\vec{E} = 0$ and $\vec{B}=0$ we get 
\begin{equation}
\frac{\partial f}{\partial\tau} + \frac{d x^{i}}{d\tau} \frac{\partial f}{\partial x^{i}} + h_{ij}' \frac{q^{i} q^{j}}{2 q} \frac{\partial f}{\partial q} = 0.
\label{VL16}
\end{equation}
For a massless particle $ g_{\alpha\beta} P^{\alpha} P^{\beta} =0$ and $x^{i} = \tau n^{i}$ and $q^{i} = n^{i} q$. Thus, Eq. (\ref{VL16}) can be written as:
\begin{equation}
\frac{\partial f}{\partial \tau} + n^{i} \frac{\partial f}{\partial x^{i}} + \frac{1}{2} h_{ij}' n^{i} n^{j} \frac{\partial f}{\partial \ln{q}} =0.
\label{V17}
\end{equation}
By perturbing  Eq. (\ref{V17}) to first order in the metric inhomogeneities 
(around a Fermi-Dirac distribution $f^{(0)}(q)$)  we obtain:
\begin{equation}
\frac{\partial f^{(1)}}{\partial \tau} + i k \mu f^{(1)} + \biggl[ \frac{\mu^2}{2} (h'+ 6 \xi') - \xi' \biggr] 
\frac{\partial f^{(0)}}{\partial \ln{q}} =0,
\label{V18}
\end{equation}
where $\mu = \hat{k}\cdot \hat{n}$ and the fluctuation of $f$ has been defined as 
$ f(q,k, \mu,\tau) = f^{(0)}(q)[ 1 + f^{(1)}(k,\mu,\tau)]$. 
By now defining the reduced phase space distribution for the neutrinos, i.e. 
\begin{equation}
{\mathcal F}_{\nu}(k, \mu, \tau) = \frac{\int q^3 d q f^{(0)}(q) f^{(1)}(k, \mu,\tau)}{\int q^3 f^{(0)}(q) dq },
\label{V19}
\end{equation}
we can easily obtain the result reported in Eq. (\ref{BZnu}). 
The quantity ${\mathcal F}_{\nu}(k,\mu,\tau)$ is the one usually expanded in Rayleigh series, 
\begin{equation}
{\mathcal F}_{\nu}(k,\mu,\tau) = \sum_{\ell} (-i)^{\ell} {\mathcal F}_{\nu\ell}(k,\tau) P_{\ell}(\mu),
\label{V20}
\end{equation}
where ${\mathcal F}_{\nu\ell}$ are the multipole moments. In the present discussion, the initial
conditions of the neutrino hierarchy are set by solving for the lowest multipoles, i.e. monopole, dipole 
and quadrupole \cite{gk2}.

\renewcommand{\theequation}{B.\arabic{equation}}
\setcounter{equation}{0}
\section{Magnetized adiabatic mode}
\label{APPB}
As discussed in the bulk of the paper, the initial conditions for the full Boltzmann hierarchy are given 
well before equality, i.e. in a regime where the ions, the electrons and the photons are all coupled together 
for different physical reasons. Given the largeness of the Coulomb and Thompson rates in Hubble 
units, it is practical to set initial conditions for the MHD fluid (i.e. the baryon fluid in the conventional 
terminology). The structure of the magnetized adiabatic mode has been derived, in the synchronous 
gauge, in Ref. \cite{gk2}. Previous analyses in different gauges can be found in \cite{mg1,mg2,mg3}
and also in \cite{mg5} (also in the synchronous gauge).  The magnetized adiabatic 
mode is here reported for completeness:
\begin{eqnarray}
 \xi(k,\tau) &=& - 2 C(k) + \biggl[\frac{4 R_{\nu} + 5}{6 ( 4 R_{\nu} + 15)} C(k) + \frac{R_{\gamma} ( 4 \sigma_{\mathrm{B}}(k) - R_{\nu} \Omega_{\mathrm{B}}(k))}{ 6 ( 4 R_{\nu} + 15)}\biggr] k^2 \tau^2,
\label{S1}\\
h(k,\tau) &=& - C(k) k^2 \tau^2 - \frac{1}{36} \biggl[ \frac{8 R_{\nu}^2 - 14 R_{\nu} - 75}{(2 R_{\nu} + 25)(4 R_{\nu} + 15)} C(k) 
\nonumber\\
&+& \frac{R_{\gamma} ( 15 - 20 R_{\nu})}{10( 4 R_{\nu} + 15) ( 2 R_{\nu} + 25)} (R_{\nu}\Omega_{\mathrm{B}}(k) - 4 \sigma_{\mathrm{B}}(k))\biggr] k^4 \tau^4,
\label{S2}\\
\delta_{\gamma}(k,\tau) &=& - R_{\gamma} \Omega_{\mathrm{B}}(k) - \frac{2}{3} \biggl[ C(k) - \sigma_{\mathrm{B}}(k) + \frac{R_{\nu}}{4} \Omega_{\mathrm{B}}(k)\biggr] k^2 \tau^2,
\label{S3}\\
\delta_{\nu}(k,\tau) &=& - R_{\gamma} \Omega_{\mathrm{B}}(k) - \frac{2}{3} \biggl[ C(k) + \frac{R_{\gamma}}{4 R_{\nu}}\biggl( 4\sigma_{\mathrm{B}}(k) - R_{\nu} \Omega_{\mathrm{B}}(k)\biggr)\biggr] k^2 \tau^2,
\label{S4}\\
\delta_{\mathrm{c}}(k,\tau) &=& - \frac{3}{4}R_{\gamma} \Omega_{\mathrm{B}}(k) - \frac{C(k)}{2} k^2 \tau^2,
\label{S5}\\
\delta_{\mathrm{b}}(k,\tau) &=& - \frac{3}{4}R_{\gamma} \Omega_{\mathrm{B}}(k) - \frac{1}{2} \biggl[ C(k) - \sigma_{\mathrm{B}}(k)+ \frac{R_{\nu}}{4} \Omega_{\mathrm{B}}(k) \biggr] k^2\tau^2,
\label{S6}\\
\theta_{\gamma\mathrm{b}}(k,\tau) &=& \biggl[ \frac{R_{\nu}}{4} \Omega_{\mathrm{B}}(k) - \sigma_{\mathrm{B}}\biggr] 
k^2 \tau -\frac{1}{36} \biggl[ 2 C(k) + \frac{R_{\nu} \Omega_{\mathrm{B}}(k) - 4 \sigma_{\mathrm{B}}(k)}{2}\biggr] k^4\tau^3,
\label{S7}\\
\theta_{\nu}(k,\tau) &=& \biggl[ \frac{R_{\gamma}}{R_{\nu}} \sigma_{\mathrm{B}}(k) - \frac{R_{\gamma}}{4} \Omega_{\mathrm{B}}(k)\biggr] k^2 \tau
- \frac{1}{36}\biggl[\frac{2 ( 4 R_{\nu} + 23)}{4 R_{\nu} + 15} C(k) 
\nonumber\\
&+& \frac{R_{\gamma}( 4 R_{\nu} + 27)}{2 R_{\nu} ( 4 R_{\nu} + 15)}( 4 \sigma_{\mathrm{B}}(k) - R_{\nu} \Omega_{\mathrm{B}}(k))\biggr] k^4 \tau^3,
\label{S8}\\
 \theta_{\mathrm{c}}(k,\tau) &=& 0,
\label{S9}\\
\sigma_{\nu}(k,\tau) &=& - \frac{R_{\gamma}}{R_{\nu}} \sigma_{\mathrm{B}}(k) + \biggl[ \frac{4 C(k)}{3( 4 R_{\nu} + 15)} + \frac{R_{\gamma}( 4 \sigma_{\mathrm{B}}(k) - R_{\nu} \Omega_{\mathrm{B}})}{ 2 R_{\nu}(4 R_{\nu} + 15)}\biggr] k^2 \tau^2.
\label{S10}
\end{eqnarray}
Note that, as thoroughly discussed in \cite{gk2,gk3} the curvature fluctuations on comoving 
orthogonal hypersurfaces can be simply expressed in terms of $\xi$ as 
\begin{equation}
{\mathcal R} = \xi + \frac{{\mathcal H} \xi'}{{\mathcal H}^2 - {\mathcal H}'}.
\label{S11}
\end{equation}
The power spectra of curvature inhomogeneities will be assigned as:
\begin{equation}
\langle {\mathcal R}(\vec{k}) {\mathcal R}(\vec{p}) \rangle = \frac{2\pi^2}{k^3} 
{\mathcal P}_{\mathcal R}(k) \delta^{(3)}(\vec{k} - \vec{p}).
\label{S12}
\end{equation}
i.e. in full agreement with the way the magnetic power spectra have been assigned in Eq. (\ref{F0}).
The two-point function of Eq. (\ref{F0}) is completely specified only if the magnetic power 
spectrum is given in explicit terms. We will choose both for the power spectrum of curvature 
perturbations and for the magnetic power spectrum the following power-law 
parametrization:
\begin{equation}
{\mathcal P}_{{\mathcal R}}(k) = {\mathcal A}_{{\mathcal R}} \biggl(\frac{k}{k_{\mathrm{p}}}\biggr)^{n_{\mathrm{s}} -1},\qquad 
{\mathcal P}_{{\mathcal R}}(k) = A_{\mathrm{B}} \biggl(\frac{k}{k_{\mathrm{L}}}\biggr)^{n_{\mathrm{B}} -1},
\label{S13}
\end{equation}
where $k_{\mathrm{p}}$ and $k_{\mathrm{L}}$ are the two pivot scales, $n_{\mathrm{s}}$ and $n_{\mathrm{B}}$ are the two 
spectral indices. In the conventional $\Lambda$CDM scenario the amplitude 
of curvature perturbations is customarily given in terms of ${\mathcal A}_{{\mathcal R}}$ which is,by definition,
the amplitude of the power spectrum evaluated at the pivot scale $k_{\mathrm{p}}$. In the 
case of the magnetic fields we will assign not the amplitude of the magnetic power spectrum but 
rather the regularized field. This technical aspect will be recalled in the following Appendix since it is required in the estimate of the angular power spectrum of Faraday rotation. 
\renewcommand{\theequation}{C.\arabic{equation}}
\setcounter{equation}{0}
\section{Power spectra of Faraday rotation}
\label{APPC}
If the magnetic field is not homogeneous, the Faraday Rotation rate will be, itself, not a homogenous quantity \cite{rev3}. 
Here the power spectrum will be computed. This calculation was also discussed 
before in the literature (see, in particular, \cite{far8} and also \cite{far5}). 
The notations employed in the 
present paper are slightly different due to the fact that we are presenting here a full calculation of the magnetized polarization observables. 
The Faraday rotation rate can be usefully expanded in ordinary (i.e. scalar) spherical harmonics:
\begin{equation}
\mathrm{F}(\hat{n}) = \sum_{\ell\, m} f_{\ell\,m} Y_{\ell\,m}(\hat{n}).
\label{F1}
\end{equation}
Since the geometry is isotropic the ensemble average of the $f_{\ell\,m}$ must not depend 
upon $m$ but only upon $\ell$ and, consequently, on symmetry ground we can expect that 
\begin{equation}
\langle f_{\ell \, m}^{*} f_{\ell' \, m'}\rangle = C_{\ell}^{(\mathrm{F})} \delta_{\ell \,\ell'} \delta_{m \,m'}.
\label{F2}
\end{equation}
By using the addition theorem of ordinary spherical harmonics  the two-point function of the Faraday rotation rate can be 
written as 
\begin{equation}
\langle \mathrm{F}(\hat{n}) \mathrm{F}(\hat{m}) \rangle = \sum_{\ell} \frac{(2\ell + 1)}{4\pi} C_{\ell}^{(\mathrm{F})} P_{\ell}(\hat{n}\cdot \hat{m}).
\label{F3}
\end{equation}
Equations (\ref{F2}) and (\ref{F3}) are only meaningful if 
$C_{\ell}^{\mathrm{F}}$ is computed in terms of the magnetic power spectrum.
To achieve this goal it is practical to expand the magnetic field not in 
scalar spherical harmonics but rather in vector spherical harmonics \cite{var,bie}.
In the latter framework it is possible to find the vector analog of the well known Rayleigh 
expansion. As thoroughly discussed in \cite{bie} this technique is 
rather well established both in the study of electromagnetic processes \cite{bie} as well
as in nuclear physics \cite{blatt}. The vector analog of the Rayleigh 
expansion can be written, in the present case, as:
\begin{equation}
\vec{B}(\vec{k}) e^{ - i k \mu \tau_{0}} =\sum_{\ell\, m} \sum_{\alpha} g_{\ell \, m}^{(\alpha)}(k,\mu) \vec{Y}_{\ell\, m}^{(\alpha)}(\hat{n}),
\label{F4}
\end{equation}
where, as usual $\mu = \hat{k}\cdot\hat{n}$. In Eq. (\ref{F4}) $\alpha$ is the polarization and 
$\vec{Y}_{\ell\, m}^{(\alpha)}$ are the vector spherical harmonics or, for short, vector 
harmonics in what follows. They can be written as \cite{var,bie}:
\begin{equation}
\vec{Y}_{\ell\, m}^{(1)}(\hat{n}) = \frac{ \vec{\nabla}_{\hat{n}} Y_{\ell \, m}(\hat{n})}{\sqrt{\ell (\ell + 1)}}, 
\qquad \vec{Y}_{\ell\, m}^{(0)}(\hat{n}) = -\frac{i(\hat{n} \times \vec{\nabla}_{\hat{n}} )Y_{\ell \, m}(\hat{n})}{\sqrt{\ell (\ell + 1)}},\qquad \vec{Y}^{(-1)}_{\ell\, m} = \hat{n} Y_{\ell\, m}(\hat{n}),
\label{F5}
\end{equation}
where $Y_{\ell\, m}(\hat{n})$ are the usual spherical harmonics. Sometimes $\vec{Y}_{\ell\, m}^{(1)}(\hat{n}) $ and 
$\vec{Y}_{\ell\, m}^{(0)}(\hat{n})$ are called, respectively, the electric and the magnetic multipoles. We shall 
avoid this terminology which could be potentially confusing.

 Equation (\ref{F4}) holds for 
a generic vector, we do know that the magnetic field is transverse and, therefore, the expansion
coefficients are more than needed. Furthermore, it should be borne in mind that, in the Faraday 
rotation rate $\mathrm{F}(\hat{n})$ the magnetic field appears to be projected over 
the direction $\hat{n}$. The two previous observations immediately demand that:
\begin{equation}
\hat{n} \cdot \vec{Y}^{(1)}_{\ell\, m}(\hat{n}) = \hat{n}\cdot \vec{Y}^{(0)}_{\ell\, m}(\hat{n}) =0, \qquad 
\hat{k}\cdot \vec{B}(\vec{k})=0.
\label{F6}
\end{equation}
As a consequence of Eq. (\ref{F6}), the multipolar expansion of Eq. (\ref{F4})  simplifies:
\begin{equation}
\hat{n} \cdot \vec{B}(\vec{k}) e^{- i k \mu \tau_{0}} = \sum_{\ell\, m}g_{\ell\, m}(k,\mu) \,\hat{n}\cdot \vec{Y}_{\ell\, m}^{(-1)}(\hat{n}),
\label{F7}
\end{equation}
where $g_{\ell\, m}(k,\mu)$ in Eq. (\ref{F7}) is the only non-vanishing coefficient of the multipolar 
expansion and it is given by \cite{var}:
\begin{eqnarray}
g_{\ell\, m}(k,\mu) &=& \frac{4 \pi (-i)^{\ell -1}}{2 \ell + 1} \{ \sqrt{\ell (\ell + 1)} [ j_{\ell + 1}(k \tau_{0}) + 
j_{\ell -1}(k\tau_{0})] \vec{B}(\vec{k}) \cdot \vec{Y}_{\ell\, m}^{(1)*}(\hat{k}) 
\nonumber\\
&-& 
[(\ell + 1) j_{\ell +1}(k\tau_{0}) - \ell j_{\ell -1}(k\tau_{0})]   \vec{B}(\vec{k}) \cdot \vec{Y}_{\ell\, m}^{(-1)*}(\hat{k})\}.
\label{F8}
\end{eqnarray}
Once more, the magnetic field is transverse (see, e.g.,  Eq. (\ref{F6})). Thus, since 
$\vec{Y}_{\ell\, m}^{(-1)*}(\hat{k})$ is proportional to $\hat{k}$, the last  term in Eq. (\ref{F8}) 
is identically zero. Recalling the well known recurrence relations 
of (spherical) Bessel functions \cite{abr,grad}, 
\begin{equation}
j_{\ell + 1}(k \tau_{0}) + j_{\ell -1}(k\tau_{0}) = \frac{2 \ell + 1}{k\tau_{0}} j_{\ell}(k\tau_{0}),
\label{F9}
\end{equation}
Eq. (\ref{F7})  becomes:
\begin{equation}
\hat{n} \cdot \vec{B}(\vec{k}) e^{- i k \mu \tau_{0}} = 4 \pi \sum_{\ell\, m} \sqrt{\ell (\ell +1)}  \frac{j_{\ell}(k\tau_{0})}{k\tau_{0}} \vec{B}(\vec{k}) \cdot \vec{Y}_{\ell \, m}^{(1)}*(\hat{k}).
\label{F10}
\end{equation}
From the result of Eq. (\ref{F10}) the $f_{\ell\, m}$ of Eqs. (\ref{F1}) and (\ref{F2}) can be swiftly 
determined and they are:
\begin{equation}
f_{\ell\, m} = \frac{4\pi (-i)^{\ell -1}}{(2\pi)^{3/2}} {\mathcal A} \sqrt{\ell (\ell +1)}\int d^{3}k \frac{j_{\ell}(k\tau_{0})}{k\tau_{0}} \vec{B}(\vec{k}) \cdot \vec{Y}_{\ell\, m}^{(1)}(\hat{k}).
\label{F11}
\end{equation}
Using Eq. (\ref{F0}) together with the orthonormality condition of vector harmonics, i.e. 
\begin{equation}
\int d \Omega_{\hat{k}}  \vec{Y}_{\ell m}^{(1)}(\hat{k}) 
\vec{Y}^{(1)*}_{\ell' m'}(\hat{k}) = \delta_{\ell \ell'} \delta_{m m'},
\label{F12}
\end{equation}
the angular power spectrum of Faraday rotation can be simply expressed as:
\begin{equation}
C_{\ell}^{(\mathrm{F})} = 4\pi {\mathcal A}^2 \ell (\ell +1) \int \frac{d k}{k} {\mathcal P}_{B}(k) \frac{j_{\ell}^2(k\tau_{0})}{k^2 \tau_{0}^2}
\label{F13}
\end{equation}
Since, by definition,   
\begin{equation}
j_{\ell}(k\tau_{0}) = \sqrt{\frac{\pi}{2 k\tau_{0}}} J_{\ell + 1/2}(k\tau_{0}), \qquad 
{\mathcal P}_{\mathrm{B}}(k) = A_{\mathrm{B}} \biggl(\frac{k}{k_{\mathrm{L}}}\biggr)^{n_{\mathrm{B}}-1},
\label{F14}
\end{equation}
Eq. (\ref{F13}) can also be written as:
\begin{eqnarray}
C_{\ell}^{(\mathrm{F})} &=& 2\pi^2 \ell (\ell + 1) {\mathcal A}^2 A_{\mathrm{B}} \biggl(\frac{k_{0}}{k_{\mathrm{L}}}\biggr)^{n_{\mathrm{B}} -1} {\mathcal I}(n_{\mathrm{B}}, \ell),
\label{F15}\\
{\mathcal I}(n_{\mathrm{B}},\ell) &=& \int_{0}^{\infty} x_{0}^{n_{\mathrm{B}} - 5} J_{\ell + 1/2}^2(x_{0}) = 
\frac{1}{2 \sqrt{\pi}} \frac{\Gamma\biggl(\frac{5 - n_{\mathrm{B}}}{2}\biggr) \Gamma\biggl( \ell + \frac{n_{\mathrm{B}}}{2} - \frac{3}{2}\biggr)}{\Gamma\biggl(\frac{6 - n_{\mathrm{B}}}{2}\biggr) \Gamma\biggl(\frac{7}{2} + \ell - \frac{n_{\mathrm{B}}}{2}\biggr)},
\label{F16}
\end{eqnarray}
where we used that $x_{0} = k \tau_{0}$ and that $k_{0} = \tau_{0}^{-1}$.  Note that 
the integral of Eq. (\ref{F16}) converges for $-1 < n_{\mathrm{B}} < 5$ so there is no need 
of an ultraviolet cut-off \footnote{This is also what happens, for instance, in the case 
of the explicit calculation of the Sachs-Wolfe plateau where a cut-off is not usually imposed 
because the integral is convergent in the physical range of spectral indices.}.
To make the expressions even more explicit we will adopt the following strategy:
\begin{itemize}
\item{} first we will trade $A_{\mathrm{B}}$ for the regularized magnetic field intesity 
$B_{\mathrm{L}}$;
\item{} second we will refer the frequency of the channel $\nu$ to the frequency 
of the maximum of the CMB emission.
\end{itemize}
Complying with the first step we have that:
\begin{eqnarray}
&& A_{\mathrm{B}} = \frac{(2\pi)^{n_{\mathrm{B}} -1}}{\Gamma\biggl(\frac{n_{\mathrm{B}} -1}{2}\biggr)}B_{\mathrm{L}}^2,\qquad n_{\mathrm{B}} > 1,
\label{F17}\\
&& A_{\mathrm{B}} = \frac{1 - n_{\mathrm{B}}}{2} \biggl(\frac{k_{0}}{k_{\mathrm{L}}}\biggr)^{1 - n_{\mathrm{B}}} B_{\mathrm{L}}^2,\qquad n_{\mathrm{B}} < 1.
\label{F18}
\end{eqnarray}
Equation (\ref{F17}) adopting a Gaussian window function, while Eq. (\ref{F18}) is derived by using a 
step function regularization cutting, effectively, all the modes with $k < k_{0}$ \cite{gk2}. 

The maximum of the CMB emission can be easily derived by maximizing the energy spectrum of the CMB, i.e. 
\begin{equation}
\Omega_{\gamma}(x) = \frac{1}{\rho_{\mathrm{c}}} \frac{d\rho_{\gamma}}{d\ln{k}}= 
\frac{15}{\pi^4} \Omega_{\gamma0} \frac{x^4}{e^{x}  -1}, \qquad 
\Omega_{\gamma 0} = 2.47 \times 10^{-5},
\label{F19}
\end{equation}
where $ x = k/T_{\gamma 0}$. By taking the derivative of $\Omega_{\gamma}(x)$ we can find the maximum:
\begin{equation}
\frac{ d \Omega_{\gamma}}{dx} = -  \frac{15}{\pi^4}\Omega_{\gamma 0} \frac{ x^3 [ 4 e^{x} ( x - 4)]}{(e^{x} -1)} =0, \qquad x_{\mathrm{max}} = 3.920
\label{F20}
\end{equation}
From Eq. (\ref{F20}) we do have that 
\begin{equation}
\nu_{\mathrm{max}} = \frac{x_{\mathrm{max}}}{2\pi} T_{\gamma 0} = 222.617 \, \mathrm{GHz}.
\label{F21}
\end{equation}
Consequently, the final formulas for $ C_{\ell}^{(\mathrm{F})}$ can be expressed as:
\begin{eqnarray}
&& C_{\ell}^{(\mathrm{F})} = \overline{{\mathcal C}}\, {\mathcal Z}^{(\mathrm{F})}_{\ell}(n_{\mathrm{B}}), \qquad 
n_{\mathrm{B}} > 1,
\label{F24}\\
&& C_{\ell}^{(\mathrm{F})} = \overline{{\mathcal C}} \,\overline{{\mathcal Z}}^{(\mathrm{F})}_{\ell}(n_{\mathrm{B}}), \qquad 
n_{\mathrm{B}} < 1,
\label{F25}
\end{eqnarray}
where $\overline{{\mathcal C}}$, $ {\mathcal Z}^{(\mathrm{F})}_{\ell}(n_{\mathrm{B}})$ and $\overline{{\mathcal Z}}^{(\mathrm{F})}_{\ell}(n_{\mathrm{B}})$ turn out to be:
\begin{eqnarray}
&& \overline{{\mathcal C}} = \frac{3 \pi^{9/2}}{10 e^2} x_{\mathrm{max}}^{-4} \overline{\Omega}_{\mathrm{BL}} 
\biggl(\frac{\nu}{\nu_{\mathrm{max}}}\biggr)^{-4} = 30.03 \,\,\overline{\Omega}_{\mathrm{BL}}
\biggl(\frac{\nu}{\nu_{\mathrm{max}}}\biggr)^{-4}, \qquad \overline{\Omega}_{\mathrm{BL}} = \frac{B_{\mathrm{L}}^2}{8\pi 
\overline{\rho}_{\gamma}},
\label{F26}\\
&& {\mathcal Z}^{(\mathrm{F})}_{\ell}(n_{\mathrm{B}}) = \biggl(\frac{k_{0}}{k_{\mathrm{L}}}\biggr)^{n_{\mathrm{B}} -1} \frac{\ell (\ell + 1) (2\pi)^{n_{\mathrm{B}} -1}}{\Gamma\biggl(\frac{n_{\mathrm{B}} -1}{2}\biggr)}  \frac{\Gamma\biggl(\frac{5 - n_{\mathrm{B}}}{2}\biggr) \Gamma\biggl( \ell + \frac{n_{\mathrm{B}}}{2} - \frac{3}{2}\biggr)}{\Gamma\biggl(\frac{6 - n_{\mathrm{B}}}{2}\biggr) \Gamma\biggl(\frac{7}{2} + \ell - \frac{n_{\mathrm{B}}}{2}\biggr)},
\label{F27}\\
&& \overline{{\mathcal Z}}^{(\mathrm{F})}_{\ell}(n_{\mathrm{B}}) = \biggl(\frac{1 -n_{\mathrm{B}}}{2} \biggr)  \frac{\ell(\ell + 1)\Gamma\biggl(\frac{5 - n_{\mathrm{B}}}{2}\biggr) \Gamma\biggl( \ell + \frac{n_{\mathrm{B}}}{2} - \frac{3}{2}\biggr)}{\Gamma\biggl(\frac{6 - n_{\mathrm{B}}}{2}\biggr) \Gamma\biggl(\frac{7}{2} + \ell - \frac{n_{\mathrm{B}}}{2}\biggr)}.
\label{F28}
\end{eqnarray}

\begin{equation}
C_{\ell}^{(\mathrm{BB})} = N_{\ell}^2 \sum_{\ell_{1},\ell_{2}} N_{\ell_{2}}^2 {\mathcal K}(\ell,\ell_{1}, \ell_{2})^2 \frac{( 2\ell_{1} + 1) ( 2 \ell_{2} +1)}{4\pi ( 2 \ell + 1)} C_{\ell_{2}}^{(\mathrm{EE})}  C_{\ell_{1}}^{(\mathrm{F})} [ {\mathcal C}^{\ell 0}_{\ell_{1} 0 \ell_{2} 0}]^2,
\label{CBBF}
\end{equation}
where:
\begin{eqnarray}
&&{\mathcal K}(\ell,\ell_{1}, \ell_{2})= - \frac{1}{2}[L^2 + L_{1}^2 + L_{2}^2 
- 2 L_{1} L_{2} - 2 L_{1} L + 2 L_{1} - 2 L_{2} - 2 L],
\label{FOR1}\\
&& L = (\ell + 1)\ell,\qquad L_{1} = (\ell_{1} + 1)\ell_{1},\qquad
L_{2} = (\ell_{2} + 1)\ell_{2},
\label{FOR2}\\
&& N_{\ell}= \sqrt{\frac{2 (\ell -2)!}{(\ell + 2)!}},\qquad  N_{\ell_2}= \sqrt{\frac{2 (\ell_{2} -2)!}{(\ell_{2} + 2)!}},
\label{FOR3}
\end{eqnarray}
In Eq.  (\ref{FOR1})
${\mathcal C}^{\ell\, 0}_{\ell_{1}\, 0\, \ell_{2}\,0}$ is given by:
\begin{eqnarray}
&&{\mathcal C}^{\ell\, 0}_{\ell_{1}\, 0\, \ell_{2}\,0}=0,\qquad \ell + \ell_{1} + \ell_{2} = 2 n + 1,
\nonumber\\
&& {\mathcal C}^{\ell\, 0}_{\ell_{1}\, 0\, \ell_{2}\,0}= \frac{(-1)^{n - \ell} \sqrt{2 \ell +1} n!}{( n -\ell_{1})! (n - \ell_{2})! (n -\ell)!} 
\sqrt{\frac{(2n - 2 \ell_{1})! ( 2 n - 2 \ell_{2})! ( 2n - 2\ell)!}{(2 n + 1)!}},\,\,\,\,\ell + \ell_{1} + \ell_{2} = 2 n,
\nonumber
\end{eqnarray}
where $n$ is a positive integer. This form of the relevant Clebsch-Gordon coefficient
 is given by \cite{var} (see, in particular, page 251) and it has been 
also used in \cite{far5} where Eq. (\ref{CBBF}) has been firstly derived.
The Clebsch-Gordon coefficient of the previous equation then vanishes unless  $|\ell_{1} - \ell_{2}| \leq \ell < \ell_{1} + \ell_{2}$ 
(triangle inequality) and unless $\ell_{1} + \ell_{2} + \ell$ is an even integer. In the two degenerate cases (i.e. $\ell = \ell_1 + \ell_{2}$ and $\ell = \ell_{1} - \ell_{2}$) 
the expressions become, respectively:
\begin{eqnarray}
&& {\mathcal C}^{\ell_{1} + \ell_{2}\,\,0}_{\ell_{1}\,0\,\ell_{2}\,0} = \frac{(\ell_{1} + \ell_{2})!}{\ell_{1}! \ell_{2}!} \sqrt{\frac{(2\ell_{1})! ( 2 \ell_{2})!}{( 2\ell_{1} + 2 \ell_{2})!}} , 
\nonumber\\
&&{\mathcal C}^{\ell_{1} - \ell_{2}\,\,0}_{\ell_{1}\,0\,\ell_{2}\,0} = (- 1)^{\ell_{2}} \frac{\ell_{1}!}{\ell_{2}! (\ell_{1} - \ell_{2})!}
\sqrt{\frac{( 2\ell_{1})! ( 2\ell_{1} - 2\ell_{2} +1}{(2\ell_{1} + 1)!}}.
\nonumber
\end{eqnarray}
It should be borne in mind that, for some applications, it useful to deal directly with the amplitude 
of the magnetic power spectrum $A_{\mathrm{B}}$. Following the conventions expressed 
by Eq. (\ref{F0}), $A_{\mathrm{B}}$ has dimensions of an energy density. In terms of $A_{\mathrm{B}}$ 
we shall then have 
\begin{eqnarray}
C_{\ell}^{\mathrm{F}} &=& \overline{{\mathcal C}}_{0} \ell (\ell +1)  \frac{\Gamma\biggl(\frac{5 - n_{\mathrm{B}}}{2}\biggr) \Gamma\biggl( \ell + \frac{n_{\mathrm{B}}}{2} - \frac{3}{2}\biggr)}{\Gamma\biggl(\frac{6 - n_{\mathrm{B}}}{2}\biggr) \Gamma\biggl(\frac{7}{2} + \ell - \frac{n_{\mathrm{B}}}{2}\biggr)}\biggl(\frac{k_{0}}{k_{\mathrm{L}}}\biggr)^{n_{\mathrm{B}} -1}
\nonumber\\
\overline{{\mathcal C}}_{0} &=& 2.863 \times 10^{-6} \biggl(\frac{\sqrt{A_{\mathrm{B}}}}{\mathrm{nG}}\biggr)^{2} \biggl(\frac{\nu_{\mathrm{max}}}{\overline{\nu}}\biggr)^{4}.
\label{ABal}
\end{eqnarray}
In terms of the parametrization introduced in Eqs. (\ref{F0}) and (\ref{S13}) it is clear that $A_{\mathrm{B}}$ has exactly 
the dimensions of a magnetic energy density.

\end{appendix}

\end{document}